\documentclass[journal]{IEEEtran}

\IEEEoverridecommandlockouts

\usepackage[colorlinks=true,allcolors=blue]{hyperref}
\usepackage{amsmath,graphicx}
\usepackage{amssymb,amsfonts,mathtools}
\usepackage{algorithmic}
\usepackage{diagbox}
\usepackage{textcomp}
\usepackage{comment}
\usepackage{xcolor}
\usepackage{paralist}
\usepackage{symbol_maosheng}
\usepackage{multirow}
\usepackage{float}
\usepackage[caption=false, font=footnotesize]{subfig}
\usepackage{color,colortbl}
\definecolor{Gray}{gray}{0.9}
\usepackage{array}
\usepackage[export]{adjustbox}
\usepackage[inline]{enumitem}
\DeclarePairedDelimiter\floor{\lfloor}{\rfloor}
\makeatletter
\newcommand{\thickhline}{%
    \noalign {\ifnum 0=`}\fi \hrule height 1pt
    \futurelet \reserved@a \@xhline
}
\renewcommand\baselinestretch{1}
\newcolumntype{"}{@{\hskip\tabcolsep\vrule width 1pt\hskip\tabcolsep}}
\makeatother
   
\usepackage{etoolbox}

\makeatletter
\patchcmd{\@makecaption}
  {\scshape}
  {}
  {}
  {}
\makeatother

\newtheorem{lem}{Lemma}
\newtheorem{prop}{Proposition}
\newtheorem{cor}{Corollary}

\usepackage{setspace}

\title{Simplicial Convolutional Filters}

\author{Maosheng~Yang$^*$,
        Elvin~Isufi$^*$,
        Michael~T.~Schaub$^\dagger$ and 
        Geert~Leus$^\ddagger $
\thanks{Preliminary results were presented at EUSIPCO 2021~\cite{yang2021finite}. $^*$Dept. of Intelligent Systems, Delft University of Technology (m.yang-2, e.isufi-1@tudelft.nl). $^\dagger$Dept. of Computer Science, RWTH Aachen University (schaub@cs.rwth-aachen.de). $^\ddagger$Dept. of Microelectronics, Delft University of Technology (g.j.t.leus@tudelft.nl). MY is supported by the TU Delft AI Labs Programme. MTS received funding from the Ministry of Culture and Science (MKW) of the German State of North Rhine-Westphalia ("NRW Rückkehrprogramm"). }}
 
\begin{document}
\maketitle
\begin{abstract}
We study linear filters for processing signals supported on abstract topological spaces modeled as simplicial complexes, which may be interpreted as generalizations of graphs that account for nodes, edges, triangular faces, etc.
To process such signals, we develop simplicial convolutional filters defined as matrix polynomials of the lower and upper Hodge Laplacians. 
First, we study the properties of these filters and show that they are linear and shift-invariant, as well as permutation and orientation equivariant.
These filters can also be implemented in a distributed fashion with a low computational complexity, as they involve only (multiple rounds of) simplicial shifting between upper and lower adjacent simplices.
Second, focusing on edge-flows, we study the frequency responses of these filters and examine how we can use the Hodge-decomposition to delineate gradient, curl and harmonic frequencies.
We discuss how these frequencies correspond to the lower- and the upper-adjacent couplings and the kernel of the Hodge Laplacian, respectively, and can be tuned independently by our filter designs.
Third, we study different procedures for designing simplicial convolutional filters and discuss their relative advantages.
Finally, we corroborate our simplicial filters in several applications: to extract different frequency components of a simplicial signal, to denoise edge flows, and to analyze financial markets and traffic networks. 
\end{abstract}

\section{Introduction}\label{sec:introduction}
Methods to process signals supported on non-Euclidean domains modeled as graphs have attracted substantial research interest recently.
Most of these graph signal processing (GSP) methods focus on signals supported on nodes, e.g., temperature measurements in weather stations network or EEG signals in a brain network~\cite{shuman2013emerging, ortega2018graph}. 
Using linear shift operators that couple node signals to each other via the edges of a graph, e.g., in terms of an adjacency or a Laplacian matrix, we can design graph filters to process such node signals~\cite{shuman2013emerging, ortega2018graph, sandryhaila2013discrete, sandryhaila2014discrete}.

However, we often encounter signals that are naturally associated with edges or sets of nodes in real-world applications.
For example, blood flow between different areas in the brain \cite{huang2018graph}, water flow in a hydrological network, data flow in a communication network, or traffic flow in a road network~\cite{leung1994traffic, schaub2018flow}.
We typically model these signals as a flow over the edges of a network. Edge flows have also been used in statistical ranking, to describe financial markets, to analyze games, etc. \cite{jiang2011statistical, candogan2011flows, gebhart2021go, mock2021political}. 
Similarly, we may even encounter signals supported on sets of nodes~\cite{huang2015metrics, bick2021higher}. For instance, in a co-authorship network, the number of publications with more than two authors can be seen as such a signal~\cite{patania2017shape}.  

In these cases, rather than focusing on utilizing the relationships between the nodes to process node signals, it can be fruitful to study relations between the node-relationships (edges, higher-order edges) themselves. 
In the case of edge-signals, we want to understand couplings between edges, e.g., mediated through a common incident node (lower adjacency) or because these edges contribute to a triadic relation (upper adjacency).
To account for such relationships, we can model a (network) system as a simplicial complex (SC).
Using this representation we can analyze signals associated to subsets of nodes, i.e., simplicial signals, with algebraic tools via so-called Hodge Laplacians~\cite{lim2015hodge,grady2010discrete,barbarossa2020}, which generalize the familiar graph Laplacians. 

In addition, the Hodge Laplacian admits a Hodge decomposition, which allows for an intuitive physical interpretation of signals supported on SCs \cite{lim2015hodge,grady2010discrete}.
Namely, the Hodge decomposition states that any edge flow can be decomposed into gradient (curl-free), curl (divergence-free) and harmonic components, respectively. For instance, a water flow may contain a non-cyclic component which can be seen as the potential difference between water stations, a locally cyclic component with non-zero curl and a harmonic component being flow-conservative \cite{schaub2018flow}. 

Previous works \cite{barbarossa2020, schaub2021} have established a framework to analyze simplicial signals and focused mostly on low-pass filtering applications. However, general linear filters for simplicial signals have not been considered in detail. 
In this paper, we propose a \emph{simplicial convolutional filter} via the shift-and-sum operation as a matrix polynomial of the Hodge Laplacians to enable a flexible simplicial signal filtering.
Our filter accounts for lower and upper adjacencies in an SC, e.g., the relationships between edges via a common node or a common 2-simplex, and allows to separately filter the three signal components provided by the Hodge decomposition. 

\smallskip\noindent\textbf{Contributions.}
Our three main contributions include:

\textit{1) Simplicial convolution}. 
We study simplicial shifting via the Hodge Laplacians as a basic  operation to propagate signals locally using both lower- and upper-connectivities in an SC. Leveraging this shifting and the shift-and-sum operation, we develop simplicial convolutional filters by aggregating multi-step shifted signals. Their local shifting operation allows a distributed implementation of the filter with a cost linear in the number of simplices. 
We show such filters are linear, shift-invariant, and equivariant to permutations of the labeling and the orientation of simplices. 

\textit{2) Filtering in the spectral domain}. 
Leveraging the simplicial Fourier transform (SFT) \cite{barbarossa2020}, we show that the principles of the convolutional theorem apply to the proposed filter, i.e., the filter output in the frequency domain operates as a point-wise multiplication between the filter frequency response and the SFT of the input signal. We further show how the simplicial frequencies  {act as measures of signal variations w.r.t. the lower and upper adjacencies} and divide them into gradient, curl and harmonic frequencies.
 {More precisely, the eigenmodes associated to these frequencies span the subspaces provided by the Hodge decomposition.} 
Ultimately, this implies that the proposed filter can regulate signals independently in the three subspaces provided by the Hodge decomposition.

\textit{3) Filter Design}. To implement a desired frequency response, we first consider a standard least-squares (LS) approach to design the filter. To avoid the eigenvalue computation, we then consider a grid-based universal design. 
As both strategies may suffer from numerical instability, we propose a numerically more stable Chebyshev polynomial design.

\smallskip\noindent\textbf{Related works.}
The idea of processing of signals defined on manifolds and topological spaces has been discussed in various areas, such as geometry processing~\cite{botsch2010polygon}, and topological data analysis~\cite{wasserman2018topological}. 
For an introduction that is geared more towards a signal processing perspective see~\cite{grady2010discrete,robinson2014topological, barbarossa2020}.

Filtering of simplicial signals has been partly approached from a regularization perspective. 
The works in \cite{schaub2018flow, schaub2021} proposed a regularized optimization framework based on (simplified variants of) the Hodge Laplacian, to promote flow conservation of the resulting estimated edge flows. 
The solution is a low-pass simplicial filter.
The same regularizer was used in~\cite{jia2019graph, schaub2021} to perform edge flow interpolation by exploiting the divergence-free and curl-free behaviors of real-world flows. 
However, these assumptions do not always hold and the filters arising from the considered regularized optimization problems have limited degrees of freedom. 

Filtering simplicial signals has also been analyzed in the Fourier domain, akin to how graph filters are analyzed via the graph Fourier transform \cite{shuman2013emerging}. 
The analogous SFT, defined via the eigendecomposition of the Hodge Laplacian was described in \cite{barbarossa2020}.
The eigenvectors provide a simplicial Fourier basis and the eigenvalues carry a notion of frequency. 

In parallel,  {signal processing techniques have been extended from SCs to cell complexes~\cite{roddenberry2022signal,sardellitti2022topological,sardellitti2021topological}. 
However, the cell filters are essentially of the same form as the simplicial convolutional filters discussed here and in the preliminary conference version of this article~\cite{yang2021finite}. 
Indeed, all the filters studied in the current paper can be generalized to cell complexes, by substituting the appropriate incidence matrices.}
Different neural network architectures on SCs have also been developed to learn from the simplicial data, e.g., \cite{yang2021simplicial, ebli2020simplicial, bodnar2021weisfeiler, bodnar2021weisfeiler_cell, bunch2020simplicial,roddenberry2021principled, roddenberry2019hodgenet}. 
Importantly, the linear operation in these different neural convolutional layers may be understood as a simplicial filter (as discussed here) with different filter parameters.

\smallskip\noindent\textbf{Outline.} We begin by introducing some preliminaries in Section \ref{sec:background}. Then we propose the simplicial convolutional filter in Section \ref{sec:simplicial-filter} and investigate its properties. In Section \ref{sec:spectral-analysis}, we introduce the simplicial Fourier transform. We then analyze the simplicial filter in the spectral domain and study the notion of simplicial frequency. Different filter design methods are discussed in Section \ref{sec:filter-design}. Finally, we use simplicial filters for subcomponent extraction and edge flow denoising, and consider applications to financial market and transportation networks in Section \ref{sec:application-numerical-experiments}.

\section{Simplicial Complexes, Signals and Hodge Laplacians} \label{sec:background}
In this section, we review SCs \cite{lim2015hodge, munkres2018elements} and signals supported on simplices \cite{barbarossa2020, schaub2021}. 
We also introduce the Hodge Laplacians as an algebraic representation of a simplicial complex, that acts as a shift operator for simplicial signals.

\smallskip\noindent\textbf{Simplicial complexes.}
Given a finite set of vertices $\ccalV$, a $k$-simplex $\ccalS^k$ is a subset of $\ccalV$ with cardinality $k+1$. 
A subset of a $k$-simplex $\ccalS^k$ with cardinality $k$ is called a \emph{face} of $\ccalS^k$; hence, $\ccalS^k$ has $k+1$ faces. A \emph{coface} of a $k$-simplex is a simplex $\ccalS^{k+1}$ that includes it. 
A simplicial complex $\ccalX$ is a finite collection of simplices with an inclusion property: for any $\ccalS^k\in\ccalX$ all its faces $\ccalS^{k-1} \subset \ccalS^{k}$ are also part of the simplicial complex, i.e., $\ccalS^{k-1} \in \ccalX$. The order of an SC is the largest order of its simplices. For an SC of order $K$, we collect the $k$-simplices into a set $\ccalX^k=\{\ccalS_1^k,\dots,\ccalS^k_{N_k}\}$ where $N_k = |\ccalX^k|$ is the number of $k$-simplices \cite{lim2015hodge,barbarossa2020}. 

\begin{figure}[!t] 
  \vspace{-3mm}
  \centering
  \subfloat[A simplicial complex.\label{1a}]{%
       \includegraphics[width=0.35\linewidth]{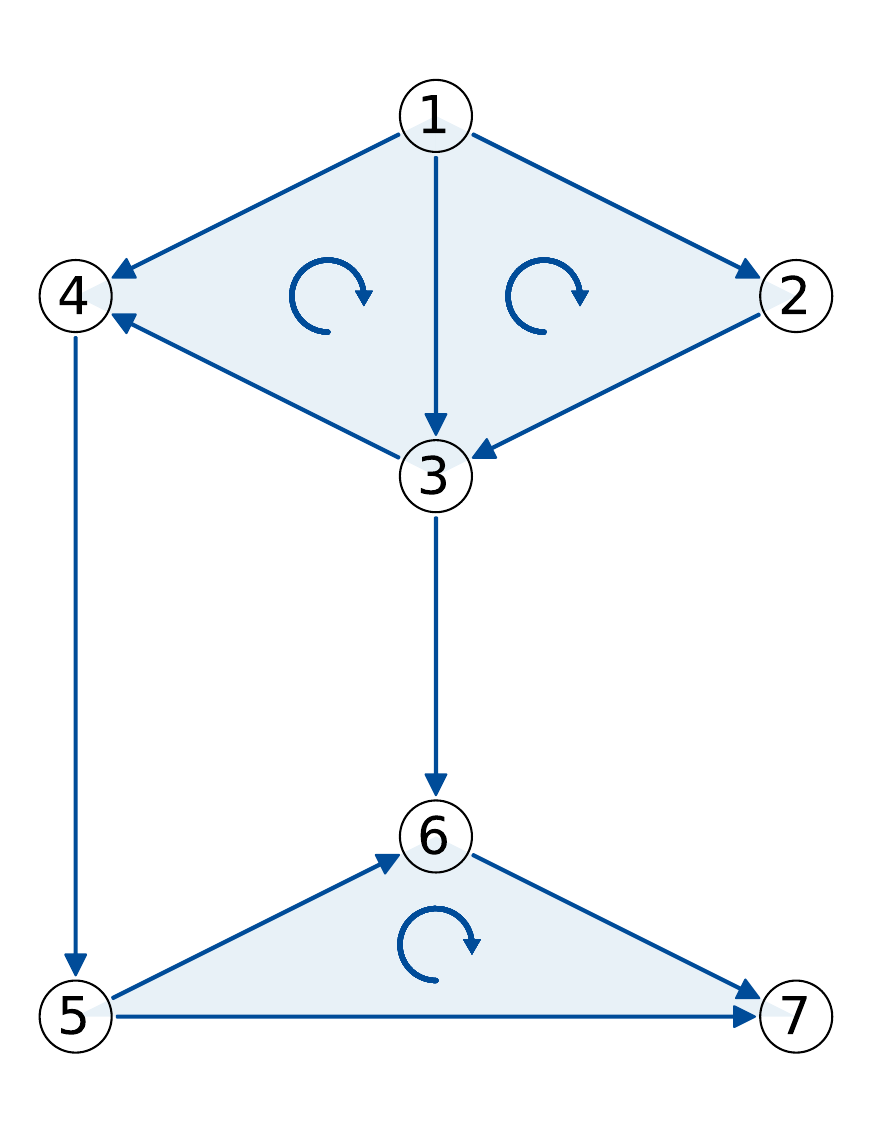}}
  \subfloat[A random edge flow.\label{1b}]{%
        \includegraphics[width=0.35\linewidth]{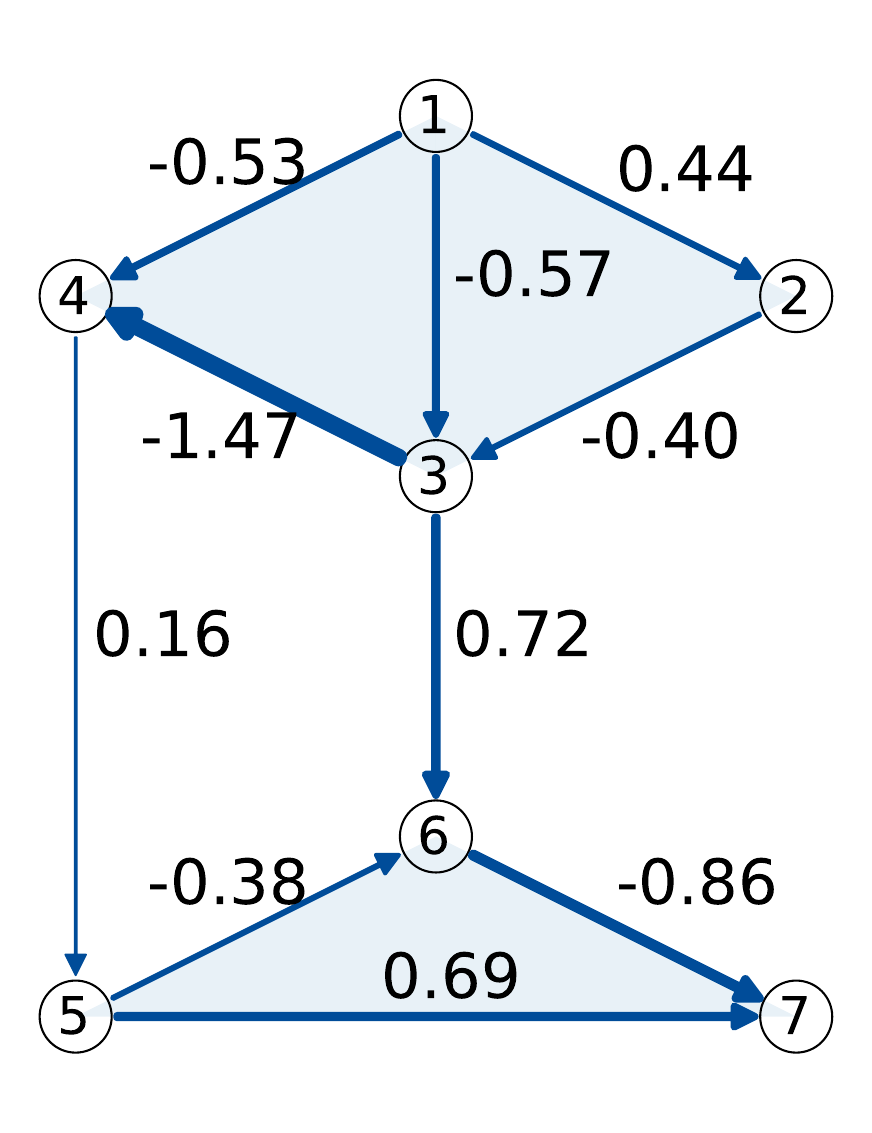}}
  \caption{Simplicial Complexes and Signals. (a): An SC of order 2 containing seven nodes, ten edges and three 2-simplex (the shaded filled triangles). Reference orientations of the simplices are indicate by corresponding arrows (the reference orientation of a node is trivial). (b): An arbitrary edge flow, where a negative flow indicates that the actual flow direction is opposite to the reference orientation and the magnitude is denoted by the edge width.}
  \label{fig:1} 
\end{figure}

Based on its geometric realizations, we call a $0$-simplex a node, a $1$-simplex an edge and a $2$-simplex a (filled) triangle. 
Note that an ``empty triangle'' formed by three vertices and three pairwise relations between them is not a $2$-simplex. 
Henceforth, we refer to 2-simplices as triangles for simplicity. 
A graph is an SC of order $1$ with $N_0$ nodes and $N_1$ edges. 
Fig. \ref{1a} shows an SC of order $2$ including nodes, edges and triangles where edge $\{5,6\}$ has nodes $\{5\}$ and $\{6\}$ as its faces and triangle $\{5,6,7\}$ as its coface.

Two $k$-simplices in an SC are \emph{lower adjacent} if they have a common face and are \emph{upper adjacent} if they are both faces of a common $(k+1)$-simplex. 
Thus, for the simplex $\ccalS_i^k$, we define its \emph{lower neighborhood} $\ccalN^k_{i,{\ell}}$ as the set of its lower adjacent $k$-simplices and its \emph{upper neighborhood} $\ccalN^k_{i,{\rm{u}}}$ as the set of its upper adjacent $k$-simplices. 
For the $i$th edge $\{5,6\}$ in Fig. \ref{1a}, we have that $\ccalN^1_{i,{{\ell}}}=\{\{4,5\},\{3,6\},\{5,7\},\{6,7\}\}$, and the upper neighborhood is $ \ccalN^1_{i,{{\rm u}}} = \{\{5,7\},\{6,7\}\}$.

\smallskip\noindent\textbf{Simplicial signals.} For computational purposes, we fix an (arbitrary) reference orientation (see \cite{lim2015hodge}  {and \cite[p. 5]{schaub2020random}} for more details) for each simplex according to the lexicographical ordering of its vertices. 
Given this reference orientation for $k$-simplices, we define a $k$-\emph{simplicial signal} $\bbs^k = [s^k_1,\dots,s^k_{N_k}]^\top\in\setR^{N_k}$ by attributing the value $s^k_i$ to the $i$th $k$-simplex $\ccalS_i^k$. 
If the signal value $s_i^k$ is positive, then the corresponding signal is aligned with the reference orientation; opposite otherwise.
Fig.~\ref{1b} illustrates an arbitrary edge flow on an SC, in which some flows are aligned with and other are opposite the reference orientation (negative).
For convenience, henceforth, we denote a node signal by $\bbv=[v_1,\dots,v_{N_0}]^\top \in \setR^{N_0}$ and an edge flow by $\bbf=[f_1,\dots,f_{N_1}]^\top \in \setR^{N_1}$.

\smallskip\noindent\textbf{Hodge Laplacians.}
We can describe the relationships between $(k-1)$-simplices and $k$-simplices by the $k$th incidence matrix $\bbB_k \in \setR^{N_{k-1}\times N_k}$, which maps each $k$-simplex to the $(k-1)$-simplices that are its faces; cf. \cite{barbarossa2020, munkres2018elements} for more details. Specifically, $\bbB_1$ is the node-edge incidence matrix, and $\bbB_2$ is the edge-triangle incidence matrix.
By definition, incidence matrices have the property  \cite{lim2015hodge,barbarossa2020}
\begin{equation} \label{eq.boundary-condition}
    \bbB_k \bbB_{k+1} = \mathbf{0}.
\end{equation}
Upon defining the incidence matrices, we can describe an SC $\ccalX$ of order $K$ via the Hodge Laplacians
\begin{equation}
    \bbL_k = \bbB_k^\top\bbB_k + \bbB_{k+1}\bbB_{k+1}^\top, \quad k = 1,\dots,K-1,\\
\end{equation} 
with the graph Laplacian $\bbL_0 = \bbB_1\bbB_1^\top$ and $\bbL_K = \bbB_{K}^\top\bbB_K$. 
The $k$th-\emph{Hodge Laplacian} $\bbL_k$ contains the \emph{lower Laplacian} $\bbL_{k,\ell} \triangleq \bbB_k^\top \bbB_k$ and the \emph{upper Laplacian} $\bbL_{k,\rm{u}} \triangleq \bbB_{k+1}\bbB_{k+1}^\top$. 
The lower Laplacian encodes the lower adjacency relationships between simplices through faces while the upper one encodes the upper adjacency relationships through cofaces. In particular, $\bbL_{1,\ell}$ encodes the edge adjacencies through their incident nodes and $\bbL_{1,\rm{u}}$ through the common triangles that they form.

\section{Simplicial convolutional filters} \label{sec:simplicial-filter}
In this section, we propose a  simplicial convolutional filter based on the Hodge Laplacian. 
We study its basic building block, the simplicial shifting, and show how its local characteristics make it amenable to a distributed implementation.
We then look into the properties of shift-invariance and permutation and orientation equivariance in the simplicial domain.  

Given the $k$th-Hodge Laplacian $\bbL_k$, we define a \emph{simplicial convolutional filter} to process a $k$-simplicial signal $\bbs^k$ as 
\begin{equation}\label{eq.sf-k}
  \bbH_{k} = h_0 \bbI + \sum_{l_1=1}^{L_1} \alpha_{l_1}(\bbB_k^\top\bbB_k)^{l_1} + \sum_{l_2=1}^{L_2}\beta_{l_2}(\bbB_{k+1}\bbB_{k+1}^\top)^{l_2},
\end{equation} 
where $\bbH_k:=\bbH(\bbL_{k,{\ell}},\bbL_{k,{\rm{u}}})$ is a matrix polynomial of the lower and upper Hodge Laplacians with filter coefficients $h_0$, $\balpha = [\alpha_1,\dots,\alpha_{L_1}]^\top$, $\bbeta = [\beta_1,\dots,\beta_{L_2}]^\top$ and filter orders $L_1,L_2$. When $k=0$, we obtain the graph convolutional filter $\bbH_0:=\bbH(\bbL_{0})$ built upon the graph Laplacian $\bbL_0$ \cite{shuman2013emerging,sandryhaila2013discrete, sandryhaila2014discrete}.

In the following we will see that assigning two different sets of coefficients to the lower and upper Laplacian parts in $\bbH_{k}$ enables the filter to treat lower and upper adjacencies differently which results in a more flexible control of the frequency response. 
Instead, the filter $\bbH_k=\sum_{l=0}^{L} h_l\bbL_k^{l}$, which is equivalent to setting $L_1=L_2=L$ and $\balpha=\bbeta$ in \eqref{eq.sf-k}, cannot differentiate between the two types of adjacencies and loses some expressive power.

When applying $\bbH_k$ to a $k$-simplicial signal $\bbs^k$, it generates an output $\bbs^k_{\rm o}=\bbH_k \bbs^k$ which is a \emph{shift-and-sum} operation where the filter $\bbH_k$ first shifts the signal $L_1$ times over the lower neighborhoods and $L_2$ times over the upper neighborhoods, and then sums the shifted results according the corresponding coefficients.
This is analogous to the convolutions of graph signals, images and time series \cite{shuman2013emerging}.  {For ease of exposition, we study the filtering process of an edge flow $\bbf$ via an edge filter $\bbH_1$ hereafter.}

\smallskip\noindent\textbf{Simplicial shifting and local implementation.} Consider an edge filter $\bbH_1$ applied to an edge flow $\bbf$ with an output
\begin{equation} \label{eq.f-o}
      \bbf_{\rm{o}} 
    = \bbH_1 \bbf = h_0\bbf + \sum_{l_1=1}^{L_1} \alpha_{l_1}\bbL_{1,\ell}^{l_1} \bbf  + \sum_{l_2=1}^{L_2}\beta_{l_2}\bbL_{1,\rm{u}}^{l_2} \bbf,
\end{equation}
where the basic operation consists of applying different powers of the lower/upper Hodge Laplacian to the edge flow. 
This basic operation is denoted \emph{simplicial shifting}. 
Let us first consider the one-step lower shifting $\bbf_{\ell}^{(1)}\triangleq\bbL_{1,\ell}\bbf$ and one-step upper shifting $\bbf_{\rm{u}}^{(1)}\triangleq\bbL_{1,\rm{u}}\bbf$. We can express the one-step shifted results on the $i$th edge, $[\bbf_{\ell}^{(1)}]_i$ and $[\bbf_{\rm{u}}^{(1)}]_i$, as 
\begin{equation} \label{eq.shifting-per-edge}
  \begin{aligned}
    [\bbf_{\ell}^{(1)}]_i & =   \sum_{j\in\{\ccalN^1_{{\ell},i} \cup \;i\}} [\bbL_{1,{\ell}}]_{ij}[\bbf]_j, \\
    [\bbf_{\rm{u}}^{(1)}]_i & =   \sum_{j\in\{\ccalN^1_{{\rm u},i} \cup \;i\}} [\bbL_{1,{\rm u}}]_{ij}[\bbf]_j, 
  \end{aligned}
\end{equation}
which are the weighted linear combinations of the edge flows on the lower and upper neighborhoods, $\ccalN^1_{{\ell},i}$ and $\ccalN^1_{{\rm u},i}$, of edge $i$. 
This implies that one-step shifting is a \emph{local operation} in the edge space within the direct lower/upper neighborhoods. 

\begin{figure}[!t] 
  \vspace{-3mm}
  \centering
  \subfloat[An indicator flow $\bbf$\label{fig:shifting_1}]{%
       \includegraphics[width=0.33\linewidth]{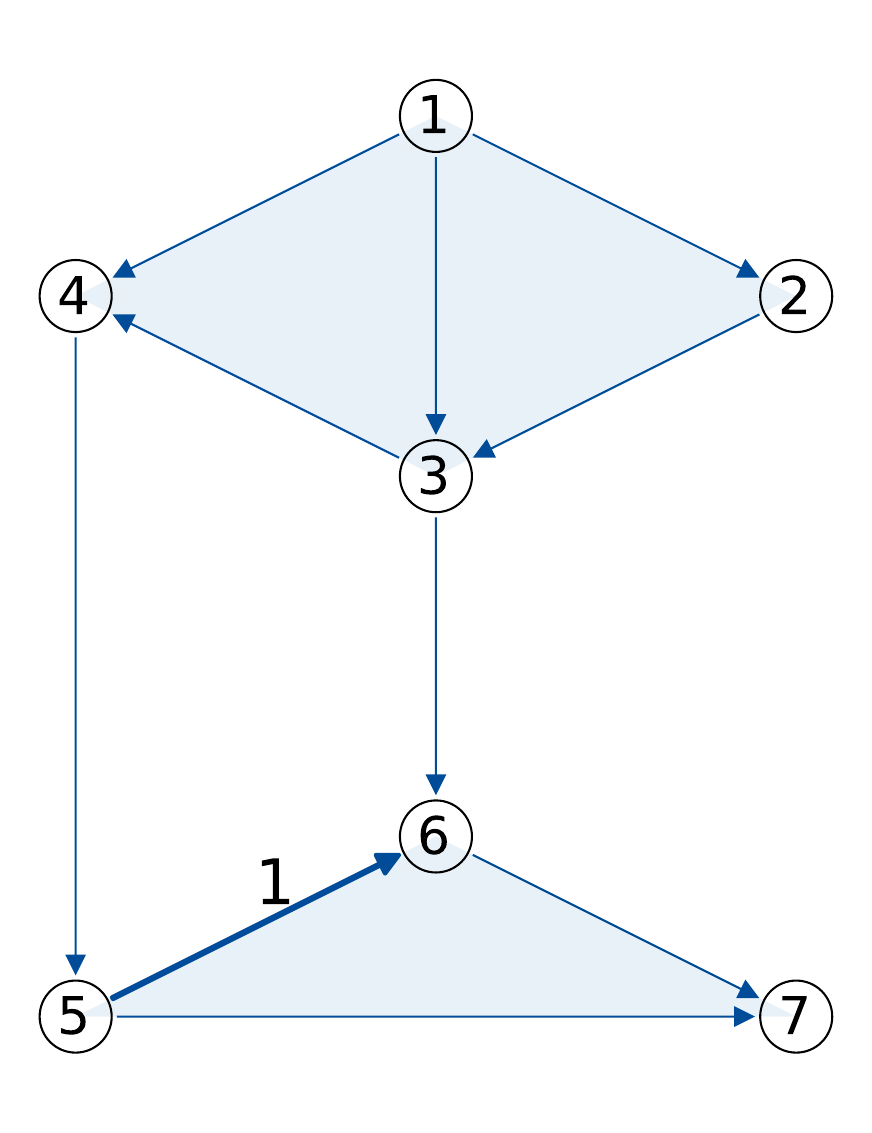}}
  \subfloat[$\bbL_{1,\ell} \bbf$\label{fig:shifting_2}]{%
        \includegraphics[width=0.33\linewidth]{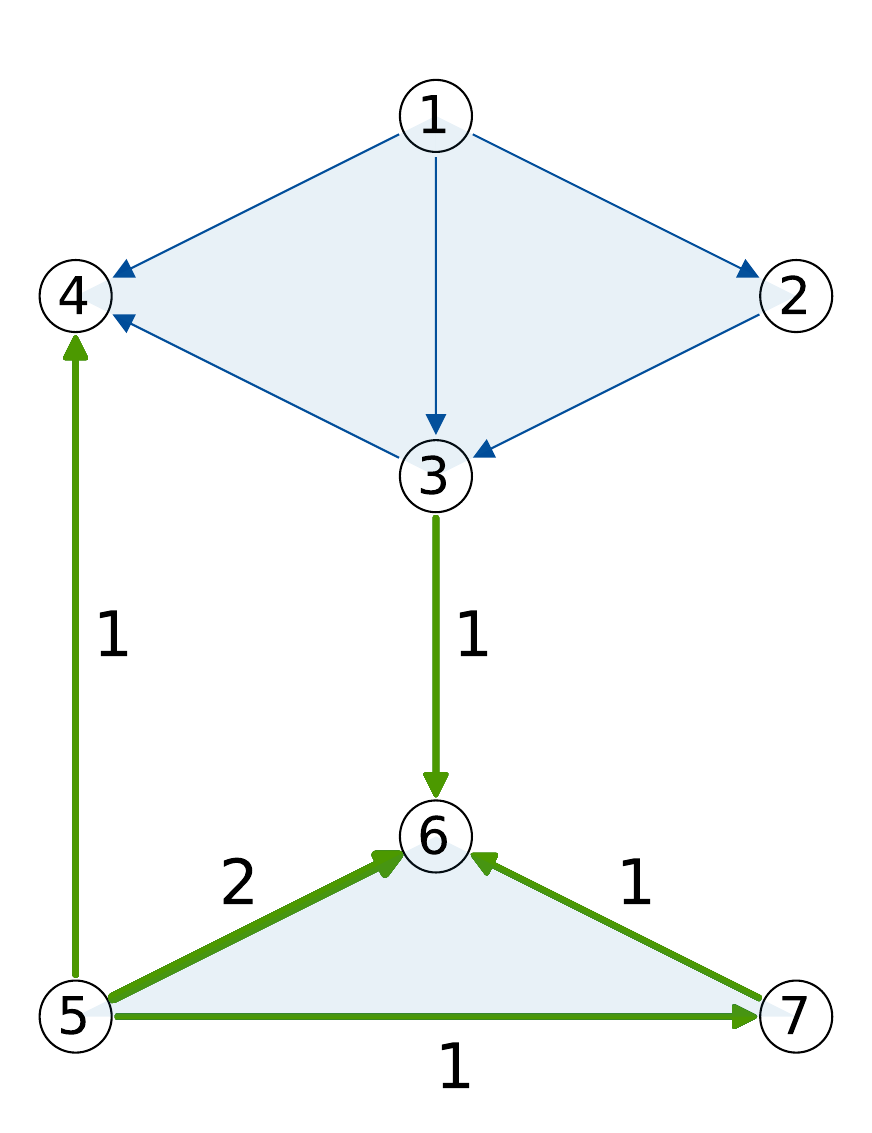}}
  \subfloat[$\bbL^2_{1,\ell} \bbf$\label{fig:shifting_3}]{%
        \includegraphics[width=0.33\linewidth]{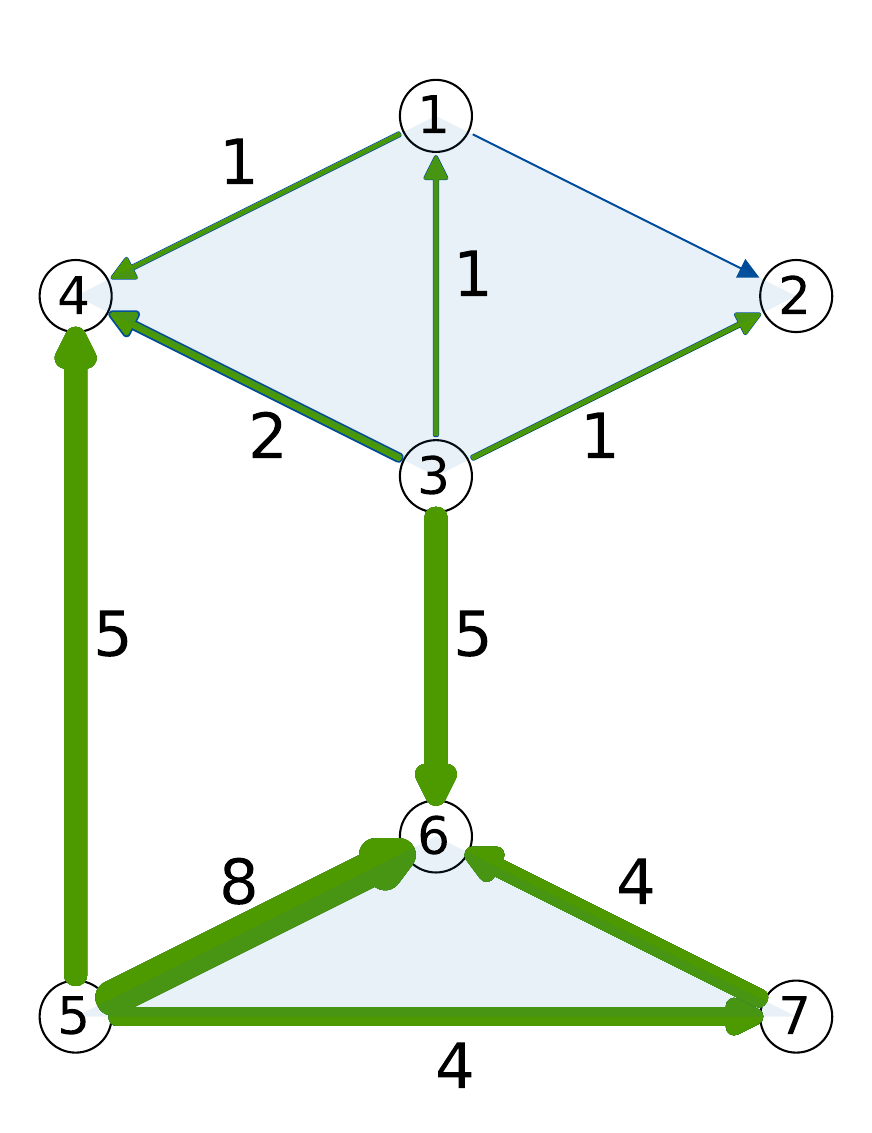}}
  \vspace{-4mm}
  \subfloat[$\bbL_{1,\rm{u}} \bbf$\label{fig:shifting_4}]{%
       \includegraphics[width=0.33\linewidth]{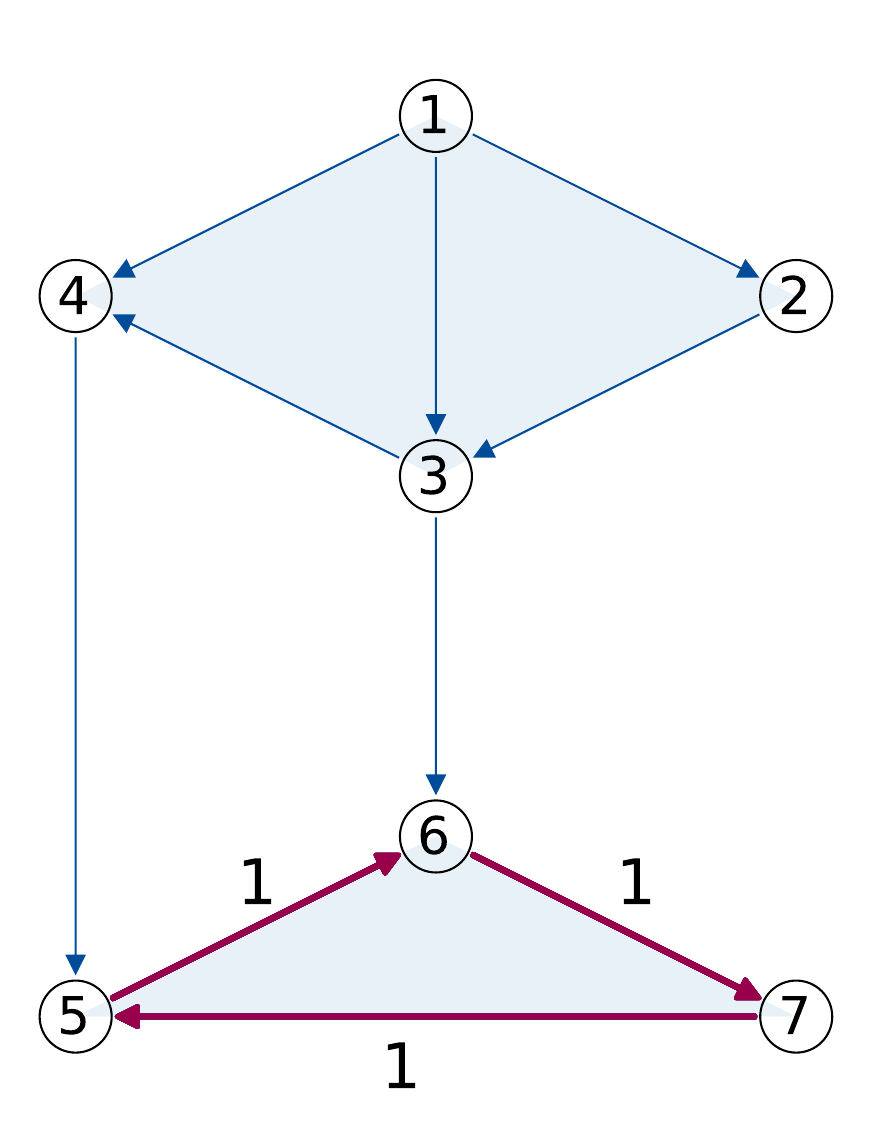}}
  \subfloat[$\bbL^2_{1,\rm{u}} \bbf$\label{fig:shifting_5}]{%
        \includegraphics[width=0.33\linewidth]{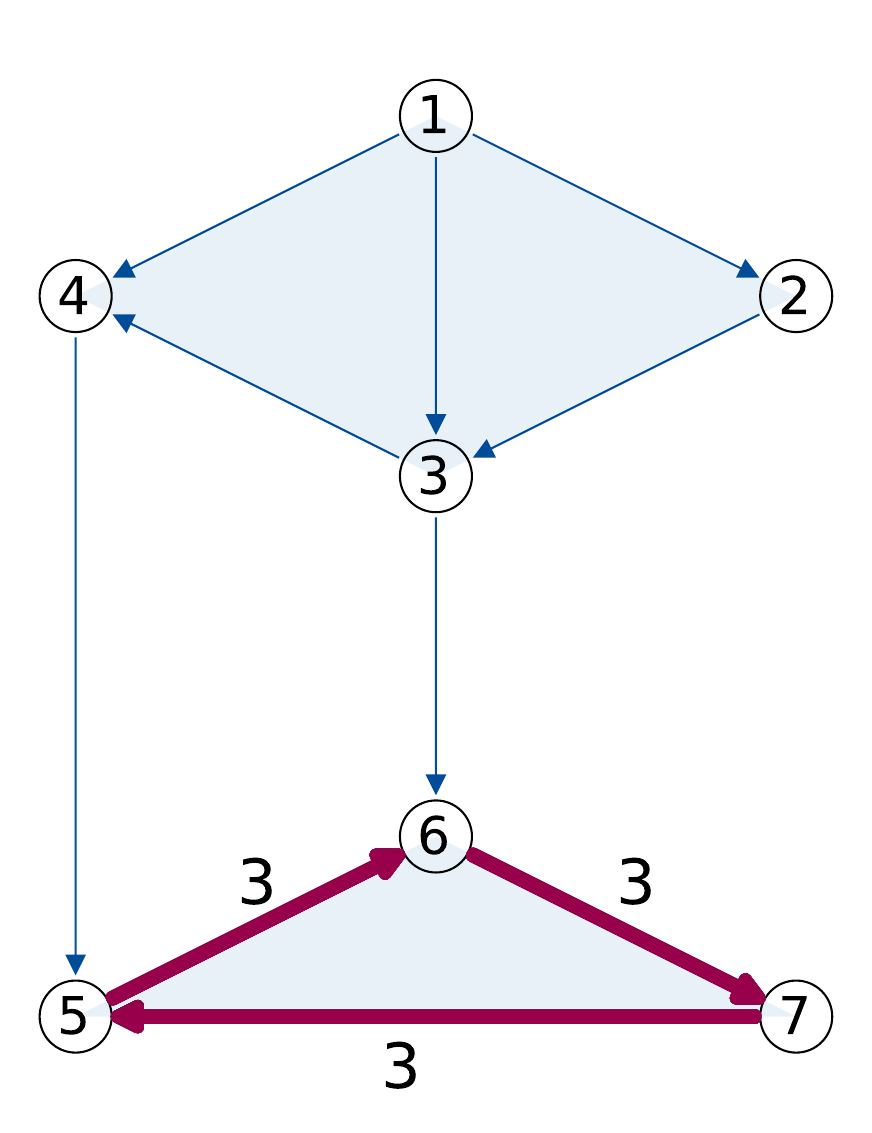}}
  \subfloat[$\bbL^2_{1,\ell} \bbf$ + $\bbL^2_{1,\rm{u}} \bbf$\label{fig:shifting_6}]{%
        \includegraphics[width=0.33\linewidth]{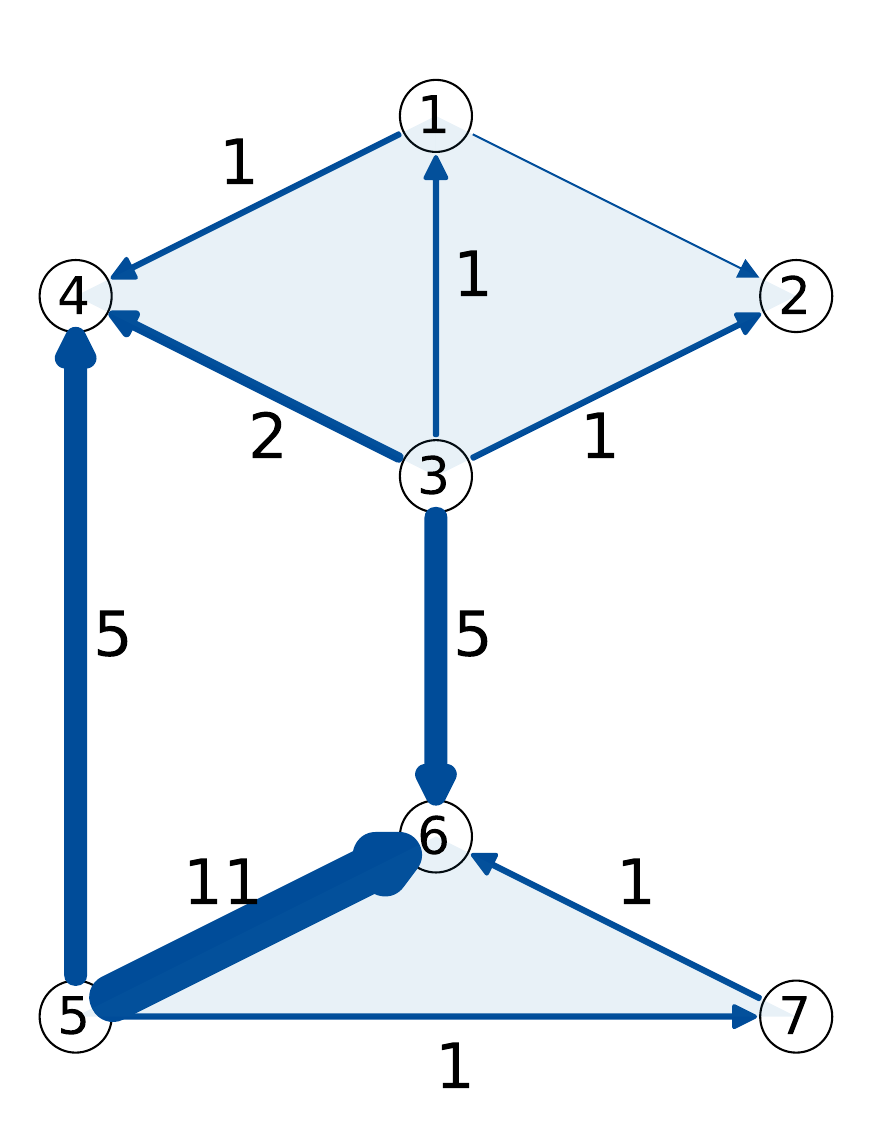}}
  \caption{Simplicial shifting. (a): An edge flow indicator $\bbf$ of edge $\{5,6\}$. (b): One-step lower shifting $\bbL_{1,\ell} \bbf$. Edge $\{5,6\}$ and its direct lower neighboring edges (green) update their flows by aggregating information from their lower neighbors and themselves. (c): Two-step lower shifting $\bbL_{1,\ell}^2\bbf$. Lower neighboring edges (green) update their flows through faces within two hops away from edge $\{5,6\}$, which can be obtained by one-step shifting $\bbL_{1,\ell}\bbf_{\ell}$. (d): One-step upper shifting $\bbL_{1,\rm{u}}\bbf$. Edge $\{5,6\}$ and its upper neighbors (red) update their flows through local information aggregation. (e): Two-step upper shifting $\bbL_{1,\rm{u}}^2\bbf$. The output is localized within the one-hop upper neighborhood, as there is no upper neighboring edge two hops away from $\{5,6\}$. (f): Two-step shifting result $\bbL_1^2\bbf$, as the sum of (c) and (e). } 
  \label{fig:shifting_illustration} 
  \vspace{-6mm}
\end{figure}

\begin{figure}[!t]
  \includegraphics[width=0.55\textwidth,right]{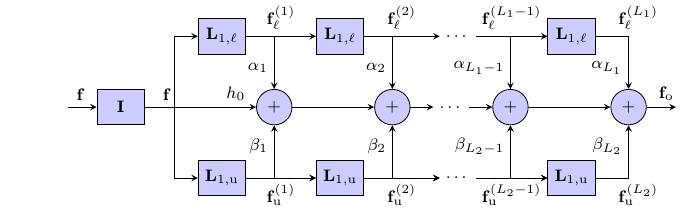}
  \vspace{-4mm}
  \caption{Simplicial convolutional filtering is a shift-and-sum operation.} 
  \vspace{-3mm}
  \label{fig:simplicial-filtering-operation-illustration}
\end{figure}

Consider now the $l$-step lower shifting of an edge flow $\bbf$, 
$  \bbf_{\ell}^{(l)} \triangleq \bbL_{1,\ell}^l \bbf = \bbL_{1,\ell} \bbf_{\ell}^{(l-1)}$,
where the second equality indicates that the $l$-step lower shifting can be computed as a one-step shifting of the previously shifted result, $\bbf_{\ell}^{(l-1)}$. 
Accordingly, the $l$-step upper shifting follows $\bbf_{\rm{u}}^{(l)} \triangleq \bbL_{1,\rm{u}}^l \bbf = \bbL_{1,\rm{u}}\bbf_{\rm{u}}^{(l-1)}$. Thus, the simplicial shifting allows a recursive implementation. 
For example, the two-step shifted results, $\bbf_{\ell}^{(2)}$ and $\bbf_{\rm{u}}^{(2)}$, can be computed from $\bbf_{\ell}^{(1)}$ and $\bbf_{\rm{u}}^{(1)}$ via another shifting. 
Each edge thus collects the flows  {from} its lower and upper neighbors two hops away. 
Fig. \ref{fig:shifting_illustration} illustrates such shifting operations.
Likewise, the $l$-step shifted results $\bbf_{\ell}^{(l)}$ and $\bbf_{\rm{u}}^{(l)}$ contain the information  {up to} the $l$-hop lower and upper neighborhoods. Finally, we can express the output \eqref{eq.f-o} as 
\begin{equation} \label{eq.f-o-2}
    \bbf_{\rm{o}} =  h_0\bbf^{(0)} + \sum_{l_1=1}^{L_1} \alpha_{l_1} \bbf^{(l_1)}_{\ell} + \sum_{l_2=1}^{L_2} \beta_{l_2} \bbf^{(l_2)}_{\rm{u}},
\end{equation}
which is a weighted linear combination of lower and upper shifted simplicial signals after different steps. 
Fig. \ref{fig:simplicial-filtering-operation-illustration} illustrates such a shift-and-sum operation. 

Thus, the simplicial convolutional filter admits a \emph{simplicial locality}, i.e., the output $\bbf_{\rm o}$ is localized in the $L_1$-hop lower neighborhood and $L_2$-hop upper neighborhood of an edge. 
If two edges are lower adjacent more than $L_1$ hops away or upper adjacent more than $L_2$ hops away, $\bbH_1$ does not mix signals defined on such edges.  {Note, however, if two edges are lower or upper adjacent at distance $d$, a $d$-step lower or upper shifting via $\bbL_{1,\ell}^{d}$ or $\bbL_{1,\rm{u}}^d$ does not necessarily cause an interaction between them: the aggregation might be cancelled out by the combination of the filter coefficients and the topology of the SC which leads to positive and negative entries in $\bbL_{1,\ell}$ or $\bbL_{1,\rm{u}}$.}

As a local operation within the simplicial neighborhood, the simplicial shifting allows for a distributed filter implementation, in which each edge updates its information only by a direct communication with its lower and upper neighbors. 
The communication complexity mainly comes from operation \eqref{eq.shifting-per-edge}, which is of order $\ccalO(D_{\ell})$ with $D_{\ell}:=\max\{\vert\ccalN_{{\ell},i}^1\vert_{i=1}^{N_1}\}$ for the lower simplicial shifting at each edge and $\ccalO(D_{\rm u})$ with $D_{\rm{u}}:=\max\{\vert\ccalN_{{\rm u},i}^1\vert_{i=1}^{N_1}\}$ for the upper simplicial shifting, i.e., the maximum number of lower and upper neighbors among all edges, respectively. Then, the total communication cost of the distributed implementation for each edge is $\ccalO(D_{\ell}L_1 + D_{\rm u}L_2)$ due to the $L_1$ lower shifting steps and $L_2$ upper ones. 

\smallskip\noindent\textbf{Linearity and shift invariance.} 
Simplicial convolutional filters are linear operators that are invariant to shifts.
\begin{prop} \label{prop.linearity-shift-invariance}
  The simplicial filter $\bbH_k$ [cf. \eqref{eq.sf-k}] is linear and shift-invariant. Specifically, in the edge space, given two edge flows $\bbf_1$ and $\bbf_2$ and a simplicial filter $\bbH_1$, we have 
  \begin{equation}
    \begin{aligned}
      \text{Linearity}: & \,\, \bbH_1 (a\bbf_1 + b\bbf_2) = a\bbH_1\bbf_1 +b\bbH_1\bbf_2, \\
      \text{Shift-invariance}: &  \,\, \bbL_{1,\ell} \big(\bbH_1\bbf_1 \big) = \bbH_1 \big(\bbL_{1,\ell} \bbf_1 \big), \\
      &  \,\, \bbL_{1,\rm{u}} \big(\bbH_1\bbf_1 \big) = \bbH_1 \big(\bbL_{1,\rm{u}} \bbf_1 \big).
    \end{aligned}
  \end{equation}
\end{prop}
\begin{IEEEproof}
  See  Appendix \ref{proof.linearity-shift-invariance}.
\end{IEEEproof}

The shift-invariance implies that applying a lower or upper Hodge Laplacian to the simplicial output is equivalent to applying them to the input signal prior to filtering. 
Consequently, it holds that  $\bbH_1 \bbH_1' \bbf = \bbH_1' \bbH_1 \bbf$ for any two filters $\bbH_1$ and $\bbH_1'$.

\smallskip\noindent\textbf{Equivariance.}
In an SC, the labeling and the reference orientation of the simplices should not affect the filter output. 
We show that this is indeed the case.
We can model the simplex relabeling of simplices by a set of permutation matrices  {\cite{roddenberry2021principled}}
\begin{equation} \label{eq.permutation-set}
   \ccalP= \{ \bbP_k \in \{0,1\}^{N_k\times N_k}: \bbP_k \bb1 = \bb1, \bbP_k^\top \bb1 = \bb1, k \geq 0 \}.  
\end{equation}
Let $\bar{\ccalP} = (\bbP_0,\bbP_1,\dots) \subset \ccalP$ denote a sequence of label permutations in an SC. 
After relabeling the simplices by $\bar{\ccalP}$, signal $\bbs^k$ becomes $\bbP_k\bbs^k$, i.e., a reordering of the entries of $\bbs^k$.
Similarly, the incidence matrix $\bbB_k$ becomes $\bar{\bbB}_k=\bbP_{k-1}\bbB_k\bbP_k$, i.e., a reordering of the rows and columns of $\bbB_k$. 
Likewise, for $\bbL_k$, we get $\bar{\bbL}_k=\bbP_k\bbL_k\bbP^\top_k$.

\begin{prop}[Permutation equivariance] \label{prop.permutation-equivariance}
  Consider the Hodge Laplacians $\bbL_k$ and $\bar{\bbL}_k = \bbP_k\bbL_k\bbP^\top$ for a permutation sequence $\bar{\ccalP}$. For the simplicial signals $\bbs^k$ and $\bar{\bbs}^{k}=\bbP_k\bbs^k$, with $\bar{\bbH}_k:= \bbH(\bar{\bbL}_k)$ [cf. \eqref{eq.sf-k}],
  the simplicial filter outputs $\bbs^k_{\rm{o}}:=\bbH_k\bbs^k$ and $\bar{\bbs}^{k}_{\rm{o}}:=\bar{\bbH}_k\bar{\bbs}^{k}$ satisfy
  \begin{equation}
    \bar{\bbs}^{k}_{\rm{o}}:=\bar{\bbH}_k\bar{\bbs}^{k} = \bar{\bbH}_k(\bbP_k\bbs^k) = \bbP_k \bbH_k\bbs^k :=  \bbP_k\bbs_{\rm{o}}^k.
  \end{equation}
\end{prop}

\begin{IEEEproof}
See  Appendix \ref{proof.permutation-equivariance}.
\end{IEEEproof}

A new reference orientation of a $k$-simplex leads to a multiplication by $-1$ of  {the columns (or rows)} of the incidence matrices $\bbB_k$ and $\bbB_{k+1}$ where the $k$-simplex appears and it also flips the sign of the corresponding simplicial signal. This can be modeled by a diagonal matrix $\bbD_k$ from the set 
\begin{equation} \label{eq.orientation-set}
  \ccalD = \{ \bbD_k = \diag(\bbd_k): \bbd_k \in  \{\pm 1\}^{N_k}, k\geq 1, \bbd_0 = \bb1  \},
\end{equation}
where $\bbd_0=\bb1$, as the orientation of the nodes is trivial {\cite{roddenberry2021principled}}. 
Denote a sequence of orientation changes by $\bar{\ccalD}= (\bbD_0, \bbD_1, \dots) \subset \ccalD$. 
A $k$-simplicial signal $\bbs^k$ becomes $\bbD_k\bbs^k$ after an orientation change by $\bar{\ccalD}$. Accordingly, the incidence matrix $\bbB_k$ becomes $\bar{\bbB}_k = \bbD_{k-1}\bbB_k\bbD_k$ and the Hodge Laplacian $\bbL_k$ becomes $\bar{\bbL}_k = \bbD_k\bbL_k\bbD_k$. 

\begin{prop}[Orientation equivariance] \label{prop.orientation-equivariance}
  Consider the Hodge Laplacians $\bbL_k$ and $\bar{\bbL}_k = \bbD_k\bbL_k\bbD_k$ for a sequence of orientation changes $\bar{\ccalD}$. For the simplicial signals $\bbs^k$ and $\bar{\bbs}^{k}=\bbD_k\bbs^k$, with 
  $\bar{\bbH}_{k} = \bbH(\bar{\bbL}_k)$, 
  the simplicial filter outputs $\bbs^k_{\rm{o}}:=\bbH_k\bbs^k$ and $\bar{\bbs}^{k}_{\rm{o}}:=\bar{\bbH}_k\bar{\bbs}^{k}$ 
  satisfy
  \begin{equation}
    \bar{\bbs}^{k}_{\rm{o}}:=\bar{\bbH}_k\bar{\bbs}^{k} = \bar{\bbH}_k(\bbD_k\bbs^k) = \bbD_k \bbH_k\bbs^k :=  \bbD_k\bbs_{\rm{o}}^k.
  \end{equation}
\end{prop}

\begin{IEEEproof}
See Appendix \ref{proof.orientation-equivariance}.
\end{IEEEproof} 

Intuitively, the two previous propositions state that for the  simplicial filter $\bbH_k$  the labeling and reference orientation of simplices are inconsequential for the filter output. 
These two properties have been previously reported in the context  {of} neural network on SCs \cite{roddenberry2021principled, yang2021simplicial, bodnar2021weisfeiler}.
These equivariances imply that we can learn a filter to process a given simplex by seeing only permuted and reoriented versions of it: if two parts of an SC are topologically equivalent and the simplices support corresponding flows, a simplicial convolutional filter yields equivalent outputs.

\section{Spectral Analysis of Simplicial Filters} \label{sec:spectral-analysis}
We now analyze the spectral properties of the simplicial convolutional filter.
First, we introduce the Hodge decomposition and review the simplicial Fourier transform (SFT) \cite{barbarossa2020}. 
Then, we investigate the simplicial frequency in terms of the Hodge decomposition. By defining three frequency types, we characterize the frequency response of the filter.  

\subsection{Hodge Decomposition} \label{sec:spectral-analysis-hodge-decomposition}
The Hodge decomposition in the edge space states that:
\begin{equation} \label{eq.hodge-decomp}
  \setR^{N_1}=\im(\bbB_1^\top)\oplus\im(\bbB_2)\oplus\ker(\bbL_1)
\end{equation}
where $\im(\cdot)$ and $\ker(\cdot)$ are the image and kernel of a matrix. 
This implies that the edge space is composed of three orthogonal subspaces, namely, the gradient space $\im(\bbB_1^\top)$, the curl space $\im(\bbB_2)$, and the harmonic space $\ker(\bbL_1)$ \cite{lim2015hodge,schaub2021}. 
Thus, any edge flow $\bbf\in\setR^{N_1}$ can be decomposed into three orthogonal components
$
    \bbf =  \bbf_{\rm{G}} + \bbf_{\rm{C}} + \bbf_{\rm{H}},
$
which are the \emph{gradient component} $\bbf_{\rm{G}}\in\im(\bbB_1^\top)$, the \emph{curl component} $\bbf_{\rm{C}}\in\im(\bbB_2)$, and the \emph{harmonic component} $\bbf_{\rm{H}}\in\ker(\bbL_1)$, respectively. Furthermore, the incidence matrices $\bbB_1$, $\bbB_2$ and their adjoints can be interpreted as follows~\cite{schaub2021,barbarossa2020}.

\smallskip\noindent\textbf{Divergence operator $\bbB_1$.} The incidence matrix $\bbB_1$ acts as a \emph{divergence operator}. 
By applying it to an edge flow $\bbf$, we compute the divergence of the flow, $\div(\bbf) = \bbB_1\bbf$. The $i$th entry of $\div(\bbf)$ is the netflow passing through the $i$th vertex, i.e., the difference between the total inflow and outflow at vertex $i$. 
A vertex is a source or sink if it has a nonzero netflow. 

\smallskip\noindent\textbf{Gradient operator $\bbB_1^\top$.} The adjoint operator $\bbB_1^\top$ is called the \emph{gradient operator}, which takes the difference between node signals along the oriented edges to induce an edge flow, $\bbf_{\rm{G}}=\bbB_1^\top \bbv$. 
We call $\bbf_{\rm{G}} \in \im(\bbB_1^\top)$ a \emph{gradient flow} and subspace $\im(\bbB_1^\top)$ the \emph{gradient space}, as any gradient flow can be induced from a node signal via the gradient operator. 

\smallskip\noindent\textbf{Curl adjoint $\bbB_2$.} By applying matrix $\bbB_2$ to a triangle signal $\bbt\in\setR^{N_2}$, we can induce a \emph{curl flow}, $\bbf_{\rm{C}} = \bbB_2 \bbt$, corresponding to a flow locally circling along the edges of triangles. The space $\im(\bbB_2)$ is the \emph{curl space} as any flow in it can be induced from a triangle signal. 

\smallskip\noindent\textbf{Curl operator $\bbB_2^\top$.} By applying the operator $\bbB_2^\top$ to an edge flow $\bbf$, we compute its curl as $\curl(\bbf) = \bbB_2^\top \bbf$, where the $i$th entry is the netflow circulating along the $i$th triangle, i.e., the sum of the edge flows forming the triangle. This can be seen as a rotational variation measure of the edge flow. 

The two incidence matrices and their adjoints provide  insights into the three orthogonal subspaces and signal components given by the Hodge decomposition.
\begin{enumerate}[label=\roman*)] 
  \item If edge flow $\bbf$ has zero divergence at each vertex, i.e., $\bbB_1\bbf = \mathbf{0} \Leftrightarrow \bbf\in\ker(\bbB_1)$, then it is cyclic or \emph{divergence-free}. The space $\ker(\bbB_1)$ is called the \emph{cycle space}, orthogonal to the gradient space, i.e., $\setR^{N_1} = \im(\bbB_1^\top) \oplus \ker(\bbB_1)$. A gradient flow always has nonzero divergence, while a curl flow is divergence-free due to \eqref{eq.boundary-condition}.
  \item If edge flow $\bbf$ has zero curl on each triangle, it is \emph{curl-free} and $\bbf\in\ker(\bbB_2^\top)$. The curl space $\im(\bbB_2)$ is orthogonal to the space $\ker(\bbB_2^\top)$ and we have $\setR^{N_1} = \im(\bbB_2)\oplus\ker(\bbB_2^\top)$. A gradient flow $\bbf_{\rm{G}}$ is curl-free due to \eqref{eq.boundary-condition}.
  \item The space $\ker(\bbL_1)$ is called the harmonic space. Any flow $\bbf_{\rm{H}}\in\ker(\bbL_1)$ satisfies $\bbL_1 \bbf_{\rm{H}} = \mathbf{0}$, which is harmonic, i.e., both divergence- and curl-free.
\end{enumerate} 

\begin{figure}[!t] 
  \vspace{-3mm}
  \centering
  \subfloat[$\bbf$ \label{fig:flow f}]{%
       \includegraphics[width=0.248\linewidth]{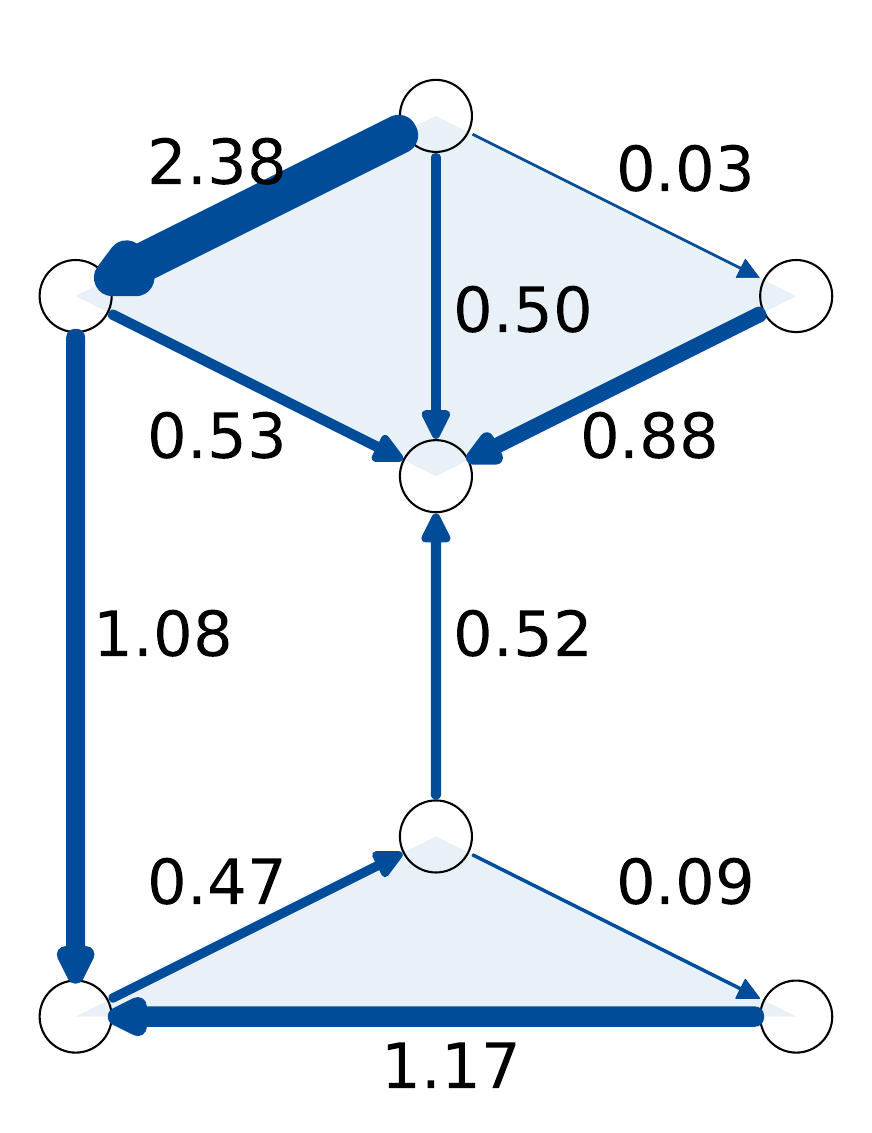}}
  \subfloat[$\bbf_{\rm{G}}$\label{fig:flow f_g}]{%
       \includegraphics[width=0.248\linewidth]{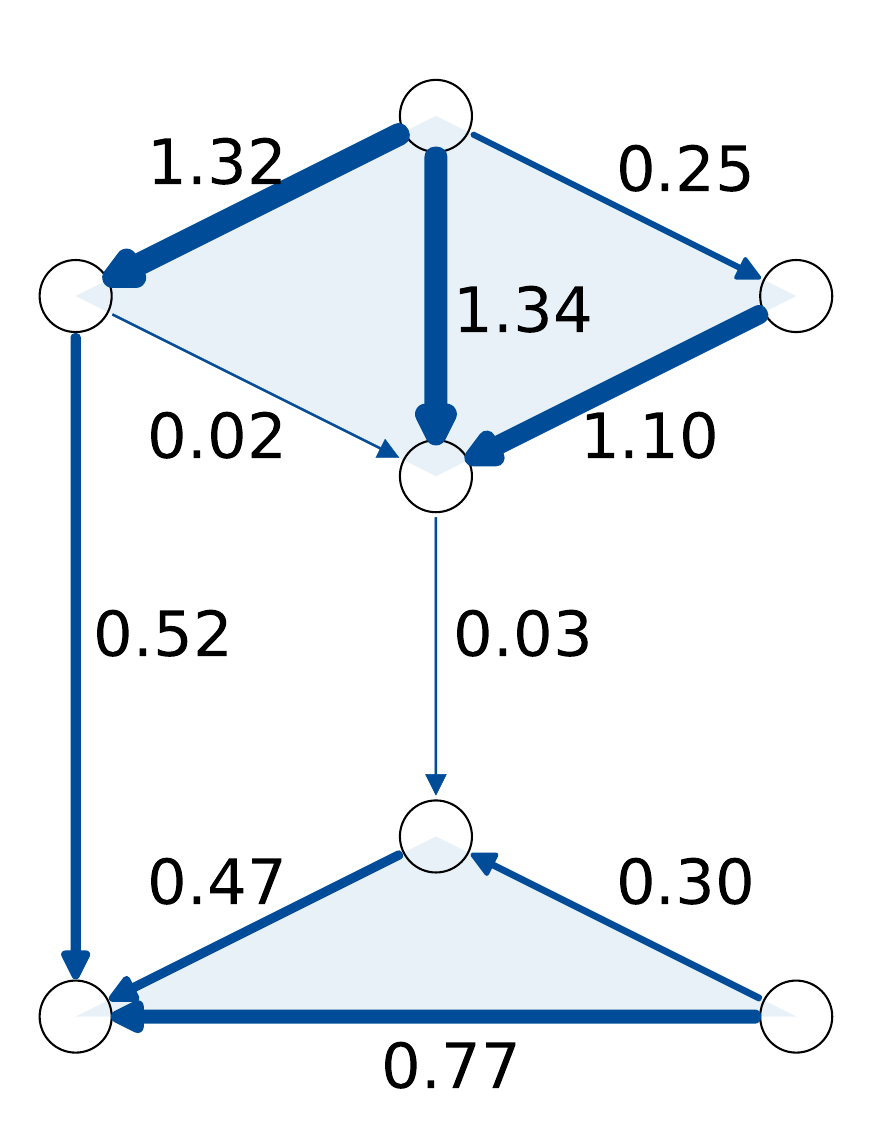}} 
  \subfloat[$\bbf_{\rm{C}}$\label{fig:flow f_c}]{%
       \includegraphics[width=0.248\linewidth]{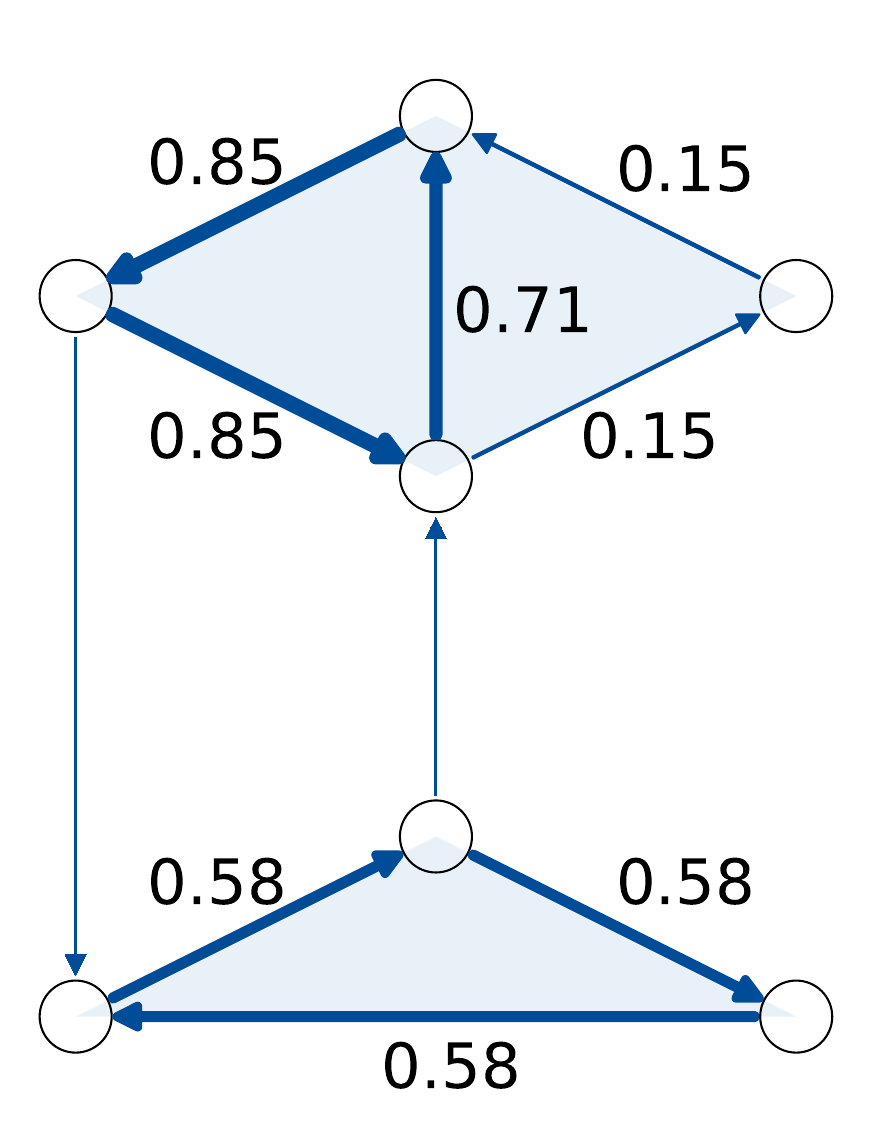}}
   \subfloat[$\bbf_{\rm{H}}$\label{fig:flow f_h}]{%
       \includegraphics[width=0.248\linewidth]{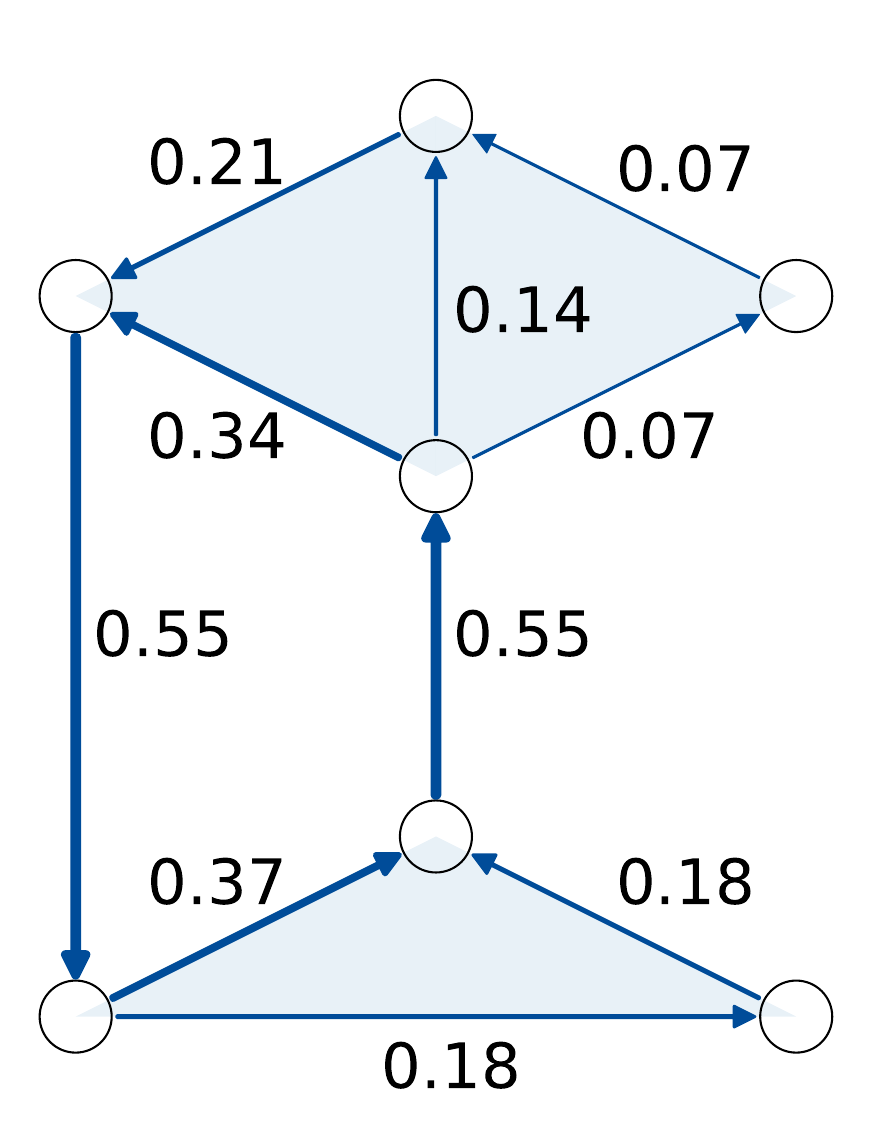}}
  \caption{Flow decomposition illustration  {(all numbers rounded to two decimal places)}. (a): A synthetic edge flow $\bbf$. (b): The gradient component $\bbf_{\rm{G}}$ has a nonzero netflow at each node, but a zero flow around each triangle. (c): The curl component $\bbf_{\rm{C}}$ has a zero netflow at each node, i.e., is divergence-free, but a nonzero flow around each triangle. (d): The harmonic component $\bbf_{\rm{H}}$ has a zero netflow at each node and zero circulation around each triangle, i.e., is divergence- and curl-free (see Section \ref{sec:subcomponent-extraction}).}
  \label{fig:flow_decomp_illustration} 
\end{figure}

The decomposition of an edge flow into its three components reveals different properties of the flow, as shown in Fig. \ref{fig:flow_decomp_illustration}. For instance, we can study the effect of an external source or sink by extracting the gradient component of the edge flow \cite{barbarossa2020}. 
In Section \ref{sec:subcomponent-extraction}, we will discuss this subcomponent extraction problem and solve it with simplicial filters. 

\subsection{Simplicial Fourier Transform} \label{sec:sft-frequency}
The Hodge Laplacians are positive semidefinite matrices and admit an eigendecomposition
\begin{equation}
 \bbL_k = \bbU_k\bLambda_k\bbU_k^\top, 
\end{equation}
where the orthonormal matrix $\bbU_k = [\bbu_{k,1},\dots,\bbu_{k,N_k}]$ collects the eigenvectors, and the diagonal matrix $\bLambda_k = \diag(\lambda_{k,1},\dots,\lambda_{k,N_k})$ the associated eigenvalues. There exists a correspondence between $\bbU_1$ and the three orthogonal subspaces given by the Hodge decomposition \eqref{eq.hodge-decomp}, detailed in the following proposition. 
\begin{prop} \label{prop.correspondence}
    Given the $1$-Hodge Laplacian of an SC $\bbL_1 = \bbL_{1,\ell} + \bbL_{1,\rm{u}}$, the following holds.
    \begin{enumerate}
        \item Gradient eigenvectors $\bbU_{\rm{G}} = [\bbu_{\rm{G},1},\dots,\bbu_{{\rm{G}},N_{\rm{G}}}] \in\setR^{N_1\times N_{\rm{G}}}$ of $\bbL_{1,\ell}$ associated with nonzero eigenvalues span the gradient space $\im(\bbB_1^\top)$ with dimension $N_{\rm{G}}$, i.e., $\im(\bbB_1^\top)=\im(\bbU_{\rm{G}})$. 
        \item Curl eigenvectors $\bbU_{\rm{C}} = [\bbu_{\rm{C},1},\dots,\bbu_{{\rm{C}},N_{\rm{C}}}] \in\setR^{N_1\times N_{\rm{C}}}$ of $\bbL_{1,\rm{u}}$ associated with nonzero eigenvalues span the curl space $\im(\bbB_{2})$ with dimension $N_{\rm{C}}$, i.e., $\im(\bbB_2)=\im(\bbU_{\rm{C}})$.
        \item Gradient eigenvectors $\bbU_{\rm{G}}$ are orthogonal to curl ones $\bbU_{\rm{C}}$. Matrix $[\bbU_{\rm{G}} \,\, \bbU_{\rm{C}}]$  forms the eigenvectors of $\bbL_1$ associated with nonzero eigenvalues, which span the space $\im(\bbL_1)$ with dimension  $N_{\rm{G}}+N_{\rm{C}}$. 
        \item Harmonic eigenvectors $\bbU_{\rm{H}} = [\bbu_{{\rm{H}},1},\dots,\bbu_{{\rm{H}},N_{\rm{H}}}] \in\setR^{N_1\times N_{\rm{H}}}$ of $\bbL_1$ associated with zero eigenvalues span the harmonic space $\ker(\bbL_1)$ with dimension $N_{\rm{H}}$, i.e., $\ker(\bbL_1)=\im(\bbU_{\rm{H}})$. Matrices $ 
        [\bbU_{\rm{H}} \,\,\bbU_{\rm{C}}]$ and $
        [\bbU_{\rm{H}} \,\, \bbU_{\rm{G}}] $ provide the eigenvectors of $\bbL_{1,\ell}$ and $\bbL_{1,\rm{u}}$ associated with zero eigenvalues, respectively.
        \item The columns of $\bbU_1$ can be ordered such that $ \bbU_1 = [
        \bbU_{\rm{H}} \,\, \bbU_{\rm{G}} \,\, \bbU_{\rm{C}} ]
         $. Matrix $\bbU_1$ forms an eigenvector basis for $\bbL_1$, $\bbL_{1,\ell}$ and $\bbL_{1,\rm{u}}$, and $N_1 = N_{\rm{H}} + N_{\rm{G}} + N_{\rm{C}}$.
    \end{enumerate}

\end{prop}

\begin{IEEEproof}
See Appendix \ref{appendix:prop1-proof}.
\end{IEEEproof}

Proposition \ref{prop.correspondence} shows that:
\begin{enumerate*}[label=\roman*)]
  \item the eigenvectors in $\bbU_1$ can fully span the three orthogonal subspaces given by the Hodge decomposition;
  \item the Hodge Laplacian $\bbL_1$ and its lower and upper counterparts, $\bbL_{1,\ell}$ and $\bbL_{1,\rm{u}}$, can be simultaneously diagonalized by $\bbU_1$; and,
  \item from $\im(\bbB_1^\top) = \im(\bbL_{1,\ell})$, we have that the image of the lower Hodge Laplacian $\bbL_{1,\ell}$ coincides with the gradient space, and from $\im(\bbB_2) = \im(\bbL_{1,\rm{u}})$, the image of $\bbL_{1\rm{u}}$ coincides with the curl space.
\end{enumerate*}
These results are applicable to the $k$-Hodge Laplacian accordingly.  See \cite[Thm. 1]{schaub2021} and \cite[Prop. 1]{barbarossa2020} for related discussions.

Thus, the lower shifting of a curl or harmonic flow leads to zero as the harmonic and curl space correspond to the null space of $\bbL_{1,\ell}$. Likewise, the upper shifting of a gradient or harmonic flow leads to zero, i.e.,
\begin{equation} \label{eq.shifting-simplex-domain}
  \bbL_{1,\ell}\bbf_{\rm{C}} = \mathbf{0},\,\, \bbL_{1,\rm{u}}\bbf_{\rm{G}} = \mathbf{0}, \,\, \bbL_{1,\ell}\bbf_{\rm{H}} = \bbL_{1,\rm{u}}\bbf_{\rm{H}} =\mathbf{0}.
\end{equation}

Given a $k$-simplicial signal $\bbs^k$, the \emph{simplicial Fourier transform (SFT)} is given by its projection onto the eigenvectors $\bbU_k$, i.e., $\tilde{\bbs}^k \triangleq \bbU^\top_k \bbs^k$. Entry $[\bbs^k]_i$ represents the weight eigenvector $\bbu_{k,i}$ has on expressing $\bbs^k$. The inverse SFT is given by $\bbs^k = \bbU_k \tilde{\bbs}^k$. For $k=0$, the SFT coincides with the GFT \cite{barbarossa2020}. As for the GFT, the eigenvalues of $\bbL_k$ carry the notion of simplicial frequencies. But in the simplex domain, this frequency notion is more involved. As we illustrate in the sequel for $k=1$, the eigenvalues in the set $\ccalQ=\{\lambda_{1,1},\dots,\lambda_{1,N_1}\}$ of $\bbL_1$ measure three types of simplicial frequencies.

\smallskip\noindent\textbf{Gradient frequency.} 
For any unit norm gradient eigenvector $\bbu_{\rm{G}}\in\bbU_{\rm{G}}$, associated to the gradient space $\im(\bbB_1^\top)$, its corresponding eigenvalue follows
\begin{equation}\label{eq.grad-freq}
  \lambda_{\rm{G}} = \bbu_{\rm{G}}^\top \bbL_1 \bbu_{\rm{G}} = \lVert \bbB_1 \bbu_{\rm{G}} \lVert_2^2 + \lVert \bbB_2^\top \bbu_{\rm{G}} \lVert_2^2 = \lVert \bbB_1 \bbu_{\rm{G}} \lVert_2^2,
\end{equation}
where the last equality is due to the fact that $\bbu_{\rm{G}}$ is curl-free, i.e., $\bbB_2^\top \bbu_{\rm{G}}=\mathbf{0}$. Thus, eigenvalue $\lambda_{\rm{G}}$ is the squared $\ell_2$-norm of the divergence $\bbB_1 \bbu_{\rm{G}}$ of the eigenvector edge flow $\bbu_{\rm{G}}$. The magnitude of $\lambda_{\rm{G}}$ measures the extent of total divergence, i.e., the nodal variation. The gradient eigenvectors associated with a large eigenvalue have a large total divergence. If the SFT $\tilde{\bbf} = \bbU_1^\top \bbf$ of an edge flow has a large weight on such an eigenvector, we say that it contains a high gradient frequency, corresponding to its large divergence. We call any eigenvalue $\lambda_{\rm{G}}$ associated to the gradient eigenvectors $\bbU_{\rm{G}}$ a \emph{gradient frequency} and collect them in the set $\ccalQ_{\rm G} = \{\lambda_{\rm{G},1},\dots,\lambda_{{\rm{G}},N_{\rm{G}}}\}$.
    
\smallskip\noindent\textbf{Curl frequency.} For any unit norm curl eigenvector $\bbu_{\rm{C}}\in\bbU_{\rm{C}}$ associated to the curl space $\im(\bbB_2)$, its corresponding eigenvalue follows
    \begin{equation} \label{eq.curl-freq}
        \lambda_{\rm{C}} = \bbu_{\rm{C}}^\top \bbL_1 \bbu_{\rm{C}} = \lVert \bbB_1 \bbu_{\rm{C}} \lVert_2^2 + \lVert \bbB_2^\top \bbu_{\rm{C}} \lVert_2^2 = \lVert \bbB_2^\top \bbu_{\rm{C}} \lVert_2^2,
    \end{equation}
where the last equality is due to the fact that $\bbu_{\rm{C}}$ is divergence-free, i.e., $ \bbB_1 \bbu_{\rm{C}}=\mathbf{0}$. Eigenvalue $\lambda_{\rm{C}}$ is the squared $\ell_2$-norm of the curl $\bbB_2^\top \bbu_{\rm{C}}$ of the eigenvector $\bbu_{\rm{C}}$. Thus, the magnitude of $\lambda_{\rm{C}}$ measures the extent of total curl, i.e., the rotational variation. Any eigenvector in the curl space corresponding to a large eigenvalue has a large total curl. If the SFT of an edge flow contains large weights on such eigenvectors, we say that it has a high curl frequency. We name any eigenvalue $\lambda_{\rm{C}}$ associated to the curl eigenvectors $\bbU_{\rm{C}}$ a \emph{curl frequency} and collect them in the set $\ccalQ_{\rm C} = \{\lambda_{\rm{C},1},\dots,\lambda_{{\rm{C}},N_{\rm{C}}}\}$.
    
\smallskip\noindent\textbf{Harmonic frequency.} The remaining eigenvalues $\lambda_{\rm{H}}$ are associated to the eigenvectors $\bbu_{\rm{H}}\in\bbU_{\rm{H}}$ which span the harmonic space $\ker(\bbL_1)$. They can be expressed as
    \begin{equation} \label{eq.harm-freq}
        \lambda_{\rm{H}} = \bbu_{\rm{H}}^\top \bbL_1 \bbu_{\rm{H}} = \lVert \bbB_1 \bbu_{\rm{H}} \lVert_2^2 + \lVert \bbB_2^\top \bbu_{\rm{H}} \lVert_2^2  = 0,
    \end{equation}
because $\bbu_{\rm{H}}$ is harmonic, i.e., both divergence- and curl-free. If the SFT of an edge flow has only nonzeros at the harmonic frequencies (which are all zeros), then it is a harmonic flow. We collect the harmonic frequencies (all zeros) in the set $\ccalQ_{\rm H} = \{\lambda_{\rm{H},1},\dots,\lambda_{{\rm{H}},N_{\rm{H}}}\}$.

\begin{figure*}[!t] 
  \centering
  \subfloat[$\lambda_{{\rm{G}},1} = 0.81$\label{fig:eig_g1}]{%
       \includegraphics[width=0.16\linewidth]{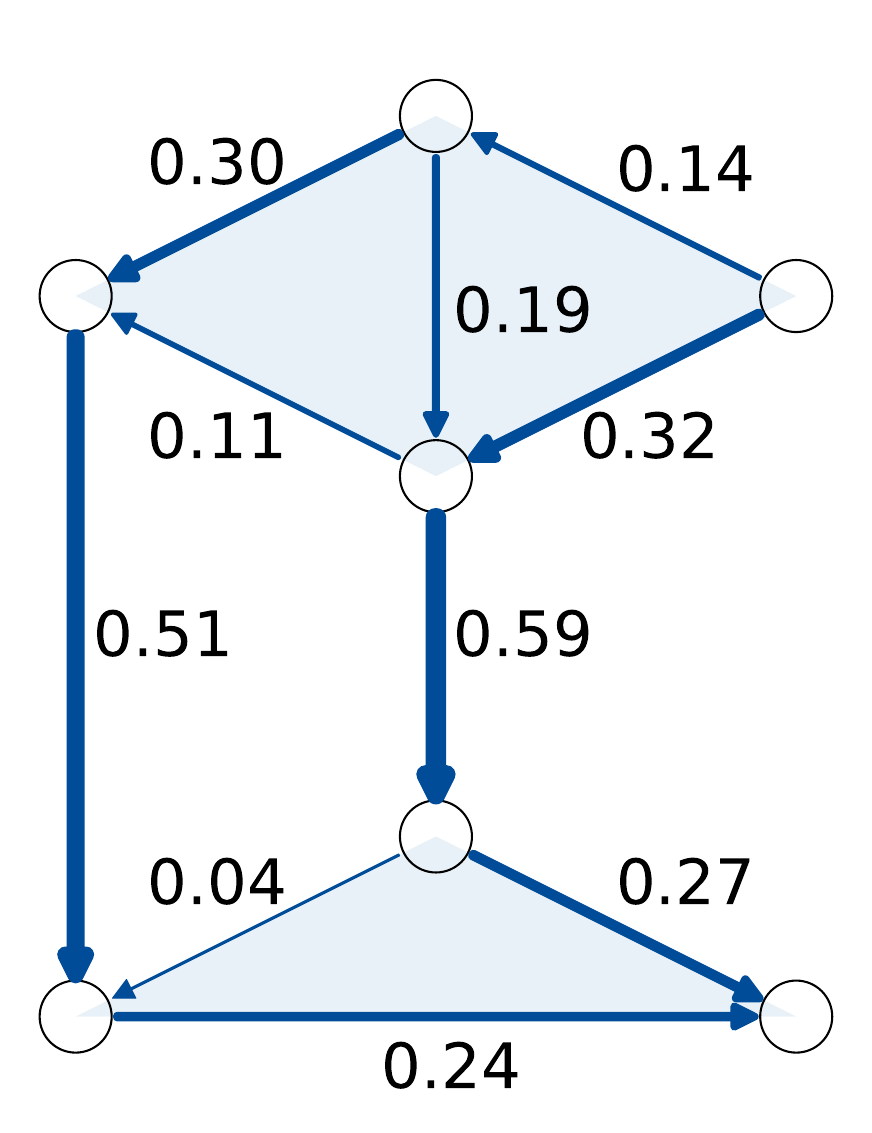}}
  \subfloat[$\lambda_{{\rm{G}},3} = 3.31$\label{fig:eig_g3}]{%
       \includegraphics[width=0.16\linewidth]{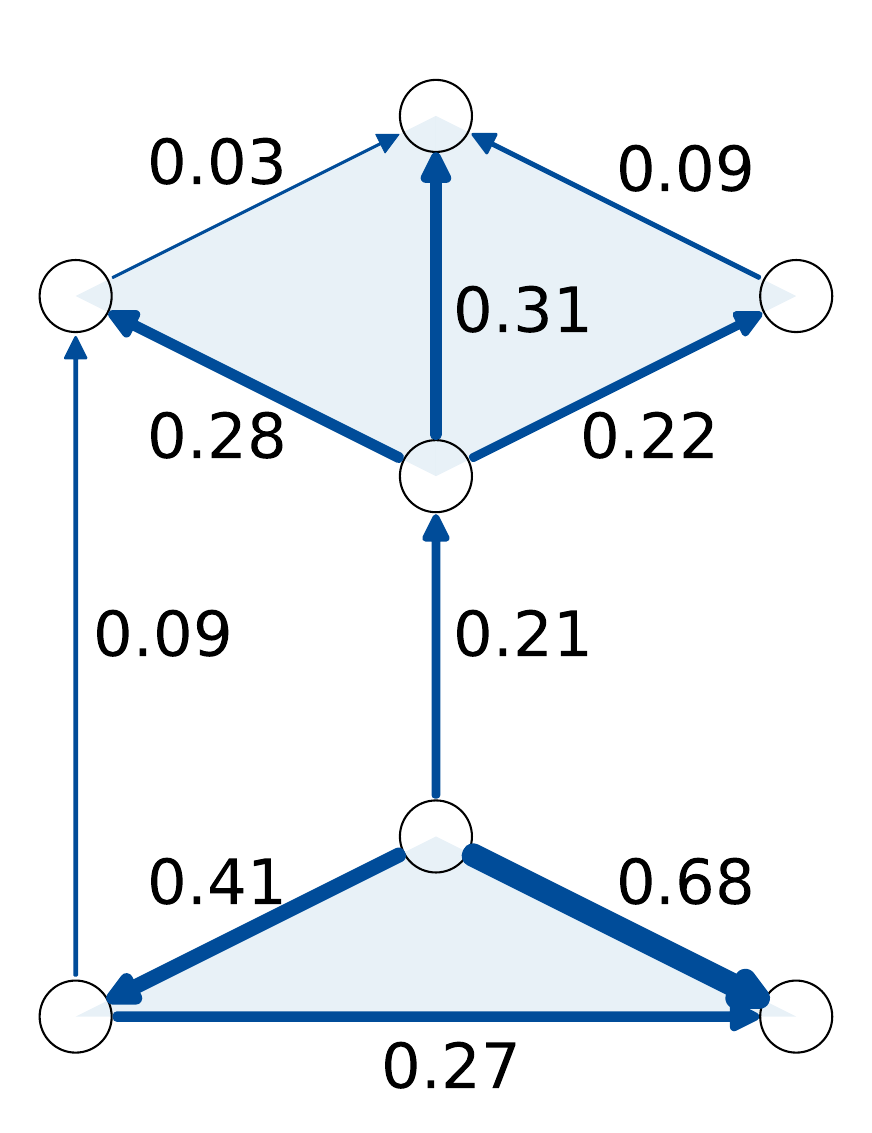}}
  \subfloat[$\lambda_{{\rm{G}},6} = 5.49$\label{fig:eig_g6}]{%
       \includegraphics[width=0.16\linewidth]{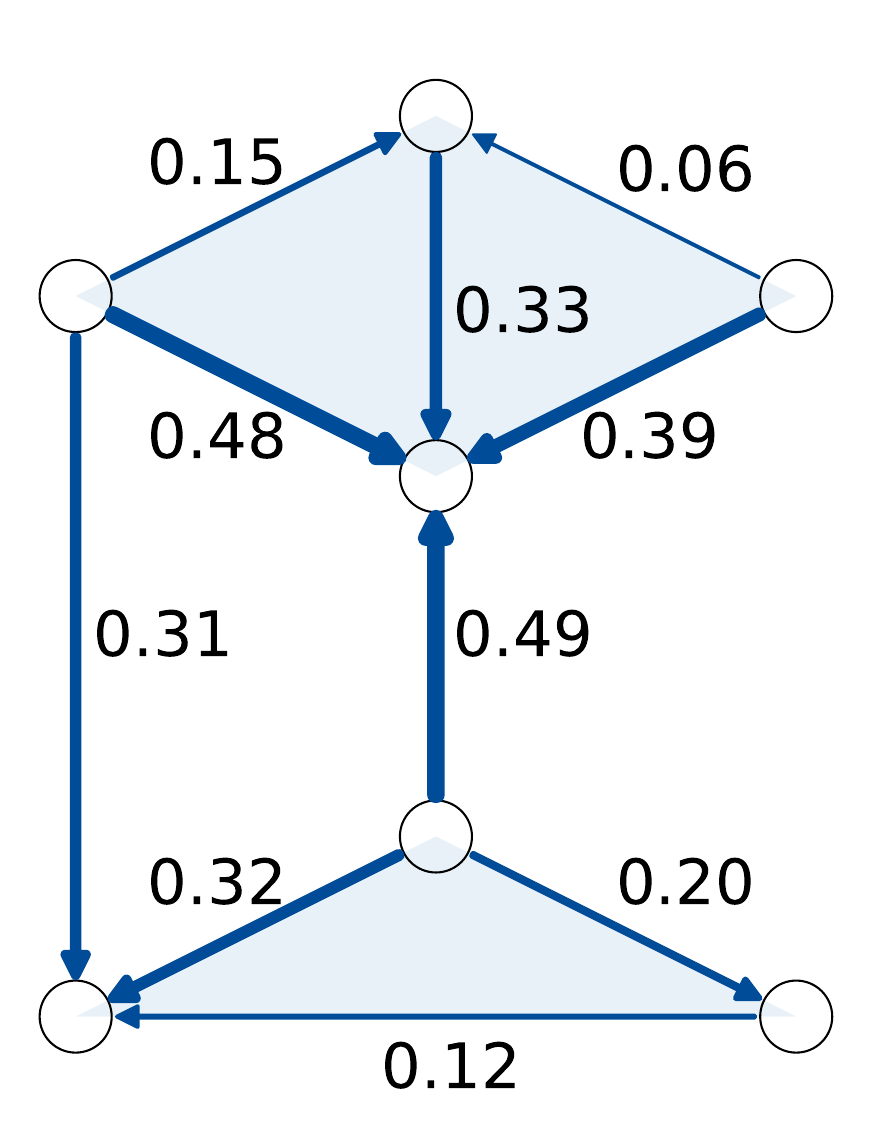}}
  \subfloat[$\lambda_{{\rm{C}},1} = 2$\label{fig:eig_c1}]{%
       \includegraphics[width=0.16\linewidth]{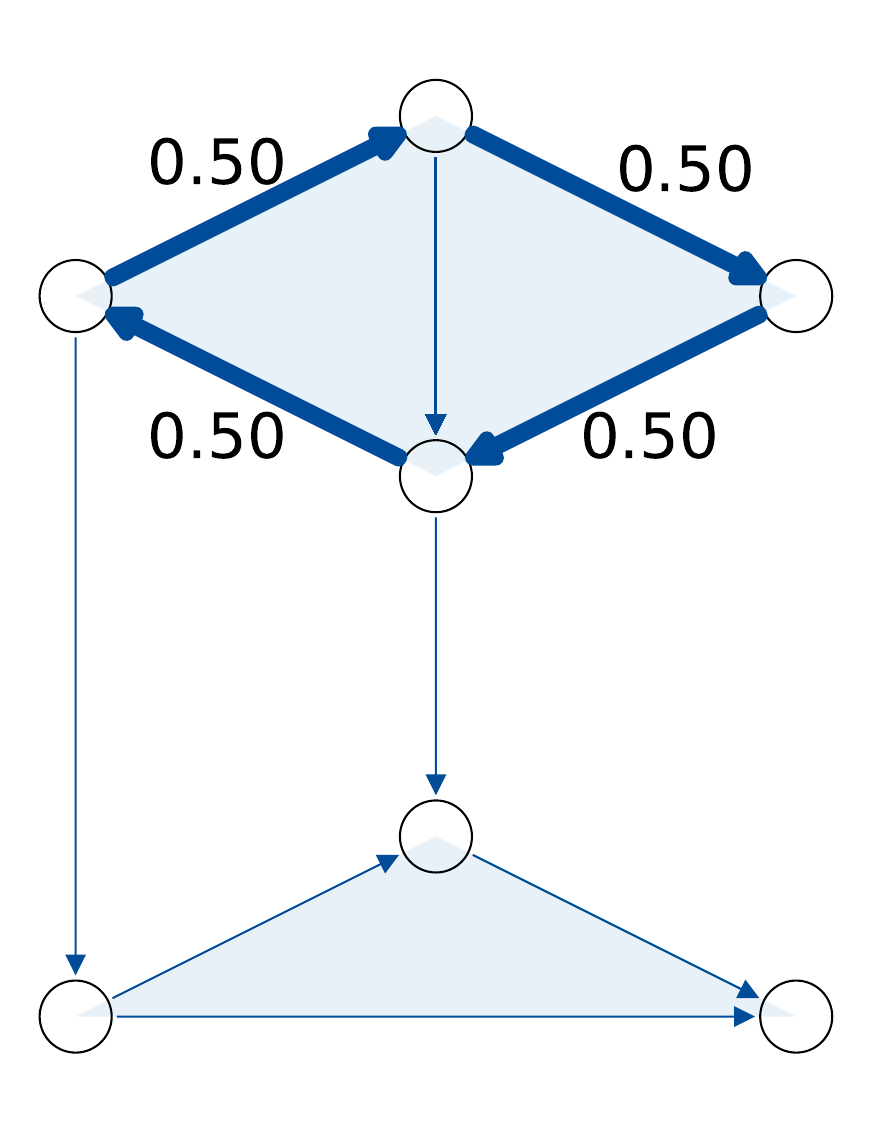}}
   \subfloat[$\lambda_{{\rm{C}},3} = 4$\label{fig:eig_c3}]{%
       \includegraphics[width=0.16\linewidth]{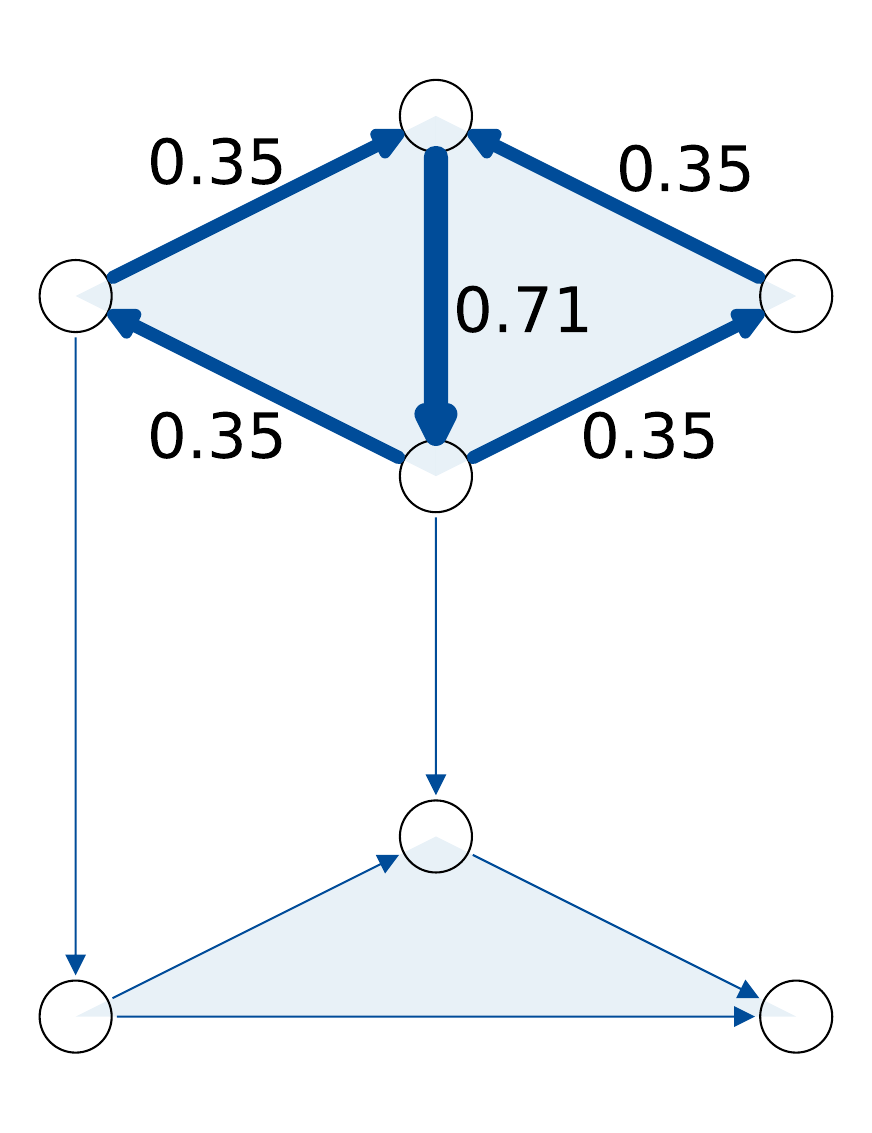}}
   \subfloat[$\lambda_{{\rm{H}},1} = 0$\label{fig:eig_h1}]{%
       \includegraphics[width=0.16\linewidth]{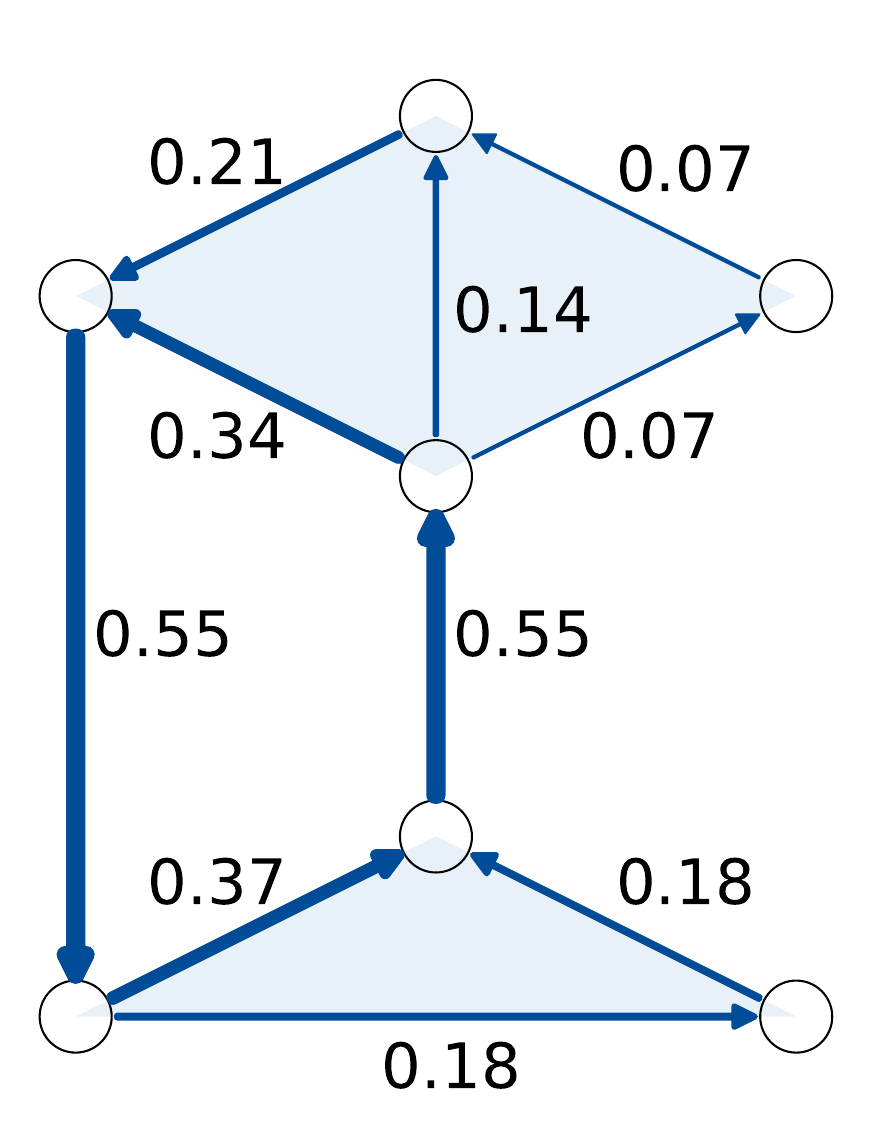}}
  \caption{Spectral analysis of the edge space in Fig. \ref{1a}. The flow value  {(all numbers rounded to two decimal places)} is indicated by the edge width and annotated next to the edge. It is zero if the edge is not annotated. If a flow orientation is opposite to the reference orientation, the corresponding flow value is negative. (a)-(c): The 1st, 3rd, and 6th eigenvectors in the gradient space $\bbU_{\rm{G}}$ with the corresponding gradient frequencies $\lambda_{\rm{G}}$. The total divergence of the eigenvector increases with the eigenvalue. (d)-(e): The 1st and 3rd eigenvectors in the curl space $\bbU_{\rm{C}}$ with the corresponding curl frequencies $\lambda_{\rm{C}}$. The total curl of the eigenvectors increases with the eigenvalue. (f): The only eigenvector in the harmonic space $\bbU_{\rm{H}}$ has frequency 0 and zero divergence and curl.}
  \label{fig:eig_vectors} 
\end{figure*}

With these three types of simplicial frequencies, low and high frequency notions in an SC are only meaningful  with respect to a certain type. A higher gradient (curl) frequency indicates respectively a larger nodal (rotational) variability. 
This is different from the frequency notion in discrete and graph signal processing. 
A zero simplicial frequency does not correspond to a constant edge flow but a globally conservative flow, i.e., divergence- and curl-free. 
Fig. \ref{fig:eig_vectors} shows examples of different eigenvectors and the associated eigenvalues, which would be the analogous of the complex exponentials in discrete-time signal processing for the edge space in Fig. \ref{1a}. 

\smallskip\noindent\textbf{Simplicial embeddings.} From Proposition \ref{prop.correspondence} and three types of simplicial frequencies, we can interpret the SFT by the following diagonalization of $\bbL_1$ 
\begin{equation} \label{eq.evd-l1-per-comp}
  \bbL_1 = \bbU_1 \blkdiag \big(\bLambda_{\rm{H}}, \bLambda_{\rm{G}}, \bLambda_{\rm{C}} \big) \bbU_1^\top
\end{equation}
with $\bbU_1 = \big[\bbU_{\rm{H}} \,\, \bbU_{\rm{G}} \,\, \bbU_{\rm{C}}\big]$ and $\blkdiag(\bbA,\bbB,\bbC)$ a block-diagonal matrix containing the square matrices $\bbA,\bbB,\bbC$ as diagonal blocks. 
Similarly,  we have
\begin{subequations}
\begin{equation} \label{eq.evd-l1l-per-comp}
  \bbL_{1,\ell}  = \bbU_1 \blkdiag \big(\mathbf{0}, \bLambda_{\rm{G}}, \mathbf{0} \big) \bbU_1^\top 
  \end{equation} \vspace{-4mm}
  \begin{equation} \label{eq.evd-l1u-per-comp}
    \bbL_{1,\rm{u}}  = \bbU_1 \blkdiag \big(\mathbf{0}, \mathbf{0}, \bLambda_{\rm{C}} \big) \bbU_1^\top 
\end{equation}
\end{subequations}
for the lower and upper Laplacians,
where $\mathbf{0}$ is an all-zero matrix of appropriate dimensions. Such insightful eigendecompositions enable us to define the following three embeddings of an edge flow $\bbf\in\setR^{N_1}$, 
\begin{equation} \label{eq.three-embeddings}
\begin{cases}
    \tilde{\bbf}_{\rm{H}} = \bbU_{\rm{H}}^\top \bbf = \bbU_{\rm{H}}^\top \bbf_{\rm{H}} \in \setR^{N_{\rm{H}}}, & \emph{harmonic embedding}\\
    \tilde{\bbf}_{\rm{G}} = \bbU_{\rm{G}}^\top \bbf = \bbU_{\rm{G}}^\top \bbf_{\rm{G}} \in \setR^{N_{\rm{G}}}, & \emph{gradient embedding}\\
    \tilde{\bbf}_{\rm{C}} = \bbU_{\rm{C}}^\top \bbf = \bbU_{\rm{C}}^\top \bbf_{\rm{C}} \in \setR^{N_{\rm{C}}}, & \emph{curl embedding}.\\
\end{cases}
\end{equation}
They follow from the orthogonality of the three components given by the Hodge decomposition. 
Equivalently, we can write the SFT of $\bbf$ as $\tilde{\bbf} = [
\tilde{\bbf}_{\rm{H}}^\top, \tilde{\bbf}_{\rm{G}}^\top , \tilde{\bbf}_{\rm{C}}^\top 
]^\top$. Each entry of an embedding represents the weight the flow has on the corresponding eigenvector (simplicial Fourier basis vector), e.g., entry $[\tilde{\bbf}_{\rm{G}}]_i$ is the SFT of $\bbf$ at the $i$th gradient frequency $\lambda_{{\rm{G}},i}$. Such an embedding provides a compressed representation of the edge flow if they present a degree of sparsity {\cite{barbarossa2020}} and allows us to differentiate different types of edge flow, e.g., to cluster trajectories and analyze ocean drift data \cite{schaub2020random, roddenberry2021hodgelets}.  

\subsection{Filter Frequency Response} \label{sec:frequency-response}
Upon defining the SFT, we can analyze the frequency response of the simplicial convolutional filter \eqref{eq.sf-k}. From the diagonalizations \eqref{eq.evd-l1l-per-comp} and \eqref{eq.three-embeddings}, we see that the lower simplicial shifting of an edge flow affects only the gradient component and the upper shifting affects only the curl component, i.e., 
\begin{equation} \label{eq.shifting-freq-domain}
  \bbL_{1,\ell}\bbf = \bbU_{\rm G} \bLambda_{\rm G} \tilde{\bbf}_{\rm G},\ \bbL_{1,\rm{u}}\bbf = \bbU_{\rm C} \bLambda_{\rm C} \tilde{\bbf}_{\rm C}.
\end{equation}
Through diagonalizing $\bbH_1$ by $\bbU_1 = \big[\bbU_{\rm{H}} \,\, \bbU_{\rm{G}} \,\, \bbU_{\rm{C}} \big] $, we can find the frequency response of $\bbH_1$ as 
\begin{equation}  \label{eq.freq-response-matrix}
  \tilde{\bbH}_1 = \bbU_1^\top \bbH_1 \bbU_1= \blkdiag \big(\tilde{\bbH}_{\rm H},  \tilde{\bbH}_{\rm G}, \tilde{\bbH}_{\rm C}   \big),
\end{equation}
where $\tilde{\bbH}_{\rm H}=h_0 \bbI$, $\tilde{\bbH}_{\rm G}=h_0\bbI + \sum_{l_1=1}^{L_1} \alpha_{l_1} \bLambda_{\rm{G}}^{l_1}$ and $\tilde{\bbH}_{\rm C}=h_0\bbI +\sum_{l_2=1}^{L_2} \beta_{l_2} \bLambda_{\rm{C}}^{l_2}$.  At an arbitrary frequency $\lambda$, the frequency response $\tilde{H}_1(\lambda)$ is given by
\begin{equation} \label{eq.freq-response}
    \begin{cases}
      \tilde{H}_{\rm H}(\lambda):=h_0, &\text{ for } \lambda\in \ccalQ_{\rm{H}}, \\
      \tilde{H}_{\rm G}(\lambda):=h_0 + \sum_{l_1 = 1}^{L_1}\alpha_{l_1}\lambda^{l_1}, &\text{ for }  \lambda \in \ccalQ_{\rm{G}}, \\
      \tilde{H}_{\rm C}(\lambda):=h_0 + \sum_{l_2=1}^{L_2}\beta_{l_2}\lambda^{l_2}, &\text{ for }  \lambda \in \ccalQ_{\rm{C}}, \\
    \end{cases}
\end{equation}
which is the filter frequency response at the harmonic, gradient and curl frequencies, respectively.  {By the definition of spectral filtering, the filter $\bbH_1$ cannot distinguish the signal components belonging to the subspace spanned by the eigenvectors associated to an eigenvalue of multiplicity greater than one. For instance, the filter cannot respond differently to multiple harmonic components, but only scale them by a factor $h_0$.}
In addition, we make the following three observations.
\begin{enumerate}
  \item Filter $\bbH_1$ controls the different frequency types independently. The coefficient $h_0$ determines the harmonic frequency response and contributes to the whole simplicial spectrum. The coefficients $\balpha$ and $\bbeta$ contribute only to the gradient and curl frequency response, respectively. This independent control on different signal subspaces corresponds to the different parameters imposed on the lower and upper adjacencies in the simplicial domain. In  {contrast}, if setting $L_1=L_2$ and $\balpha=\bbeta$ in $\bbH_1$, the filter cannot regulate the gradient and curl spaces independently and has less flexibility. 

  \item The gradient frequency response is fully determined by the matrix polynomial in $\bbL_{1,\ell}$ [cf. \eqref{eq.evd-l1l-per-comp} and \eqref{eq.freq-response-matrix}]. Thus, if $L_2=0$, $\bbH_1$ has as responses $h_0$ for $\lambda\in\ccalQ_{\rm{H}}\cup\ccalQ_{\rm{C}}$, and $\tilde{H}_{\rm{G}}=h_0 + \sum_{l_1 = 1}^{L_1}\alpha_{l_1}\lambda^{l_1}, \text{ for }  \lambda \in \ccalQ_{\rm{G}}$. This controls the gradient and non-gradient frequencies with a reduced design burden due to fewer parameters. But we give up the control on the curl frequencies.  Likewise for the case of $L_1=0$, which is beneficial when only curl components need to be tuned.
  
  \item  {At two \emph{overlapping} frequencies  $\lambda_1=\lambda_2$ with $\lambda_1\in\ccalQ_{\rm{G}}$, and $\lambda_2\in\ccalQ_{\rm{C}}$, filter $\bbH_1$ responds differently as $\tilde{H}_{\rm{G}}(\lambda_1)$ and $\tilde{H}_{\rm{C}}(\lambda_2)$ [cf. \eqref{eq.freq-response}], respectively. This cannot be realized for the filter $\bbH_1=\sum_{l=0}^{L}h_l\bbL_1^l$, i.e., setting $L_1=L_2$ and $\balpha=\bbeta$, which follows $\tilde{H}_{\rm{G}}(\lambda_1)=\tilde{H}_{\rm{C}}(\lambda_2)$ instead. In the latter case, it will preserve the unwanted curl component at $\lambda_2$ when setting $\tilde{H}_{\rm{G}}(\lambda)=1$, for $\lambda\in\ccalQ_{\rm{G}}$, when the goal is to extract the gradient component.}
\end{enumerate}

\section{Filter Design} \label{sec:filter-design}
Given a training set of input-output edge flow relations $\ccalT=\{(\bbf_1,\bbf_{\rm{o},1}),\dots,(\bbf_{|\ccalT|},\bbf_{{\rm{o}},|\ccalT|})\}$, we can learn the filter coefficients in a data-driven fashion by fitting the filtered output $\bbH_1\bbf$ to the output $\bbf_{\rm{o}}$. Specifically, consider a mean squared error (MSE) cost function and a regularizer $r(h_0,\balpha,\bbeta)$ to avoid overfitting, we formulate the problem as
\begin{equation} \label{eq.data-driven-design-1}
  \underset{h_0,\balpha,\bbeta}{\min} \frac{1}{|\ccalT|} \sum_{(\bbf_{i},\bbf_{{\rm{o}},i})\in\ccalT} \lVert \bbH_1\bbf_i - \bbf_{{\rm{o}},i} \lVert^2_2 \, + \, \gamma r(h_0,\balpha,\bbeta),
\end{equation}
with $\gamma>0$. A flow prediction based on \eqref{eq.data-driven-design-1} is detailed in \cite{yang2021finite}. 

In this section, we focus in detail on designing the simplicial filter given a desired frequency response.
Specifically, we assume a desired frequency response $g_0$ at the harmonic frequency $\lambda=0$, a gradient frequency response $g_{\rm{G}}(\lambda)$ for $\lambda\in\ccalQ_{\rm{G}}$, and a curl frequency response $g_{\rm{C}}(\lambda)$ for $\lambda\in\ccalQ_{\rm{C}}$. To design the coefficients $h_0,\balpha,\bbeta$, our goal is then to approximate the desired response by the filter frequency response $\tilde{H}_1(\lambda)$ [cf. \eqref{eq.freq-response}], which can be formulated as
\begin{equation} \label{eq.spectral-design}
  \begin{cases}
  h_0  \approx g_0,  & \text{ for } \lambda_i = 0, \\
  h_0 + \sum_{l_1=1}^{L_1} \alpha_{l_1}\lambda_i^{l_1}  \approx g_{\rm{G}}(\lambda_i) ,  & \text{ for } \lambda_i \in \ccalQ_{\rm{G}}, \\
  h_0 + \sum_{l_2=1}^{L_2} \beta_{l_2}\lambda_i^{l_2}  \approx g_{\rm{C}}(\lambda_i) ,  & \text{ for } \lambda_i \in \ccalQ_{\rm{C}}.
  \end{cases}
\end{equation} 
In the following, we first use a standard least-squares (LS) approach to solve \eqref{eq.spectral-design}. Later, we consider a universal design to avoid the eigenvalue computation when a continuous desired frequency response is given. In particular, we consider a grid-based and a Chebyshev polynomial approach.

\subsection{Least-Squares Filter Design} \label{sec:ls-filter-design}

Denote the number of distinct gradient frequencies in $\ccalQ_{\rm{G}}$ by $D_{\rm{G}}$, and that of distinct curl frequencies in $\ccalQ_{\rm{C}}$ by $D_{\rm{C}}$. Then, the three sets of equations in \eqref{eq.spectral-design} contain respectively one, $D_{\rm{G}}$ and $D_{\rm{C}}$ distinct linear equations. Let $\bbg=[g_0, \bbg_{\rm G}^\top, \bbg_{\rm C}^\top]^\top$ collect the desired responses at distinct frequencies where $[\bbg_{\rm{G}}]_i = g_{\rm{G}}(\lambda_{{\rm{G}},i})$, for $i=1\dots,D_{\rm{G}}$, is the response at the $i$th distinct gradient frequency $\lambda_{{\rm{G}},i}$, and $[\bbg_{\rm{C}}]_i = g_{\rm{C}}(\lambda_{{\rm{C}},i})$, for $i=1\dots,D_{\rm{C}}$, at the $i$th distinct curl frequency. Then, we can obtain the filter coefficients by solving the LS problem
\begin{equation}\label{eq.ls-1}
  \underset{h_0,\balpha,\bbeta}{\min} \left\|  \begin{bmatrix} {\bf 1} \ \vline &  \begin{matrix} {\bf 0} \\ \hline \begin{matrix} 
  \bPhi_{\rm{G}} & {\bf 0} \\ {\bf 0} & \bPhi_{\rm{C}}
  \end{matrix} \end{matrix}  \end{bmatrix} 
  \begin{bmatrix}
  h_0 \\ \balpha \\ \bbeta
  \end{bmatrix}
  - \bbg
  \right\|^2_2,
\end{equation}
where $\bf 1$ ($\bf 0$) is an all-one (all-zero) matrix or vector of an appropriate dimension, 
$\bPhi_{\rm{G}}\in\setR^{D_{\rm{G}} \times L_1}$ and $\bPhi_{\rm{C}}\in\setR^{D_{\rm{C}} \times L_2}$ are Vandermonde matrices with respective entries $[\bPhi_{\rm{G}}]_{ij} = \lambda_{{\rm{G}},i}^{j}$ and $[\bPhi_{\rm{C}}]_{ij} = \lambda_{{\rm{C}},i}^{j}$. We refer to problem \eqref{eq.ls-1} as the \emph{LS design}, which can be solved either with a direct solver or with a decoupled solver, studied as follows.

\smallskip\noindent\textbf{Direct LS design.} From the Cayley-Hamilton theorem \cite{horn2012matrix} and \cite[Thm. 3]{sandryhaila2013discrete}, we know that any analytical function of a matrix can be expressed as a matrix polynomial of degree less than its minimal polynomial degree, which equals the number of distinct eigenvalues for a positive semi-definite matrix. 
Thus, we can assume the filter orders $L_1\leq D_{\rm G}$ and $L_2\leq D_{\rm C}$. Under this condition, problem \eqref{eq.ls-1} admits a unique solution which can be obtained via the pseudo-inverse of the system matrix. Furthermore, when $L_1=D_{\rm{G}}$ and $L_2=D_{\rm{C}}$, $\bPhi_{\rm{G}}$ and $\bPhi_{\rm{C}}$ are square and any two rows in them are linearly independent, the solution leads to a zero cost in \eqref{eq.ls-1}. We refer to this pseudo-inverse solution of \eqref{eq.ls-1} as the \emph{direct LS design}. 

In addition, given a desired edge operator $\bbG$, the following proposition states that it can be implemented by filter $\bbH_1$.
\begin{prop} \label{prop.perfect-implementation}
A desired linear operator $\bbG\in\setR^{N_1\times N_1}$ can be perfectly implemented by a simplicial filter $\bbH_1$ if the following three conditions hold true:
\begin{enumerate}[label=\roman*)]
    \item Matrices $\bbG$ and $\bbL_1$ are simultaneously diagonalizable, i.e., $\bbU_1$ forms an eigenvector basis for $\bbG$. Let 
    $\bbg = 
    [\bbg_{\rm{H}}^\top , \bbg_{\rm{G}}^\top , \bbg_{\rm{C}}^\top] ^\top 
    $ 
    collect the eigenvalues of $\bbG$. 
    \item If two eigenvalues of $\bbL_1$ are of the same frequency type and equal, the corresponding eigenvalues of $\bbG$ are also equal. For $\lambda_{{\rm{H}},i}=\lambda_{{\rm{H}},j}=0\in\ccalQ_{\rm{H}}$, it holds that $[\bbg_{\rm{H}}]_i = [\bbg_{\rm{H}}]_j$, for $\lambda_{{\rm{G}},i} = \lambda_{{\rm{G}},j}\in\ccalQ_{\rm G}$, $[\bbg_{\rm{G}}]_i = [\bbg_{\rm{G}}]_j$, and for $\lambda_{{\rm{C}},i} = \lambda_{{\rm{C}},j}\in\ccalQ_{\rm C}$, $[\bbg_{\rm{C}}]_i= [\bbg_{\rm{C}}]_j$.
    \item The filter orders of $\bbH_1$ fulfill $L_1\geq D_{\rm{G}}$, $L_2\geq D_{\rm{C}}$.
\end{enumerate}
\end{prop}
\begin{IEEEproof}
  See  Appendix \ref{proof.perfect-implementation}.
\end{IEEEproof}


\smallskip\noindent\textbf{Decoupled LS design.} We can reduce the  complexity of solving \eqref{eq.ls-1} by decoupling the cost function for different frequency types. 
First, we rewrite problem \eqref{eq.ls-1} as 
\begin{equation} \small \label{eq.ls-2}
  \begin{aligned}
  \underset{h_0,\balpha,\bbeta}{\min} & \left\| 
  \big[
  \mathbf{1} \,\, \bPhi_{\rm{G}}
  \big] 
  \begin{bmatrix}
  h_0 \\ \balpha
  \end{bmatrix}
  - \bbg_{\rm{G}}
  \right\|^2_2 + 
   \left\| 
  \big[
  \mathbf{1} \,\, \bPhi_{\rm{C}}
  \big]
  \begin{bmatrix}
  h_0 \\ \bbeta
  \end{bmatrix}
  - \bbg_{\rm{C}}
  \right\|^2_2 \\ &  + \left\| h_0 - g_0 \right\|^2_2 . 
  \end{aligned}
\end{equation}
To approximate a solution, we ignore the dependence of the first two terms in \eqref{eq.ls-2} on coefficient $h_0$ and solve the last term separately to estimate $h_0$. 
We then substitute the estimate $\hat{h}_0$ in the first two terms to obtain $\hat{\balpha}$ and $\hat{\bbeta}$, given by
\begin{equation} \label{eq.ls-2-solution}
   \hat{h}_0 = g_0, \,
   \hat{\balpha} = \bPhi_{\rm{G}}^\dagger (\bbg_{\rm{G}} - g_0 \mathbf{1}), \,
   \hat{\bbeta} = \bPhi_{\rm{C}}^\dagger (\bbg_{\rm{C}} - g_0 \mathbf{1}),
\end{equation}
which, referred to as the \emph{decoupled LS design}, is suboptimal compared to the direct LS design. 
The following proposition discusses this suboptimality.
\begin{prop}\label{prop.optimality}
  The decoupled LS design \eqref{eq.ls-2-solution} converges to the direct solution of \eqref{eq.ls-1} as $\lVert \bPhi_{\rm G} \bPhi_{\rm G}^\dagger - \bbI\lVert_F \rightarrow 0$ and  $\lVert\bPhi_{\rm C} \bPhi_{\rm C}^\dagger -\bbI\lVert_F \rightarrow 0$ where $\lVert \cdot \lVert_F$ is the Frobenius norm. 
\end{prop}
\begin{IEEEproof}
  See  Appendix \ref{proof.optimality}.
\end{IEEEproof}

Proposition \ref{prop.optimality} states that as the pseudo-inverses of $\bPhi_{\rm G}$ and $\bPhi_{\rm C}$ become closer to the true inverses, the decoupled solution converges to the direct solution. 
Moreover, the decoupled LS design reduces the computational cost to $\ccalO(D_{\rm G}^3 + D_{\rm C}^3)$ from $\ccalO((1+D_{\rm{G}}+D_{\rm C})^3)$ for the direct one. 
However, both designs require the computation of the eigenvalues of the Hodge Laplacians, which takes in general a computational complexity of $\ccalO(N_1^3)$  \cite{watkins2007matrix}. 
In the sequel, we consider a universal design strategy to avoid the eigenvalue computation, specifically, a grid-based design and a Chebyshev polynomial design \cite{shuman2018distributed}. 

\subsection{Grid-Based Filter Design} \label{sec:grid-based-design}
\begin{figure}[!t]
  \centering
  \includegraphics[width=0.9\linewidth,scale=0.5]{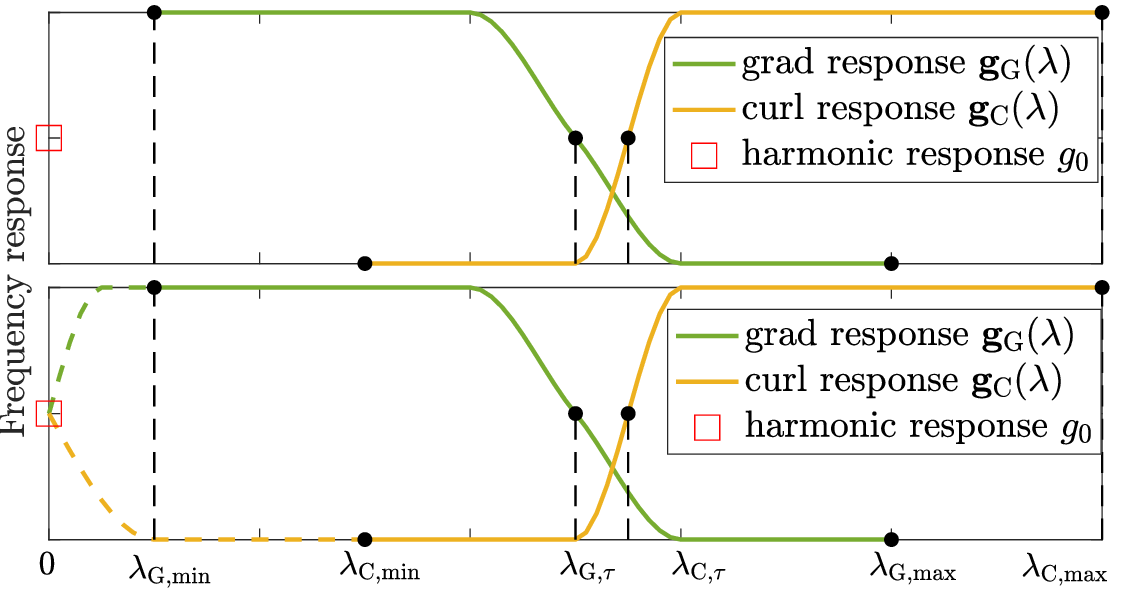}
  \caption{An example of a continuous frequency response for the grid-based (top) and Chebyshev (bottom) filter design, which promotes low gradient frequencies and the high curl frequencies. The cut-off frequencies are subscripted by $\tau$. The harmonic frequency response is given at frequency 0. For the Chebyshev polynomial design, we require that the gradient and curl are continuous functions starting from frequency 0 and $g_{\rm{G}}(0)=g_{\rm{C}}(0)=g_0$.}
  \label{fig:continuous-desired-frequency-response}
\end{figure}

The grid-based filter design aims to match the desired frequency response in a continuous interval where the exact frequencies lie such that the eigenvalue computation of $\bbL_1$ can be avoided. 
Given a harmonic frequency response $g_0$, a continuous gradient frequency response $g_{\rm G}(\lambda), \lambda \in [\lambda_{\rm G, \min}, \lambda_{\rm G, \max}]$ and a continuous curl frequency response $g_{\rm C}(\lambda), \lambda \in [\lambda_{\rm C, \min}, \lambda_{\rm C, \max}]$,  we want that
\begin{equation} \label{eq.grid-based-design}
  \begin{cases}
   h_0 - g_0 \approx 0   \\
   \int_{\lambda_{{\rm{G}},\min}}^{\lambda_{{\rm{G}},\max}} \big|h_0+\sum_{l_1=1}^{L_1}\alpha_{l_1}\lambda^{l_1} - g_{\rm{G}}(\lambda) \big|^2 {\rm{d}} \lambda \approx 0  \\
   \int_{\lambda_{{\rm{C}},\min}}^{\lambda_{{\rm{C}},\max}} \big|h_0+\sum_{l_2=1}^{L_2}\beta_{l_2}\lambda^{l_2} - g_{\rm{C}}(\lambda) \big|^2 {\rm{d}} \lambda \approx 0 , \\
  \end{cases}
\end{equation}
which is a continuous version of~\eqref{eq.spectral-design}. 
An example of the continuous frequency response is given in Fig. \ref{fig:continuous-desired-frequency-response} (top).

By sampling $M_1$ and $M_2$  (grid)-points uniformly from the intervals $[\lambda_{\rm G, \min}, \lambda_{\rm G, \max}] $ and $[\lambda_{\rm C, \min}, \lambda_{\rm C, \max}]$, the problem \eqref{eq.grid-based-design} can then be formulated as an LS problem of form \eqref{eq.ls-1} but with the sampled frequencies as the entries of the system matrix instead of the true eigenvalues.
We can again solve this LS problem either via a direct pseudo-inverse of the system matrix or via the decoupled solution method [cf.\eqref{eq.ls-2-solution}]. 
Notice that the largest true eigenvalue can be approximated by efficient algorithms, e.g., power iteration, \cite{watkins2007matrix,sleijpen2000jacobi}.
For the smallest, we can set a small value greater than 0 as the lower bound. 

\subsection{Chebyshev Polynomial Filter Design}
\label{sec:chebyshev-implementation}
Both discussed filter designs rely on solving an LS problem, which has a Vandermonde matrix as the system matrix, and suffers from numerical instability. 
To tackle this issue, we consider a Chebyshev polynomial based filter design \cite{shuman2018distributed, druskin1989two}. 
As illustrated in Fig. \ref{fig:continuous-desired-frequency-response} (bottom), consider a continuous gradient frequency response $g_{\rm G}(\lambda),\lambda\in[0, \lambda_{\rm G, \max}]$ and a continuous curl frequency response $g_{\rm C}(\lambda),\lambda\in[0, \lambda_{\rm C, \max}]$. 
We further require that $g_{\rm G}(0) = g_{\rm C}(0)=g_0$. 
That is, the continuous functions $g_{\rm G}(\lambda)$ and $g_{\rm C}(\lambda)$ are defined starting from frequency 0, at which they are equal to the harmonic frequency response $g_0$.  

As the filter $\bbH_1$ is a sum of matrix polynomials of $\bbL_{1,\ell}$ and $\bbL_{1,\rm u}$, our strategy is to first consider the Chebyshev polynomial design for each of them so to separately obtain the gradient and curl frequency responses, then sum these two polynomials to obtain the final filter. 
However, one type of frequency response could be affected unwantedly by the identity matrix term in the Chebyshev polynomial designed for the other frequency response, as detailed later. 
The requirement that $g_{\rm{G}}(0)=g_{\rm{C}}(0)=g_0$ allows a possible correction.

First, we approximate the gradient frequency response via a truncated series of shifted Chebyshev polynomials $\bbH_{\ell}:=\bbH_{\ell}(\bbL_{1,\ell})$. Let $\bar{P}_l(\lambda),\lambda\in[-1,1]$ be the $l$th Chebyshev polynomial of the first kind \cite{mason2002chebyshev}. 
We perform a transformation $P_l(\lambda):=\bar{P}_l(\frac{\lambda-\omega}{\omega})$ with $\omega:=\frac{\lambda_{{\rm{G}},\max}}{2}$ to shift the domain to $[0,\lambda_{{\rm{G}},\max}]$. 
We then approximate the operator $g_{\rm{G}}(\bbL_{1,\ell})$ that has the gradient frequency response $g_{\rm G}(\lambda)$ by $\bbH_{\ell}$ of order $L_1$
\begin{equation} \label{eq.chebyshev-1}
     \bbH_{\ell} = \frac{1}{2}c_{\ell,0} \bbI + \sum_{l_1=1}^{L_1} c_{{\ell},l_1} P_{l_1}(\bbL_{1,\ell})
\end{equation}
where we have $P_{0}(\bbL_{1,\ell})=\bbI$, $P_{1}(\bbL_{1,\ell})=\frac{2}{\lambda_{\rm{G},\max}} \bbL_{1,\ell} -  \bbI$, the $l_1$th Chebyshev term, for $l_1\geq 2$, is
\begin{equation}\label{eq.chebyshev-term}
    P_{l_1}(\bbL_{1,\ell}) = 2 P_{1}(\bbL_{1,\ell})  P_{l_1-1}(\bbL_{1,\ell}) - P_{l_1-2}(\bbL_{1,\ell}),
\end{equation}
and the Chebyshev coefficients $c_{{\ell},l_1}$ can be computed as 
\begin{equation} \label{eq.chebyshev-coeff}
  c_{{\ell},l_1} = \frac{2}{\pi} \int_{0}^{\pi} \cos(l_1\phi) g_{\rm{G}}\big( \omega (\cos\phi + 1) \big) {\rm d} \phi.
\end{equation}
The frequency response $\tilde{H}_{\ell}(\lambda)$ of $\bbH_{\ell}$ can be found as 
\begin{equation} \label{eq.chebyshev-freq-response-l}
  \begin{cases}
  \begin{aligned}
      & p_{\ell,0} := \frac{1}{2}c_{\ell,0}+\sum_{l_1=1}^{\floor{L_1/2}} (c_{{\ell},2l_1} - c_{{\ell},2l_1-1}), \text{ for } \lambda\in\ccalQ_{\rm{H}} \cup \ccalQ_{\rm{C}} \\
      &\tilde{H}_{{\ell,\rm{G}}}(\lambda) := \frac{1}{2}c_{\ell,0} + \sum_{l_1=1}^{L_1} c_{{\ell},l_1} P_{l_1}(\lambda), \text{ for } \lambda\in\ccalQ_{\rm{G}},
  \end{aligned}
  \end{cases}
\end{equation}
with coefficient $p_{\ell,0}$ on the identity term of $\bbH_{\ell}$, which is the frequency response at the harmonic and curl frequencies, associated to the kernel of $\bbL_{1,\ell}$ [cf. \eqref{eq.evd-l1l-per-comp}]. 
For a reasonably large $L_1$ we have $p_{\ell,0}\approx g_0$ and $\tilde{H}_{{\ell,\rm{G}}}(\lambda) \approx g_{\rm{G}}(\lambda), \lambda\in\ccalQ_{\rm{G}}$.

Second, to approximate the curl frequency response $g_{\rm{C}}(\lambda)$, we follow the same procedure [cf. \eqref{eq.chebyshev-1}-\eqref{eq.chebyshev-coeff}] to obtain the Chebyshev polynomial $\bbH_{\rm{u}}:=\bbH_{\rm{u}}(\bbL_{1,\rm{u}})$ of order $L_2$  
\begin{equation} \label{eq.chebyshev-u}
    \bbH_{\rm{u}} = \frac{1}{2}c_{\rm{u},0} \bbI + \sum_{l_2=1}^{L_2} c_{{\rm{u}},l_2} P_{l_2}(\bbL_{1,\rm{u}}).
\end{equation}
It has a frequency response $\tilde{H}_{\rm{u}}(\lambda)$ 
\begin{equation} \label{eq.chebyshev-freq-response-u}
  \begin{cases}
  \begin{aligned}
    &p_{\rm{u},0} := \frac{1}{2}c_{\rm{u,0}}+\sum_{l_2=1}^{\floor{L_2/2}} (c_{{\rm{u}},2l_2} - c_{{\rm{u}},2l_2-1}), \text{ for } \lambda\in\ccalQ_{\rm{H}} \cup \ccalQ_{\rm{G}} \\
    &\tilde{H}_{{\rm{u,C}}}(\lambda):=\frac{1}{2}c_{\rm{u},0} + \sum_{l_2=1}^{L_2} c_{{\rm{u}},l_2} P_{l_2}(\lambda), \text{ for } \lambda\in\ccalQ_{\rm{C}}.
  \end{aligned}
  \end{cases}
\end{equation}
with coefficient $p_{\rm{u},0}$ on the identity term  of $\bbH_{\rm{u}}$, which is the frequency response at the harmonic and gradient frequencies, associated to the kernel of $\bbL_{1,\rm{u}}$ [cf. \eqref{eq.evd-l1u-per-comp}]. 
For a reasonably large $L_2$ we have $p_{\rm{u},0}\approx g_0$ and $\tilde{H}_{{\rm{u,C}}}(\lambda) \approx g_{\rm{C}}(\lambda), \lambda\in\ccalQ_{\rm{C}}$.

Lastly, by summing $\bbH_{\ell}$ and $\bbH_{\rm{u}}$, we obtain a filter that approximates the gradient and curl frequency responses.
However, from \eqref{eq.chebyshev-freq-response-l}, we see that $\bbH_{\ell}$ generates a response $p_{\ell,0}$ at both harmonic and curl frequencies. 
This will lift up the curl frequency response unwantedly by $p_{\ell,0}$. 
Similarly, $\bbH_{\rm{u}}$ has the effect of lifting the gradient frequency response by $p_{\rm{u},0}$ [cf. \eqref{eq.chebyshev-freq-response-u}]. 
By requiring that $g_{\rm{G}}(0)=g_{\rm{C}}(0)=g_0$, we can remove this unwanted influence by subtracting a term $g_0\bbI$ from the summation.
Hence, the final Chebyshev polynomial design $\bbH_{1}$ of orders $L_1$ and $L_2$ is given by
\begin{equation} \label{eq.chebyshev-2}
  \bbH_{1} = \bbH_{\ell} + \bbH_{\rm{u}}- g_0\bbI,
\end{equation}
which has a frequency response $\tilde{H}_{1}(\lambda)$  
\begin{equation} \label{eq.chebyshev-freq-response}
  \begin{cases}
    p_{\ell,0} + p_{\rm{u},0} - g_0 ,& \text{for } \lambda\in\ccalQ_{\rm{H}} \\
    \tilde{H}_{{\rm{l,G}}}(\lambda) + p_{\rm{u},0} - g_0, & \text{for } \lambda\in\ccalQ_{\rm{G}} \\
    \tilde{H}_{{\rm{u,C}}}(\lambda) + p_{\ell,0} - g_0, & \text{for } \lambda\in\ccalQ_{\rm{C}}.
  \end{cases}
\end{equation} 
The following proposition states that the approximation error of the Chebyshev polynomial design \eqref{eq.chebyshev-2} is bounded. 

\begin{prop} \label{prop.chebyshev-bound}
  Let $\bbG$ be the desired operator corresponding to the continuous gradient and curl frequency responses $g_{\rm G}(\lambda)$ with $\lambda\in[0,\lambda_{\rm{G},\max}]$ and  $g_{\rm C}(\lambda)$ with $\lambda\in[0,\lambda_{\rm{C},\max}]$, as well as the harmonic one $g_{\rm G}(0) = g_{\rm C}(0)=g_0$. Let $\bbH_{1}$ [cf. \eqref{eq.chebyshev-2}] be a truncated series of Chebyshev polynomials of orders $L_1$ and $L_2$ with frequency response $\tilde{H}_{1}(\lambda)$ [cf. \eqref{eq.chebyshev-freq-response}]. Define
  \begin{equation}
    \begin{aligned}
      B_1(L_1):=  
     \sup_{\lambda\in[0,\lambda_{\rm{G},\max}]} \bigl\{ \big| \tilde{H}_{{\ell}}(\lambda) - g_0 - g_{\rm{G}}(\lambda) \big| \bigr\},  \\
      B_2(L_2):=   \sup_{\lambda\in[0,\lambda_{\rm{C},\max}]} \bigl\{ \big|  \tilde{H}_{{\rm{u}}}(\lambda) - g_0 - g_{\rm{C}}(\lambda) \big| \bigr\},
    \end{aligned}
  \end{equation} 
  and $B:=\max \big\{B_1(L_1),B_2(L_2)\big\}$. Then, we have that
  \begin{equation}
     \lVert \bbG - \bbH_{1} \lVert_2 := \underset{\bbf\neq\mathbf{0}}{\max}  \frac{\lVert (\bbG - \bbH_{1}) \bbf \lVert_2}{\lVert \bbf \lVert_2} \leq  B
  \end{equation}
\end{prop}
\begin{IEEEproof}
  See  Appendix \ref{proof.chebyshev-bound}.
\end{IEEEproof} 

When $g_{\rm{G}}(\cdot)$ and $g_{\rm{C}}(\cdot)$ are real analytic, a stronger bound can be found \cite{shuman2018distributed,hammond2011wavelets}. In addition, we make the following three comments.
\begin{enumerate*}[label=\roman*)]
    \item When only the gradient frequency response $g_{\rm G}(\lambda)$ is of interest, we can directly consider the Chebyshev polynomial $\bbH_{\ell}$ [cf. \eqref{eq.chebyshev-1}] with $L_2=0$. Likewise we consider $\bbH_{\rm{u}}$ when only $g_{\rm{C}}(\lambda)$ is of interest.
    \item If a frequency response $g(\lambda)$ is given on the whole spectrum, we can build a filter $\bbH=\sum_{l=0}^{L}h_l\bbL_1^{l}$ based on a Chebyshev polynomial design analogous to the graph filter case \cite{shuman2018distributed}. 
    \item The Chebyshev polynomial design requires no eigenvalue computation of $\bbL_1$. Thus, it does not suffer from numerical instability, allows to build simplicial filters with large $L_1$ and $L_2$ for an accurate design, and admits a recursive and distributed implementation due to the Chebyshev
    polynomial property \eqref{eq.chebyshev-term}. 
\end{enumerate*}


\section{Applications} \label{sec:application-numerical-experiments}
In this section, we first discuss how to use a simplicial filter for subcomponent extraction and edge flow denoising. 
We then consider analyses of financial markets, street and traffic networks.
These are similar to previous works \cite{schaub2018flow, barbarossa2020, jiang2011statistical,youn2008price,schaub2014structure}, but we directly use our here developed filters instead of employing a regularized optimization problem (which implicitly defines a low-pass filter).

To gauge the performance in estimation tasks, we use the normalized root mean square error (NRMSE), $e = \lVert \hat{\bbf} - \bbf_0 \lVert_2/ \lVert \bbf_0 \lVert_2$ with the flow estimate $\hat{\bbf}$ and the true flow $\bbf_0$. 
For filter design problems, we evaluate the spectral norm $\lVert \bbH_1 -\bbG\lVert_2$ with the designed filter $\bbH_1$ and the true operator $\bbG$. The Chebfun toolbox was used for the Chebyshev polynomial filter design \cite{Driscoll2014}.

\subsection{Subcomponent Extraction} \label{sec:subcomponent-extraction}
In pairwise ranking problems, we aim to rank alternatives by comparing their scores. The work \cite{jiang2011statistical} modeled the score differences between alternatives as edge flows on a pairwise comparison graph. The gradient component of these flows gives a global ranking and the curl components measures the inconsistency of the ranking. For further examples see \cite{candogan2011flows, gebhart2021go, mock2021political}. 
A common approach to obtain the three components of the flow is to compute \cite{lim2015hodge, jiang2011statistical, barbarossa2020}: 
\begin{equation} \label{eq.subcomponent-extraction-typical-ls-solution}
    \hat{\bbf}_{\rm{G}}  =  \bbP_{\rm{G}} \bbf, \,\,
    \hat{\bbf}_{\rm{C}}  = \bbP_{\rm{C}} \bbf, \,\,
    \hat{\bbf}_{\rm{H}}  = \bbP_{\rm{H}} \bbf = \bbf - \hat{\bbf}_{\rm{G}} - \hat{\bbf}_{\rm{C}}.
\end{equation}
where $\bbP_{\rm{G}} = \bbB_1^\top (\bbB_1\bbB_1^\top)^\dagger \bbB_1$ is the projection onto the gradient space, the curl projector is $\bbP_{\rm{C}} = \bbB_2(\bbB_2^\top\bbB_2)^\dagger \bbB_2^\top$ and the harmonic projector $\bbP_{\rm{H}} = \bbI- \bbL_1\bbL_1^\dagger$. 
Notably, we can use a (polynomial) simplicial filter $\bbH_1$ to implement these operators, too.

\begin{lem} \label{lemma.subcomponent-extraction-lemma}
  The projection operators \eqref{eq.subcomponent-extraction-typical-ls-solution} are equivalent to $\bbP_{\rm{G}}=\bbU_{\rm{G}}\bbU_{\rm{G}}^\top$, $\bbP_{\rm{G}}=\bbU_{\rm{C}}\bbU_{\rm{C}}^\top$ and $\bbP_{\rm{H}} = \bbU_{\rm{H}}\bbU_{\rm{H}}^\top$.
  As $\bbU$ requires the knowledge of global properties in this form the projections cannot be computed in a distributed way.
  
  However, for $L_1=D_{\rm{G}}$, $L_2=D_{\rm{C}}$, there exists a unique $\{h_0,\balpha,\bbeta\}$ such that $\bbH_1=\bbP_{\rm{G}}$. These coefficients can be found by solving the system \eqref{eq.ls-1} with $g_0=0,\bbg_{\rm{C}}=\mathbf{0}$ and $\bbg_{\rm{G}}=\bb1$; analogous arguments lead to distributed implementations of $\bbH_1=\bbP_{\rm{C}}$ and $\bbH_1=\bbP_{\rm{H}}$.
\end{lem}
\begin{IEEEproof}
  See Appendix \ref{proof.subcomponent-extraction-lemma}.
\end{IEEEproof}

For the gradient and curl components, this can be simplified.

\begin{cor} \label{cor.subcomponent-extraction-cor}
  For the gradient projector $\bbP_{\rm{G}}$, there exist a filter with $L_1=D_{\rm{G}},L_2=0$ and a unique $\{h_0,\balpha\}$ such that $\bbH_1=\bbP_{\rm{G}}$. The solution is $h_0=0$ and $\balpha=\bPhi_{\rm{G}}^{-1}\bb1$. For the curl projector $\bbP_{\rm{C}}$, there exist a filter with $L_1=0,L_2=D_{\rm{C}}$ and a unique $\{h_0,\bbeta\}$ such that $\bbH_1=\bbP_{\rm{C}}$. The solution is $h_0=0$ and $\bbeta=\bPhi_{\rm{C}}^{-1}\bb1$
\end{cor}
\begin{IEEEproof}
  See  Appendix \ref{proof.cor.subcomponent-extraction-cor}.
\end{IEEEproof}

Corollary \ref{cor.subcomponent-extraction-cor} shows a gradient projector can be build solely upon $\bbL_{1,\ell}$, since $\bbL_{1,\rm{u}}$ has no effect in the gradient space [cf. \eqref{eq.shifting-simplex-domain} and \eqref{eq.shifting-freq-domain}]. Similar arguments hold for the curl projector. 

\begin{figure}[!t]
  \centering
  \includegraphics[width=3.34in, scale=1]{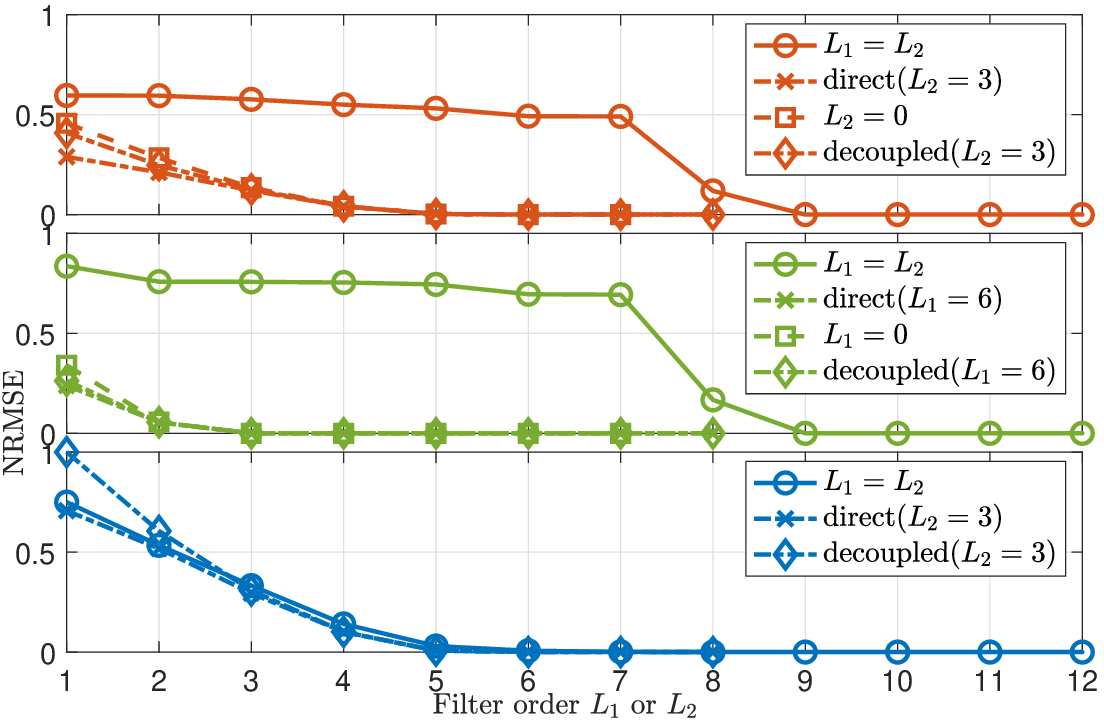}
  \caption{Subcomponent extraction performance by filters $\bbH_1$ with different parameters based on an LS design. The extraction becomes better as the filter order increases. For the gradient and curl components, setting $L_1=L_2$ and $\balpha=\beta$ worsens the performance because of the limited expressive power.  {For the general filter form}, the direct and decoupled LS designs have smaller performance difference as the filter order grows (see Proposition \ref{prop.optimality}).}
  \label{fig:subcomponent-extraction-synthetic}
\end{figure}

Fig. \ref{fig:subcomponent-extraction-synthetic} reports the performance of the subcomponent extraction based on $\bbH_1$ via an LS design. 
We generated a synthetic edge flow $\bbf = \bbU_1 \tilde{\bbf}$ with a flat spectrum $\tilde{\bbf} = \bb1$ on the SC in Fig. \ref{1a}.  We observe that as the filter order increases, the filter performs better as its expressive power increases. Setting $L_1=L_2$ with $\balpha=\bbeta$ reduces the expressive power as expected. If filter orders obey $L_1\geq 6$ and $L_3\geq 3$, then the gradient (or curl) component can be perfectly extracted, as  {shown in Fig. \ref{fig:flow_decomp_illustration} and} indicated by Lemma \ref{lemma.subcomponent-extraction-lemma} and Corollary \ref{cor.subcomponent-extraction-cor}. In the general parameter setting, the decoupled LS design performs closer to the direct LS design as the filter order grows as shown in Proposition \ref{prop.optimality}.

\subsection{Edge Flow Denoising} \label{sec:edge-flow-denoising}
Consider a noisy edge flow $\bbf = \bbf^{0} + \bepsilon$ with $\bbf^{0}$ the true edge flow and $\bepsilon$ a zero-mean white Gaussian noise. To obtain an estimate $\hat{\bbf}$, we can solve the regularized optimization problem~\cite{schaub2021, schaub2018flow}:
\begin{equation} \label{eq.regularization}
    \underset{\hat{\bbf}}{\min} \,\, \lVert \hat{\bbf} - \bbf \lVert_2^2 \, + \, \mu \hat{\bbf} \bbP \hat{\bbf},
\end{equation}
with an optimal solution $\hat{\bbf} = \bbH_{P}\hat{\bbf} := (\bbI+\mu \bbP)^{-1}\bbf$. 
For matrix $\bbP$ we have two choices:
\begin{enumerate*}[label=(\roman*)]
    \item the edge Laplacian $\bbL_{1,\ell} =  \bbB_1^\top\bbB_1$, leading to a regularizer $\lVert \bbB_1\hat{\bbf}\lVert_2^2$ to penalize the flows with a nonzero divergence \cite{schaub2018flow}; 
    \item the Hodge Laplacian $\bbL_1$, leading to a regularizer $\hat{\bbf} \bbL_1 \hat{\bbf} = \lVert \bbB_1\hat{\bbf}\lVert_2^2 + \lVert \bbB_2^\top\hat{\bbf}\lVert_2^2$ to penalize the flows with a nonzero divergence or curl \cite{schaub2021}.
\end{enumerate*}

\begin{figure}[!t]
    \centering
    \includegraphics[scale=0.48]{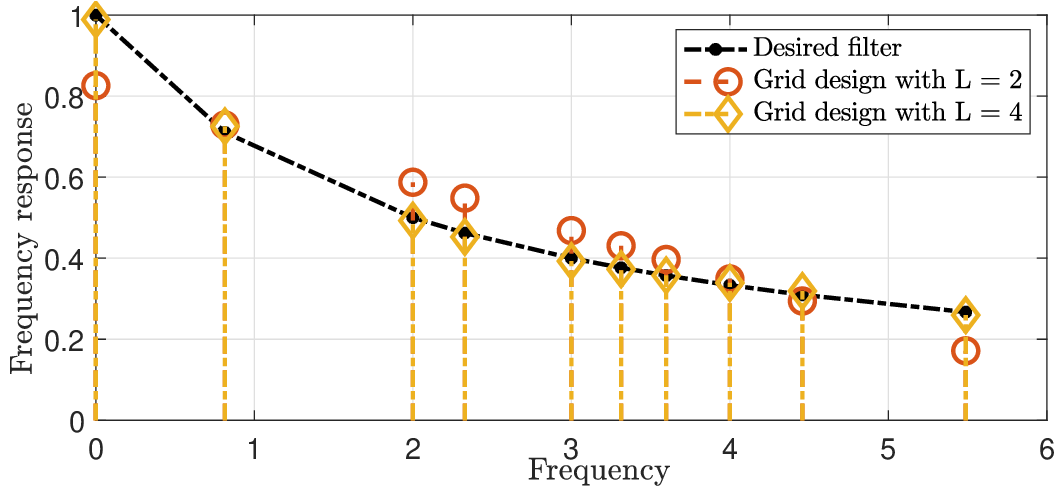}
    \caption{Frequency responses of the denoising filter $\bbH_{P}=(\bbI+0.5\bbL_1)^{-1}$ based on the grid design with different filter orders.}
    \label{fig:grid-based-design-denoising-filter}
\end{figure}

Operator $\bbH_P$ has the frequency response $\tilde{H}_P(\lambda_i) = 1/(1+\mu \lambda_i)$ with 
\begin{enumerate*}[label=(\roman*)]
    \item $\lambda_i = 0$ or $\lambda_i \in\ccalQ_{\rm{G}}$ for $\bbP = \bbL_{1,\ell}$,
    \item $\lambda_i = 0$ or $\lambda_i \in \ccalQ_{\rm{G}} \cup \ccalQ_{\rm{C}}$ for $\bbP = \bbL_1$.
\end{enumerate*}
Thus, $\bbH_P$ is a low pass filter which suppresses either the gradient frequencies or the non-harmonic frequencies. We can implement a simplicial filter to approximate $\bbH_P$. Fig. \ref{fig:grid-based-design-denoising-filter} shows the frequency responses of the filter $\bbH_P$ with $\bbP=\bbL_1$, $\mu=0.5$ based on the grid-based design, for which we considered 10 samples in the frequency interval $[0,5.488]$ and the maximal eigenvalue is estimated with 50 steps of power iterations.  {The grid-based filter design errors compared to using the true eigenvalues are negligible, 0.023 and 0.004 for $L=2$ and $L=4$, respectively.}

\begin{figure}[!t] 
  \vspace{-5mm}
\centering
\subfloat[Noise-free flow $\bbf_0$\label{fig:synthetic-denoising-a}]{%
     \includegraphics[width=0.33\linewidth]{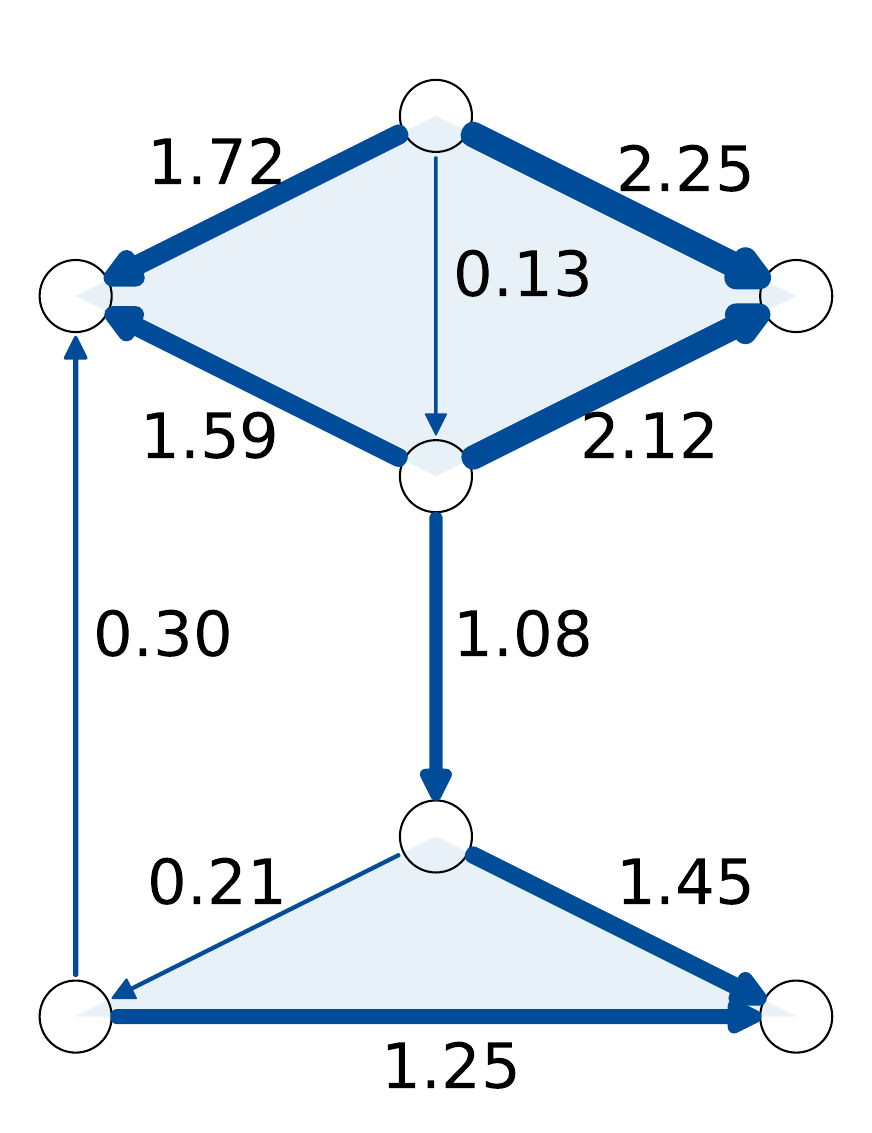}}
\subfloat[Noisy flow $\bbf$\label{fig:synthetic-denoising-b}]{%
      \includegraphics[width=0.33\linewidth]{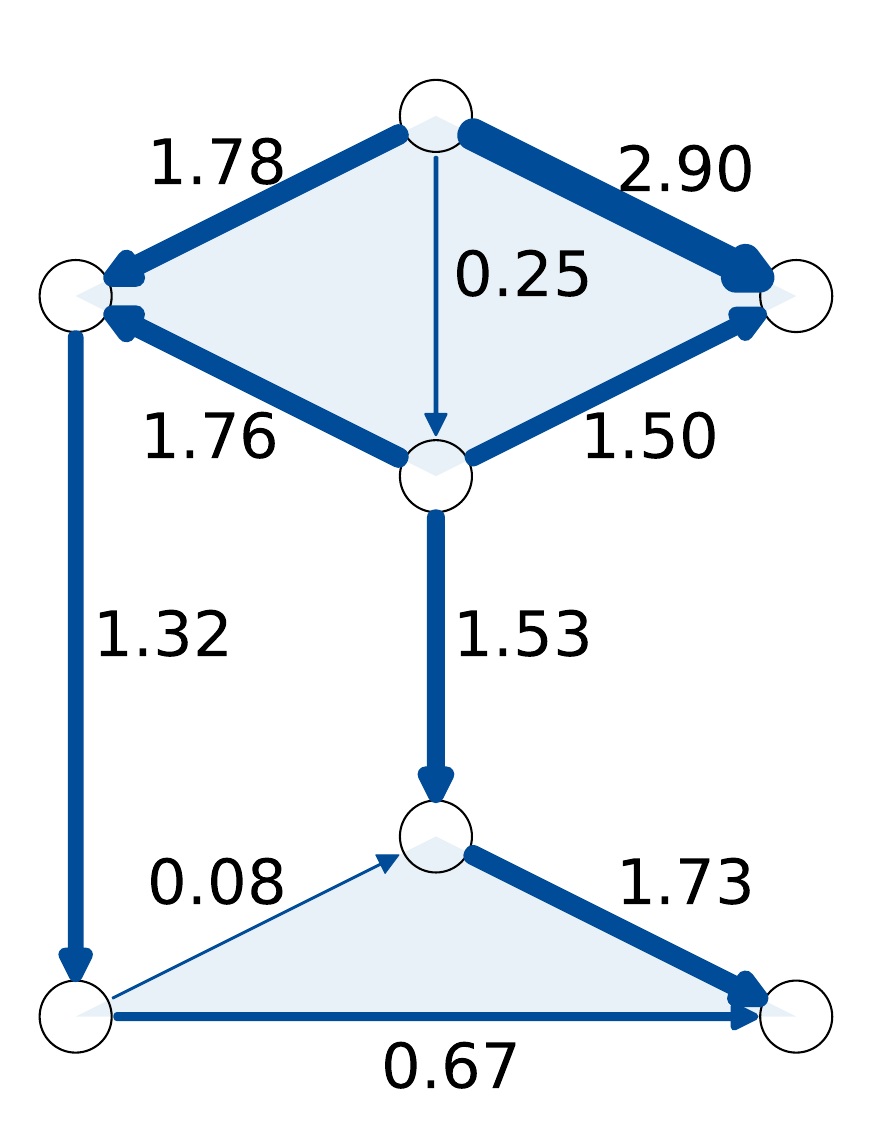}}
\subfloat[$\bbH = (\bbI+0.5\bbL_1)^{-1}$\label{fig:synthetic-denoising-c}]{%
      \includegraphics[width=0.33\linewidth]{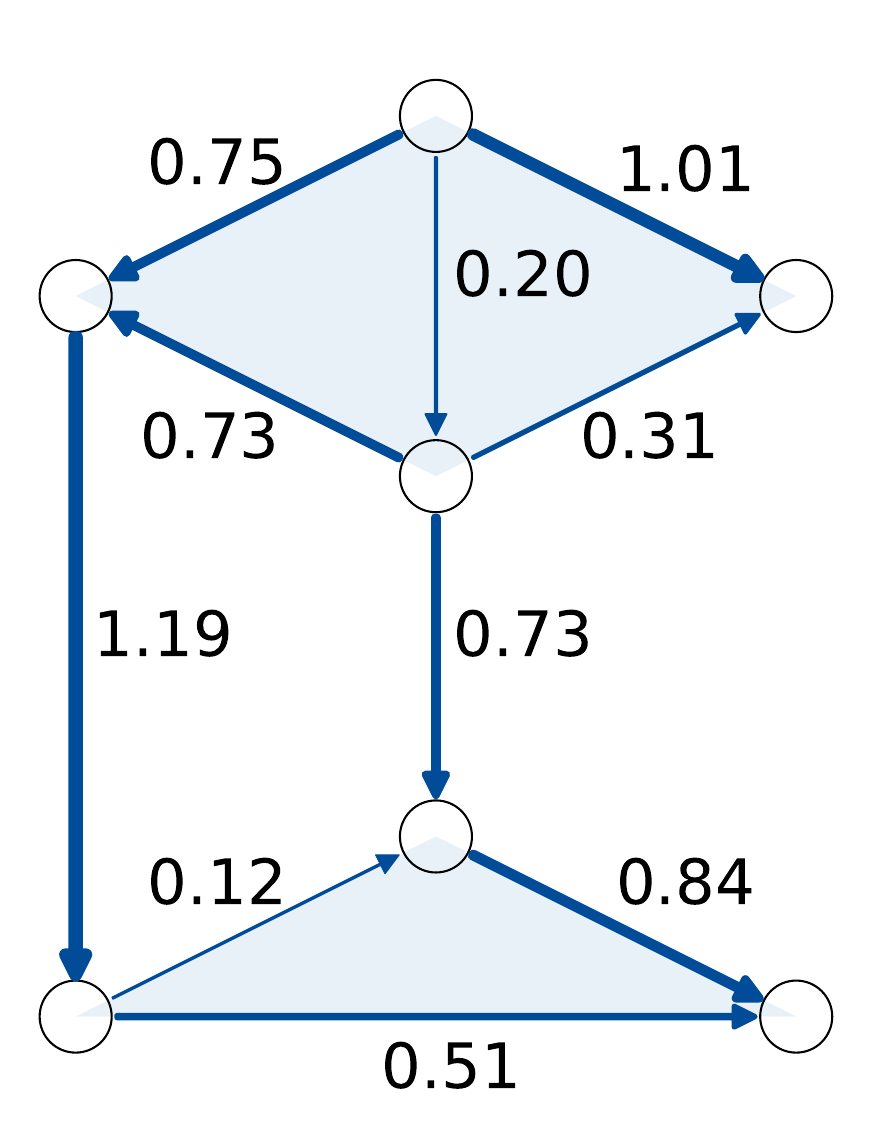}}
\vspace{-3mm}
\subfloat[$\bbH = (\bbI+0.5\bbL_{1,\ell})^{-1} \hspace{-1pt}$\label{fig:synthetic-denoising-d}]{%
     \includegraphics[width=0.33\linewidth]{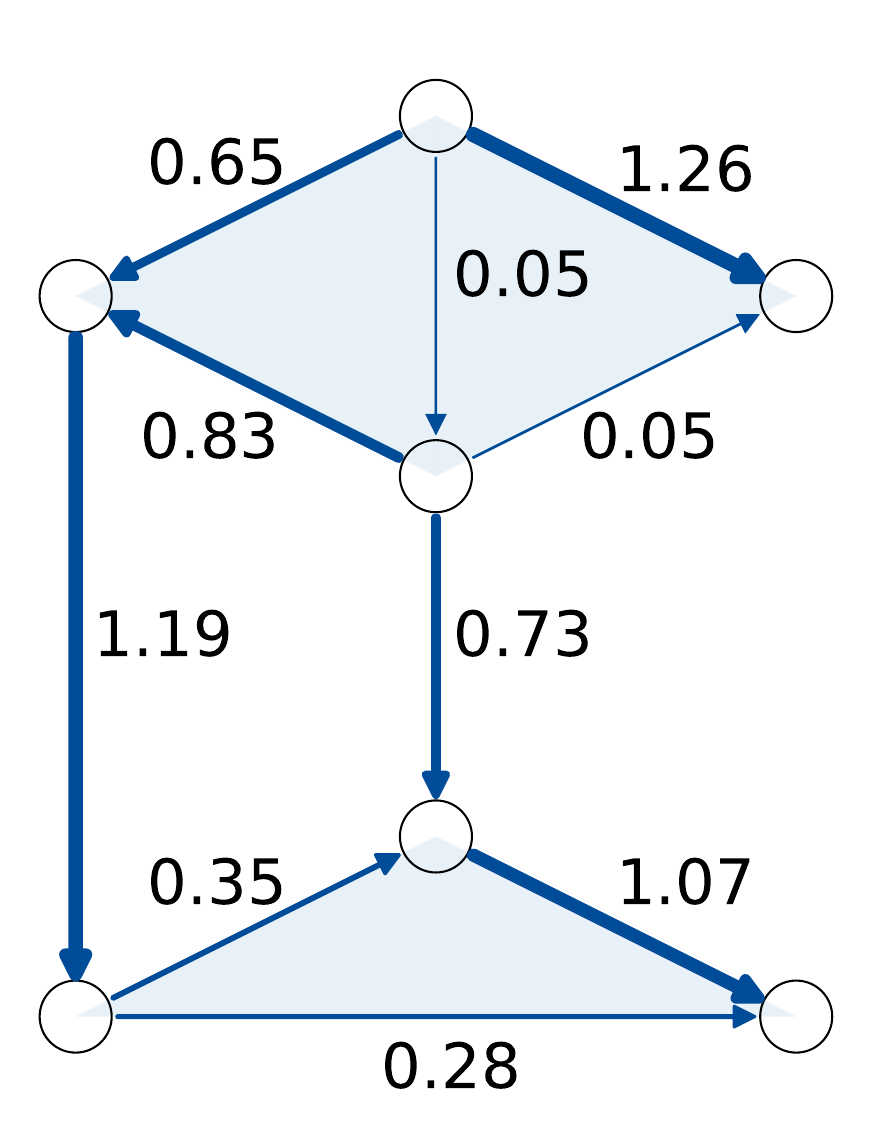}}
\subfloat[$\bbH_1$, $\balpha=\bbeta$\label{fig:synthetic-denoising-e}]{%
      \includegraphics[width=0.33\linewidth]{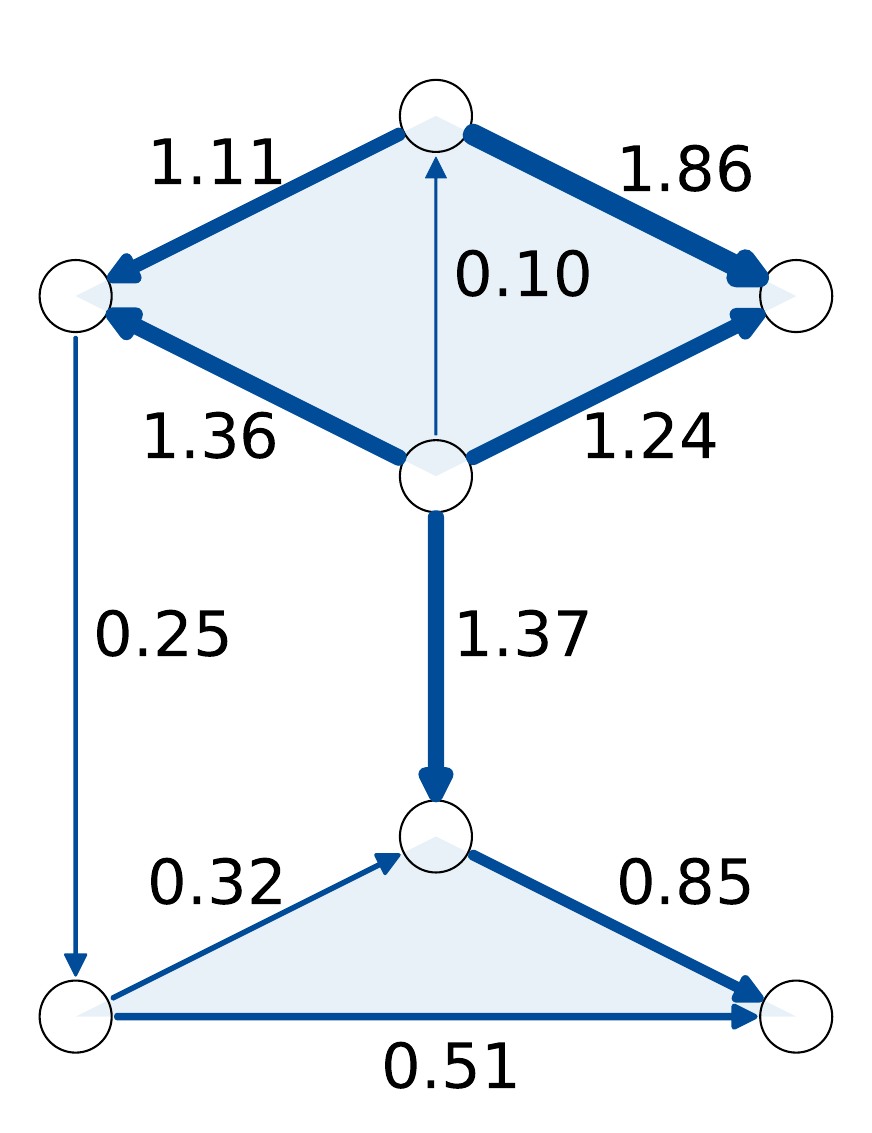}}
\subfloat[$\bbH_1$, $L_1,L_2=1$\label{fig:synthetic-denoising-f}]{%
      \includegraphics[width=0.33\linewidth]{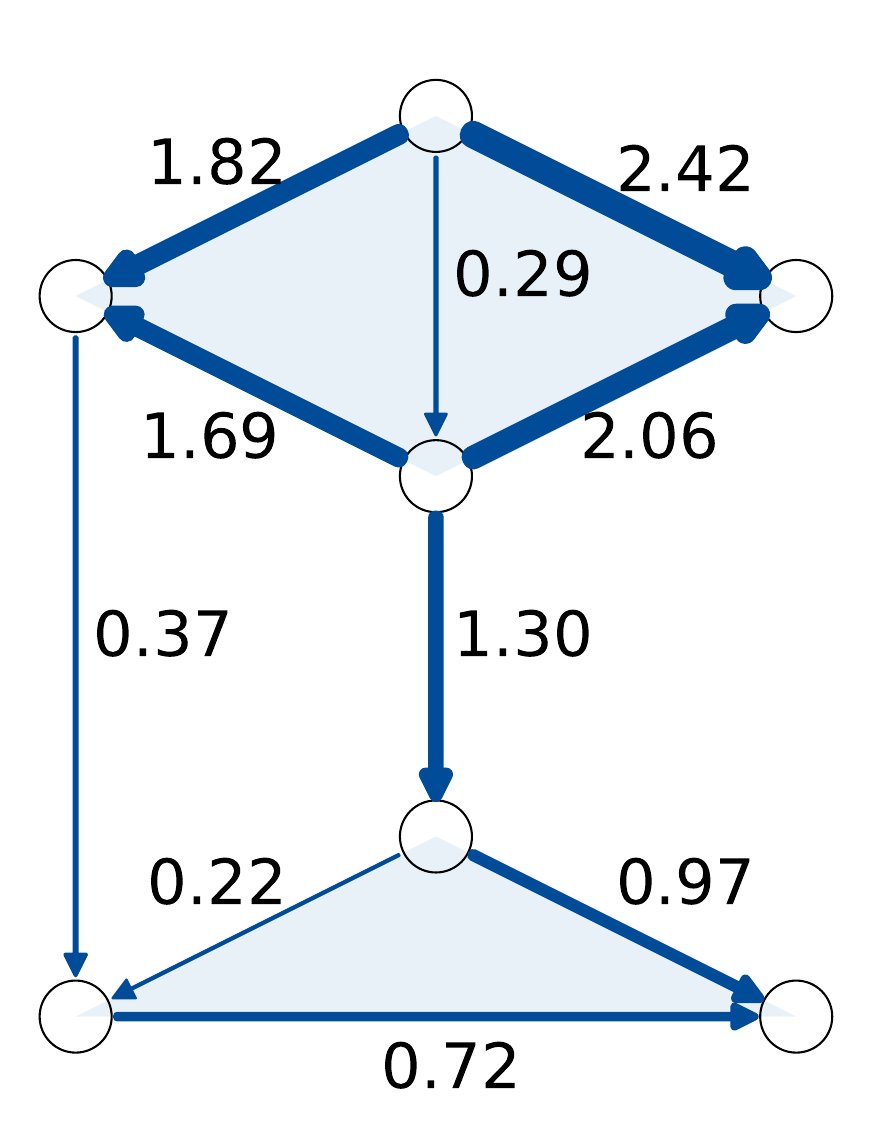}}
\caption{Gradient flow denoising. (a) Edge flow $\bbf_0$ induced by a node signal with a flat spectrum. (b) The noisy observation $\bbf$ with error $e=0.46$. (c-d) Denoising with the low-pass filter $\bbH_{P}$ with (c) $\bbP = \bbL_1$ \cite{schaub2021}, or (d) $\bbP = \bbL_{1,\ell}$ \cite{schaub2018flow} leads to an even larger error of $e=0.70$ or  $e=0.73$, respectively. (e) Denoising by a gradient based simplicial filter $\bbH_1$ with an order $L_1=L_2=4$ and $\balpha=\bbeta$, yields a much better result with error $e=0.39$. (f) Denoising by a general filter $\bbH_1$ with $\balpha\neq\bbeta$ provides an even smaller error with $e=0.23$, even for a lower filter order $L_1=L_2=1$.
}
\label{fig:synthetic-denoising} 
\end{figure}

Note that the above optimization framework relies on the assumption that the true edge flow is either divergence-free or harmonic, which is not always true for real-world flows \cite{jia2019graph, jiang2011statistical}. 
When different spectral properties of the underlying edge flow are known, or certain spectral properties of noise are known, we are able to deal with various situations for denoising by properly designing the simplicial filters. 
To illustrate this we induced a gradient flow from a node signal with a flat spectrum and added Gaussian noise with an error of $0.46$, shown in Figs. \ref{fig:synthetic-denoising-a} and \ref{fig:synthetic-denoising-b}. Fig. \ref{fig:synthetic-denoising} contrasts the denoising results of the regularized optimization methods  with our filters that preserve the gradient component based on an LS design (Corollary \ref{cor.subcomponent-extraction-cor}). 
As Figs. \ref{fig:synthetic-denoising-c} to \ref{fig:synthetic-denoising-f}, shown in this context the low-pass filters which are implicitly defined via the optimization procedures lead to large errors. 
In contrast the simplicial filters that preserve the gradient can denoise properly. Setting $\balpha=\bbeta$ again reduces the performance. 

\subsection{Currency Exchange Market}

A currency exchange market can be described as a network where the vertices represent currencies, and the edges indicate the pairwise exchange rates. 
If all pairs of currencies are exchangeable, the vertex set $\ccalV$ and edge set $\ccalE$ make up a complete graph. 
For any currencies $i,j,k \in \ccalV$, we expect an arbitrary-free condition, $r^{i/j} r^{j/k} = r^{i/k}$ with the exchange rate $r^{i/j}$ between $i$ and $j$, i.e., the exchange path $i\rightarrow j \rightarrow k$ provides no gain or loss over a direct exchange $i\rightarrow k$. If we represent the exchange rates as edge flows $f_{ij}=\log(r^{i/j})$, this can be translated into the fact that $\bbf$ is curl-free, i.e., $f_{ij} + f_{jk} + f_{ki} = 0$. Therefore, an ideal exchange edge flow is a gradient flow. 
This idea was exploited in~\cite{jiang2011statistical,jia2019graph} to assess arbitrage possibilities in exchange markets, and provide arbitrage free exchange rates, respectively.

Here, we illustrate how we can analogously remove arbitrage opportunities via a simplicial filter that preserves only the gradient component of a given exchange rate flow. 
For a complete graph, there  {are two distinct eigenvalues, zero and $N_0$,} for the lower or upper Hodge Laplacian. Then,  {based on Corollary \ref{cor.subcomponent-extraction-cor}} we can extract the gradient component via $\bbH_1 = \frac{1}{N_0}\bbL_{1,\ell}$. Similarly, filter $\bbH_1 = \frac{1}{N_0} \bbL_{1,\rm{u}}$ can extract the curl component, which indicate possible arbitrage opportunities.

\begin{table}[!t]
  \caption{Currency exchange rates captured from Yahoo!Finance, not arbitrage-free.}
  \centering
  \resizebox{\columnwidth}{!}{
  {\scriptsize
  \setlength{\tabcolsep}{0.4em}
  \begin{tabular}{c|c|c|c|c|c|c|c}
  \thickhline
    &  USD & EUR  & CNY & HKD & GBP & JPY & AUD \\ \thickhline
  1 USD & 1      & 0.8422 & 6.3739 & 7.7666  & 0.7207 & 110.1020 & 1.3377 \\
  1 EUR & 1.1873 & 1      & 7.5681 & 9.2218  & 0.8557 & 130.7314 & 1.5883 \\
  1 CNY & 0.1539 & 0.1321 & 1      & 1.2185  & 0.1131 & 17.2683  & 0.2099 \\
  1 HKD & 0.1288 & 0.1085 & 0.8207 & 1       & 0.0928 & 14.1718  & 0.1723 \\
  1 GBP & 1.3871 & 1.1685 & 8.8414 & 10.7732 & 1      & 152.6758 & 1.8557 \\
  1 JPY & 0.0091 & 0.0077 & 0.0579 & 0.0706  & 0.0066 & 1        & 0.0122 \\
  1 AUD & 0.7475 & 0.6299 & 4.7602 & 5.8001  & 0.5385 & 82.1837  & 1   \\ \hline
  \thickhline
  \end{tabular} } }
  \label{tab:forex-1}
  \end{table}

  \begin{table}[!t]
  \caption{The gradient component, arbitrage-free, provides a fair market.}
  \centering
  \resizebox{\columnwidth}{!}{
  {\scriptsize
  \setlength{\tabcolsep}{0.4em}
  \begin{tabular}{c|c|c|c|c|c|c|c}
  \thickhline
  & USD    & EUR    & CNY    & HKD     & GBP    & JPY      & AUD    \\ \thickhline
  1 USD & 1      & 0.8422 & 6.3738 & 7.7665  & 0.7208 & 110.0171 & 1.3385 \\
  1 EUR & 1.1874 & 1      & 7.5680 & 9.2216  & 0.8559 & 130.6292 & 1.5893 \\
  1 CNY & 0.1569 & 0.1321 & 1      & 1.2185  & 0.1131 & 17.2608  & 0.2100 \\
  1 HKD & 0.1288 & 0.1084 & 0.8207 & 1       & 0.0928 & 14.1656  & 0.1723 \\
  1 GBP & 1.3873 & 1.1684 & 8.8425 & 10.7746 & 1      & 152.6286 & 1.8557 \\
  1 JPY & 0.0091 & 0.0077 & 0.0579 & 0.0706  & 0.0066 & 1        & 0.0122 \\
  1 AUD & 0.7471 & 0.6292 & 4.7618 & 5.8022  & 0.5385 & 82.1919  & 1 \\ \hline 
      \thickhline
  \end{tabular} } }
  \label{tab:forex-2}
  \end{table}
  
In Table \ref{tab:forex-1}, we show a real-world exchange market of seven currencies at 2021/07/12 10:30 UTC from the Currency Converter Yahoo!Finance. 
 {We built an SC formed by the seven currencies and all the 2-, 3-cliques, where the edge flow $\bbf$ is the logarithm of the exchange rates in the upper triangular part without the diagonal in Table \ref{tab:forex-1}.}
 {If one unit of currency yields more than $0.3\%$ benefit or loss after two successive exchanges, we say the corresponding exchange rates are non-arbitrage-free.  By computing the curl $\bbB_2^\top\bbf$}, we can identify six such triangles (up to machine precision) that are not arbitrage-free., e.g., USD-JPY-AUD (1 USD would yield 1.0041 USD), EUR-JPY-AUD, and HKD-GBP-JPY.  
By applying a filter $\bbH_1 = \frac{1}{N_0}\bbL_{1,\ell}$ on the exchange rate flows, we can extract its gradient flow, leading to the arbitrage-free exchange rate flow in Table \ref{tab:forex-2}. This yields essentially the same results as solving the LS optimization problem considered in \cite{jiang2011statistical}. 
This simple example demonstrates the use of simplicial filters to generate an efficient financial market. Especially in a complete market, the form of the filters is trivial and the computational cost is much smaller compared to solving the LS problems.


\subsection{London Street Network: Fast PageRank of Edges}


{PageRank, as a ranking scheme for web pages, can be studied in terms of a random walk on a graph, which can be used to measure the centrality of a node. PageRank was extended to the edge space to assess the topological importance of an edge in \cite{schaub2020random}.} The input edge flow $\bbf$ is an indicator vector which has value one on the edge of interest and zeros on the rest, then a PageRank vector $\bpi$ follows the linear system \cite[Def. 6.2]{schaub2020random},
$
    (\gamma \bbI + \bbL_{1,\rm{n}}) \bpi = \bbf, 
$
with $\bbL_{1,\rm{n}}$ being the normalized 1-Hodge Laplacian\footnote{$\bbL_{1,\rm{n}}$ admits a similarity transformation to its symmetric version, so its eigenvalues are real and carry the simplicial frequency notion.} \cite[Def. 3.3]{schaub2020random} and $\gamma>0$. The solution is $\bpi = (\gamma \bbI + \bbL_{1,\rm{n}})^{-1} \bbf$ with the PageRank operator $\bbH_{\rm{PR}}:=(\gamma \bbI + \bbL_{1,\rm{n}})^{-1}$,  {which does not require to construct the edge space random walk matrix \cite[Thm. 3.4]{schaub2020random} compared to a power iteration implementation.} 
For an indicator edge flow $\bbf$ of edge $i$, the absolute values of the entries of $\bpi$ are the influence measures edge $i$ has on the edges and the signs the influence orientations w.r.t. the reference orientations \cite{schaub2020random}. Furthermore, the overall importance of an edge can be assessed with the $\ell_2$-norm $\lVert \bpi\lVert_2$ of its PageRank vector $\bpi$. By extracting the gradient component $\bpi_{\rm{G}}$,
we can study the importance of this edge w.r.t. the gradient space via its $\ell_2$-norm $\lVert \bpi_{\rm{G}}\lVert_2$ or relative norm $\frac{\lVert \bpi_{\rm{G}}\lVert_2}{\lVert \bpi\lVert_2}$; likewise for the curl and harmonic components.

\begin{figure}[!t]
  \centering
  \includegraphics[scale=0.44]{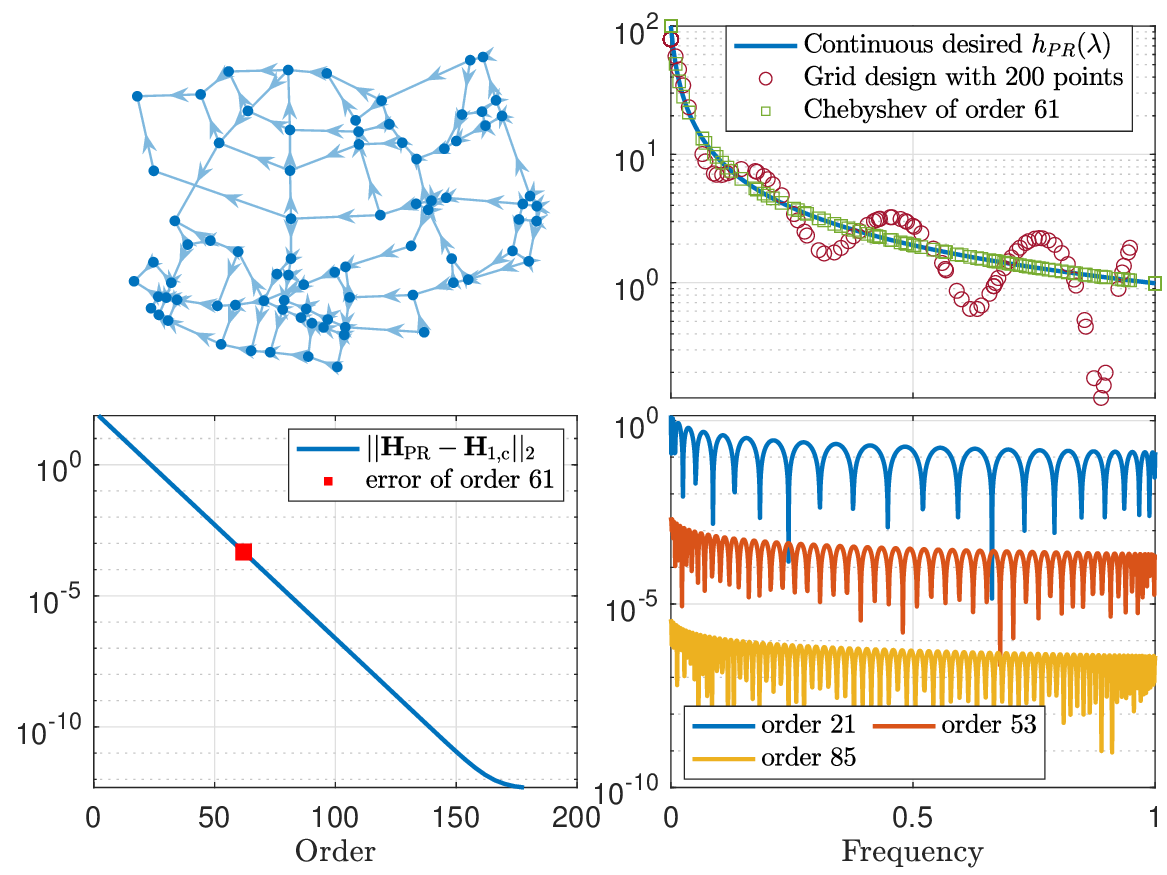}
  \caption{PageRank analysis of the street network of London. Top left: network illustration. Top right: Frequency responses of the grid-based designed filter of order 9 and the Chebyshev filter  of order 61 w.r.t. the desired continuous frequency response. Bottom left: Spectral norm error of the Chebyshev polynomial filter of different orders. Bottom right: Continuous frequency response errors of the Chebyshev filter of different orders.}
  \label{fig:london_network_analysis}
\end{figure}

\smallskip\noindent\textbf{Experiment setup.}
The operator $\bbH_{\rm{PR}}$ can be interpreted as a low pass filter which attenuates the high gradient and curl frequencies. But to implement it we need to invert a matrix. 
Instead, we propose here a faster variant via a simplicial filter $\bbH_1:=\bbH_1(\bbL_{1,\rm{n}})$ built on the normalized Hodge Laplacian. To find the filter coefficients, we considered a grid-based design and a Chebyshev polynomial with a desired response, $g(\lambda) = \frac{1}{\gamma + \lambda}$ with $\lambda \in [0,1]$ and  $\gamma = 0.01$.

We implemented the PageRank operator in the street network of London with 82 crossings (nodes), 130 streets (edges) and 12 triangles, as shown in the top left of Fig. \ref{fig:london_network_analysis} \cite{youn2008price, schaub2014structure}. We considered a grid-based design with 200 samples within the eigenvalue interval and a filter order of 9 and implemented the Chebyshev polynomials $\bbH_{1}$ of different orders. 

\smallskip\noindent\textbf{Results.} From Fig. \ref{fig:london_network_analysis}, we see that the performance of the grid-based design decreases heavily when the filter order is larger than 9 due to the numerical instability, while a Chebyshev design allows a more accurate design as shown in Fig. \ref{fig:london_network_analysis}.

\begin{figure}[!t]
  \centering
  \includegraphics[scale=0.5]{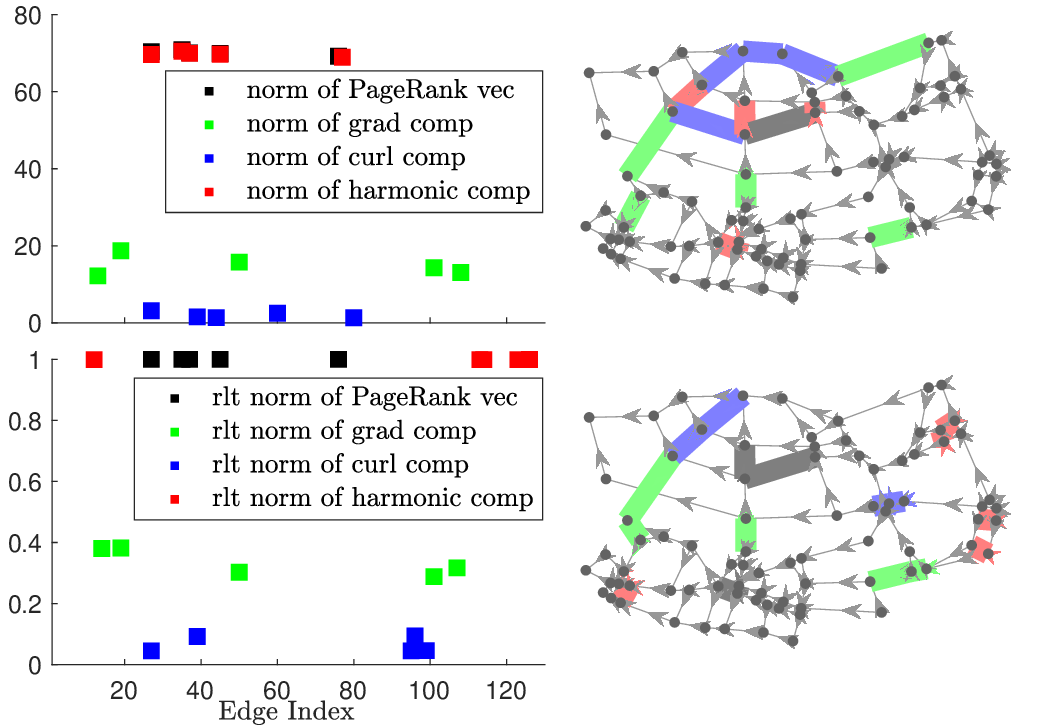}
  \caption{PageRank vectors analysis. The scattered squares are the edges whose PageRank vectors contain the top-five largest absolute (Top) and relative (Bottom) total norm, gradient, curl and harmonic norms. The left figures show the PageRank values w.r.t. edge indices. The rights figures show the highlighted edges in the network with shaded grey indicating the total norm, green the gradient norm, blue the curl norm and red the harmonic norm.}
  \label{fig:london_pr_norm_comp}
\end{figure}

We then computed the PageRank results of all edges with a Chebyshev filter of order 61 and obtained the norms of their three components in the absolute and relative senses. From Fig. \ref{fig:london_pr_norm_comp}, we can identify the most influential streets (dark grey) in the network, as the indicator flow on these streets induce a PageRank vector with the largest total norm. The streets (green) that have the biggest influence on the gradient space are the ones on which a traffic change leads to congestion on the intersections as the traffic flows on them have a large divergence. The red streets induce the most impact on the harmonic space and the blues ones on the curl space, where the traffic flows tend to induce a global or local cyclic flow, thus a small chance of congestion. The influences are measured in a relative sense in the bottom figures, and we notice that most streets would not cause a large influence on the curl space. 

\begin{figure}[!t]
  \centering
  \includegraphics[scale=0.44]{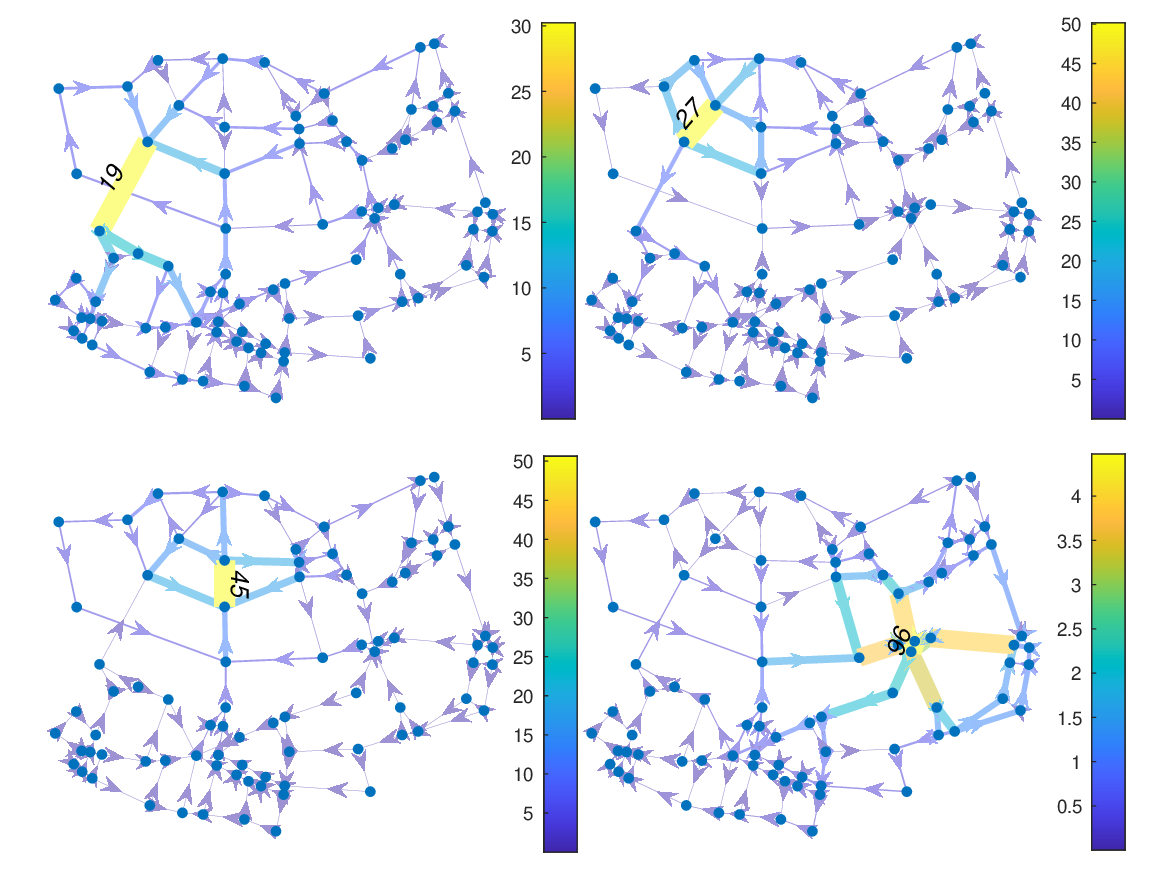}
  \caption{Examples of the PageRank vectors of four edges. The edge width and color indicates the magnitude of the PageRank result on that edge. The labeled edges are the chosen edges, also with the largest PageRank results.}
  \label{fig:london_pr_results_examples}
\end{figure}

Finally, we show the simplicial PageRank vector of four edges to assess where their influences are concentrated in Fig. \ref{fig:london_pr_results_examples}. Edge 19 (top left), sitting in the  {gradient} space, has a large influence in terms of gradient flow components on the surrounding edges to which congestion on edge 19 would spread, as shown in Fig. \ref{fig:london_pr_norm_comp}. Edge 27 (top right) has a large influence on edges that form 1-dimensional ``hole'' \cite{lim2015hodge}, containing mostly harmonic components. This may imply that a traffic change on edge 27 would less likely cause congestion. Edge 45 (bottom left), whose PageRank result has a large total norm, as seen in Fig. \ref{fig:london_pr_norm_comp}, acts similar as edge 27. Edge 96 (bottom right) induces smaller influences than the other three, but they reach further in the network. The most influenced edges are its direct upper neighbors, as also seen in the bottom (blue) of Fig. \ref{fig:london_pr_norm_comp}, where congestion would rather not happen. 

\subsection{Chicago Road Network: Gradient Component Extraction}

We now conduct the subcomponent extraction on a real-world larger network.
On the Chicago road network with 546 nodes, 1088 edges, and 112 triangles \cite{stabler2018transportation}, we perform the gradient component extraction of the measured traffic flow, which is not divergence-free, via filter $\bbH_1$ built on the lower Hodge Laplacian. 
It is challenging to perform the filter design in this setting because some simplicial frequencies are close to each other, leading to the ill-conditioning of the LS design. This can be avoided by the Chebyshev polynomial design. 
This requires
a continuous desired frequency response to perform the gradient component extraction, which ideally is an indicator function $\bb1_{\lambda>0}$ with $\lambda\in[0,\lambda_{\rm{G},\max}]$. 
Here we use the logistic function
$
   g_{\rm{G}}(\lambda)=\frac{1}{1+\exp^{-k(\lambda-\lambda_0)}}  
$
with the growth rate $k>0$ and the midpoint $\lambda_0$. 
If the smallest gradient frequency is close to 0, a large $k$ and a small $\lambda_0$ are required to achieve a good approximation of the ideal indicator function. 

We applied different filter design methods.
For the LS-based methods, we set a filter order of $9$ to avoid ill-conditioning and we considered the decoupled solver. 
Moreover, we treated the eigenvalues with a difference smaller than $0.3$ as the same for the LS design, leading to 30 ``different'' eigenvalues. 
For the grid-based design, we uniformly sampled 100 points in the interval $[0,\lambda_{\rm{G},\max}]$ with $\lambda_{\rm{G},\max}=10.8$ approximated by a 50-step power iteration. 
Lastly, we set $k=100$ and $\lambda_0=0.01$ for the logistic function in the Chebyshev polynomial design. 

\begin{figure}[!t] 
\centering
\subfloat[ \label{fig:chicago_gradient_freq_response}]{%
     \includegraphics[width=0.95\linewidth]{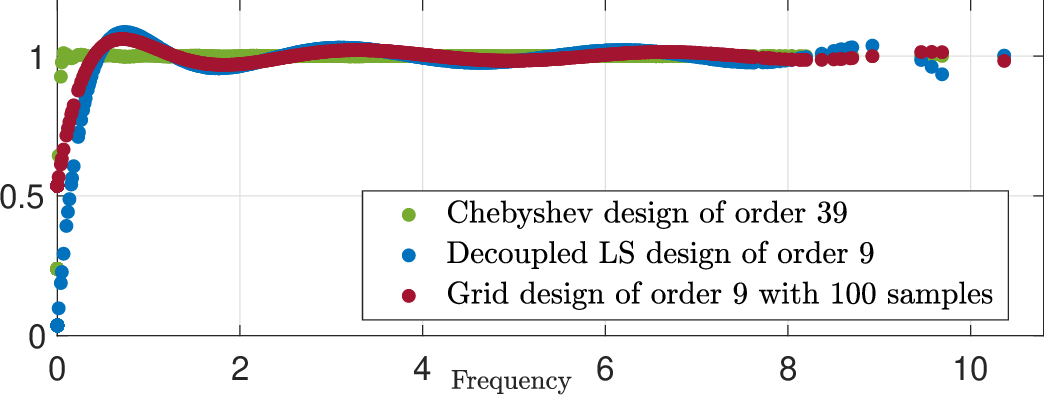}}
 \vspace{0mm}
\subfloat[\label{fig:chicago_grad_extrac_comp}]{%
      \includegraphics[width=1\linewidth]{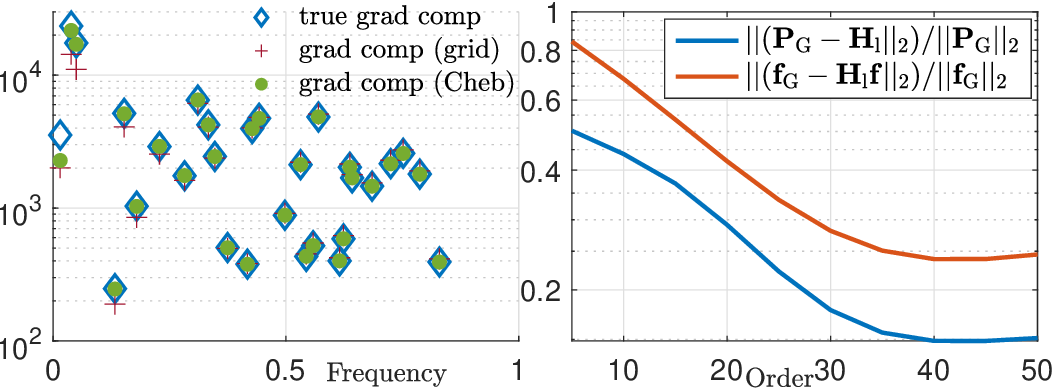}}
\caption{Chicago road network gradient component extraction. (a): Filter frequency responses with different designs. (b): Left: SFT of the extracted gradient component in frequency range $[0,1]$. Right: Approximation errors of Chebyshev filters of different orders and the extracted gradient component.}
\label{fig:chicago}
\end{figure}

As seen in Fig. \ref{fig:chicago_gradient_freq_response}, the Chebyshev polynomial of an order $39$ only has one frequency response smaller than 0.9 at the smallest gradient frequency, while at the remaining frequencies, it is able to well preserve the gradient component. The other methods have a poorer performance especially at small gradient frequencies. 
We then compared the gradient component extracted by above grid-based and Chebyshev polynomial filters. Fig. \ref{fig:chicago_grad_extrac_comp} (left) reports the SFT of the extracted flows at frequencies smaller than 1. The Chebyshev polynomial has a good extraction ability as it performs well even at the very small frequencies where the grid-based design fails. Fig. \ref{fig:chicago_grad_extrac_comp} (right) shows the filter design and the extraction errors of the Chebyshev polynomial of different orders. The extraction error cannot be further reduced because the traffic flow contains large components at the small frequencies and the logistic function, after all, is an approximate of the indicator function. 

\section{Conclusion}
We proposed a simplicial convolutional filter as a matrix polynomial of the Hodge Laplacians for simplicial signal processing. It generates an output as a linear combination of the shifted simplicial signals. This shift-and-sum operation is analogous to the convolutions of time series, images and graph signals and allows for a distributed filter implementation. In the frequency domain, the filter acts as a pointwise multiplication respecting the convolution theorem. Furthermore, the lower and upper Hodge Laplacians encode lower and upper adjacencies, respectively. For an edge flow, its lower shifting propagates the flow to its neighbors via the common incident nodes, while the upper one via the common triangles. By assigning  two different sets of coefficients on the lower and upper parts in the filter, we can differentiate the lower and upper adjacencies. Via the Hodge decomposition, we see that this corresponds to an independent control of the filter on the gradient and curl spaces in the frequency domain. To achieve a desired frequency response, different filter design approaches are considered with pros and cons. The filter provides a faster and distributed solution for subcomponent extraction, simplicial signal denoising and other tasks in exploiting the higher-order connectivities of the network. 

\bibliographystyle{IEEETtran}
\bibliography{refs}

\appendices 

\section{Proof of Proposition \ref{prop.linearity-shift-invariance}} \label{proof.linearity-shift-invariance}
Due to the distributivity of matrix-vector multiplication, we have $\bbL_{1,\ell} (a\bbf_1+b\bbf_2) = a \bbL_{1,\ell} \bbf_1 + b\bbL_{1,\ell} \bbf_2$, and $\bbL_{1,\rm{u}} (a\bbf_1+b\bbf_2) = a \bbL_{1,\rm{u}} \bbf_1 + b\bbL_{1,\rm{u}} \bbf_2$. Then, by working out the expression $\bbH_1 (a\bbf_1 + b\bbf_2)$ and using the distributivity w.r.t the addition, the linearity proof completes. Since we have $\bbL_{1,\ell}^l \bbL_{1,\ell} = \bbL_{1,\ell} \bbL_{1,\ell}^l$, and similarly for for $\bbL_{1,\rm{u}}$, and $\bbL_{1,\ell} \bbL_{1,\rm{u}} = \mathbf{0}$, the shift-invariance holds. The same proof applies to the general case with $k\neq1$.

\section{Proof of Proposition \ref{prop.permutation-equivariance}} \label{proof.permutation-equivariance}
Since the permutation matrix $\bbP_k$ is orthogonal, i.e., $\bbP_k\bbP_k^\top=\bbP_k^\top\bbP_k=\bbI$, we have that $(\bbP_k\bbL_{k,\ell}\bbP^\top_k)^{l_1} = \bbP_k\bbL_{k,\ell}^{l_1} \bbP^\top_k$, and similarly $(\bbP_k\bbL_{k,\rm{u}}\bbP^\top_k)^{l_2} = \bbP_k\bbL_{k,\rm{u}}^{l_2} \bbP^\top_k$. Thus, we can express the permuted simplicial filter as 
\begin{equation} \label{eq.permutation-equivariance-proof-1}
  \begin{aligned}
    \bar{\bbH}_k & = h_0 \bbI + \sum_{l_1=1}^{L_1} \alpha_{l_1}(\bbP_k\bbL_{k,\ell}\bbP^\top_k)^{l_1} + \sum_{l_2=1}^{L_2}\beta_{l_2}(\bbP_k\bbL_{k,\rm{u}}\bbP^\top_k)^{l_2} \\
     & =  h_0 \bbI +  \sum_{l_1=1}^{L_1} \alpha_{l_1}\bbP_k\bbL_{k,\ell}^{l_1}\bbP^\top_k +  \sum_{l_2=1}^{L_2}\beta_{l_2}\bbP_k \bbL_{k,\rm{u}}^{l_2} \bbP^\top_k \\
    & = \bbP_k \bigg( h_0 \bbI +  \sum_{l_1=1}^{L_1} \alpha_{l_1}\bbL_{k,\ell}^{l_1} + \sum_{l_2=1}^{L_2}\beta_{l_2} \bbL_{k,\rm{u}}^{l_2} \bigg) \bbP^\top_k \\
    & = \bbP_k \bbH_k \bbP^\top_k.
  \end{aligned}
\end{equation}
The output on the permuted SC can be written as $ \bar{\bbs}^{k}_{\rm{o}}:=\bar{\bbH}_k\bar{\bbs}^{k} = \bar{\bbH}_k(\bbP_k\bbs^k) = \bbP_k \bbH_k \bbP^\top_k \bbP_k\bbs^k = \bbP_k \bbH_k \bbs^k := \bbP_k\bbs^k_{\rm{o}}$. The proof completes.

\section{Proof of Proposition \ref{prop.orientation-equivariance}} \label{proof.orientation-equivariance}
The diagonal matrix $\bbD_k$ satisfies that $\bbD_k\bbD_k^\top = \bbD_k^\top\bbD_k=\bbI$. Following from that, we have that $(\bbD_k\bbL_{k,\ell}\bbD^\top_k)^{l_1} = \bbD_k\bbL_{k,\ell}^{l_1} \bbD^\top_k$, and similarly $(\bbD_k\bbL_{k,\rm{u}}\bbD^\top_k)^{l_2} = \bbD_k\bbL_{k,\rm{u}}^{l_2} \bbD^\top_k$. Following the same procedure in \eqref{eq.permutation-equivariance-proof-1}, we have 
\begin{equation}
  \begin{aligned}
    \bar{\bbH}_k & = h_0 \bbI + \sum_{l_1=1}^{L_1} \alpha_{l_1}(\bbD_k\bbL_{k,\ell}\bbD^\top_k)^{l_1} + \sum_{l_2=1}^{L_2}\beta_{l_2}(\bbD_k\bbL_{k,\rm{u}}\bbD^\top_k)^{l_2} \\
    & = \bbD_k \bbH_k \bbD^\top_k.
  \end{aligned}
\end{equation}
Thus, the filter output on the reoriented simplices can be expressed as $\bar{\bbs}^{k}_{\rm{o}}:=\bar{\bbH}_k\bar{\bbs}^{k} = \bar{\bbH}_k(\bbD_k\bbs^k) = \bbD_k \bbH_k\bbs^k :=  \bbD_k\bbs_{\rm{o}}^k$. The proof completes.

\section{Proof of Proposition \ref{prop.correspondence}} \label{appendix:prop1-proof}

We show the proof for each item.

1) First, we show that the image of $\bbL_{1,\ell}$ is equivalent to the gradient space $\im(\bbB_1^\top)$.
\begin{enumerate*}[label=(\roman*)]
    \item To show $\im(\bbL_{1,\ell})\subseteq\im(\bbB_1^\top)$: From $\bbL_{1,\ell} = \bbB_1^\top\bbB_1$, for any non-zero $\bbx\in\im(\bbL_{1,\ell})$, we have $\bbx=\bbL_{1,\ell}\bby$, and we can always find a vector $\bbz=\bbB_1\bby\in\setR^{N_0}$ such that $\bbx = \bbB_1^\top\bbz$; 
    \item To show $\im(\bbB_1^\top)\subseteq \im(\bbL_{1,\ell})$: for every non-zero $\bbx\in\im(\bbB_1^\top)$, we can find some $\bby\perp\ker(\bbB^\top_1)$ such that $\bbx=\bbB_1^\top\bby\neq\mathbf{0}$. This implies $\bby\in\im(\bbB_1)$, so there exists some $\bbz\in\setR^{N_1}$ such that $\bby=\bbB_1\bbz$ and $\bbx = \bbB_1^\top\bbB_1\bbz = \bbL_{1,\ell}\bbz$, and hence $\im(\bbB_1^\top)\subseteq \im(\bbL_{1,\ell})$.  
\end{enumerate*}
Combining (i) and (ii), we have that $\im(\bbL_{1,\ell}) = \im(\bbB_1^\top)$.

Second, we show that the eigenvectors $\bbU_{\rm{G}}$ of $\bbL_{1,\ell}$ associated with nonzero eigenvalues span the image of $\bbL_{1,\ell}$. As $\bbL_{1,\ell}$ is positive semidefinite (PSD), thus, diagonalizable, the geometric multiplicity of every eigenvalue equals to the algebraic multiplicity. That is, all the eigenvectors are linearly independent and form an eigenbasis. Then, matrix $\bbU_{\rm{G}}$ has a full column rank. Furthermore, for any $\bbx\in\im(\bbL_{1,\ell})$, we have $\bbx = \bbL_{1,\ell}\bby = \bbU_{\rm{G}}\bLambda_{\rm{G}}\bbU_{\rm{G}}^\top \bby$, i.e., $\bbx\in\im(\bbU_{\rm{G}})$ and $\im(\bbL_{1,\ell})\subseteq \im(\bbU_{\rm{G}})$. Due to $\dim\,\im(\bbL_{1,\ell}) = \rank(\bbL_{1,\ell}) = N_{\rm{G}}$, matrix $\bbU_{\rm{G}}$ spans the image of $\bbL_{1,\ell}$ and the gradient space $\im(\bbB_1^\top)$.
    
2) The proof of 2) follows similarly as the proof of 1).
    
3) For arbitrary eigenvectors $\bbu_{\rm{G}}$ in $\bbU_{\rm{G}}$ and $\bbu_{\rm{C}}$ in $\bbU_{\rm{C}}$, we have $\bbu_{\rm{G}} = \frac{1}{\lambda_{\rm{G}}} \bbL_{1,\ell} \bbu_{\rm{G}}$ and $\bbu_{\rm{C}} = \frac{1}{\lambda_{\rm{C}}} \bbL_{1,\rm{u}} \bbu_{\rm{C}}$. Thus, from \eqref{eq.boundary-condition}, their inner product follows $\bbu_{\rm{G}}^\top\bbu_{\rm{C}} = \frac{1}{\lambda_{\rm{G}}\lambda_{\rm{C}}} \bbu_{\rm{G}}^\top \bbL_{1,\ell} \bbL_{1,\rm{u}} \bbu_{\rm{C}} = 0$, i.e., $\bbU_{\rm{G}}\perp\bbU_{\rm{C}}$. Moreover, from the definition of $\bbL_1$, matrix
    $[\bbU_{\rm{G}} \, \bbU_{\rm{C}}]$ 
    contains the eigenvectors of $\bbL_1$ associated with nonzero eigenvalues. The Hodge decomposition indicates that $\im(\bbB_1^\top)\oplus \im(\bbB_2) = \im(\bbL_1)$. By combining with 1) and 2), we have that $\im(\bbB_1^\top)=\im(\bbU_{\rm{G}})$ and $\im(\bbB_2)=\im(\bbU_{\rm{C}})$. 

 4) As $\bbL_1$ is PSD, the eigenvectors associated to zero eigenvalues are linearly independent. Any vector $\bbx\in\ker(\bbL_1)$ follows $\bbL_1\bbx = \mathbf{0}$, which means $\bbx$ is an eigenvector associated with an eigenvalue $0$, i.e., $\ker(\bbL_1) = \im(\bbU_{\rm{H}})$ with dimension $N_{\rm{H}}$.  Moreover, we have $\ker(\bbL_1) = \ker(\bbL_{1,\ell}) \cap \ker(\bbL_{1,\rm{u}})$ from the definition of $\bbL_1$, then the columns of $\bbU_{\rm{H}}$ can be used as eigenvectors of $\bbL_{1,\ell}$ or $\bbL_{1,\rm{u}}$ associated with zero eigenvalues. From 3) and $\ker(\bbL_1) = \im(\bbU_{\rm{H}})$, we have $\bbU_{\rm{H}}\perp \bbU_{\rm{G}}$ and $\bbU_{\rm{H}}\perp \bbU_{\rm{C}}$. Thus, matrix $ 
    [\bbU_{\rm{H}} \,\,\bbU_{\rm{C}}]$ (or $
    [\bbU_{\rm{H}} \,\, \bbU_{\rm{G}}] $) provides the eigenvectors of $\bbL_{1,\ell}$ (or $\bbL_{1,\rm{u}}$) associated with zero eigenvalues, and the proof completes.

 5) From 3) and 4), we have that matrix $\bbU_1$ collects all eigenvectors of $\bbL_1$. From 1), 2), and 4), we have that $\bbU_1$ provides all eigenvectors for $\bbL_{1,\ell}$ and $\bbL_{1,\rm{u}}$.

\section{Proof of Proposition \ref{prop.perfect-implementation}} \label{proof.perfect-implementation}
With condition i), we can eigendecompose the operator $\bbG$ as $\diag(\bbg) = \bbU_1^\top\bbG\bbU_1$, then the equivalence between $\bbG$ and $\bbH_1$ can be achieved through a set of linear equations in the spectral domain, i.e., $H_1(\lambda_i) = g_i$, for all $i=1,\dots,N_1$. Based on condition ii), this set of linear equations is equivalent to linear system \eqref{eq.ls-1} and \eqref{eq.ls-2} with $1+D_{\rm{G}} + D_{\rm{C}}$ equations. With the filter order requirement in condition iii), the Vandermonde matrices $\bPhi_{\rm G}$ and $\bPhi_{\rm C}$ have full row rank. Thus, there exist at least one solution to problem \eqref{eq.ls-1} and the proof completes.

\section{Proof of Proposition \ref{prop.optimality}} \label{proof.optimality}
The cost function $J$ in problem \eqref{eq.ls-2} is convex w.r.t. variables, $h_0$, $\balpha$ and $\bbeta$. Thus, we could find the optimality condition by setting the gradients of the cost function w.r.t. the three variables to zeros, given by
 \begin{equation} \label{eq.cond}
  \begin{cases}
    \begin{aligned} 
      \nabla_{h_0} J = h_0 & -  g_0 + (h_0\bb1 + \bPhi_{\rm G} \balpha - \bbg_{\rm G})^\top \bb1 \\
      & + (h_0\bb1 + \bPhi_{\rm C} \bbeta - \bbg_{\rm C})^\top \bb1 = 0 ,
    \end{aligned} \\
    \nabla_{\balpha} J = \bPhi_{\rm G}^\top (h_0\bb1 + \bPhi_{\rm G} \balpha - \bbg_{\rm G}) = \mathbf{0} ,\\    
    \nabla_{\bbeta} J = \bPhi_{\rm C}^\top (h_0\bb1 + \bPhi_{\rm C} \bbeta - \bbg_{\rm C}) = \mathbf{0}.
  \end{cases}
\end{equation}

First, consider the case where we have that $\lVert \bPhi_{\rm G} \bPhi_{\rm G}^\dagger - \bbI\lVert_F = 0$ and  $\lVert\bPhi_{\rm C} \bPhi_{\rm C}^\dagger -\bbI\lVert_F = 0$, i.e., $\bPhi_{\rm G} \bPhi_{\rm G}^\dagger = \bbI$ and  $\bPhi_{\rm C} \bPhi_{\rm C}^\dagger = \bbI$, then the solution \eqref{eq.ls-2-solution} results in that $(\hat{h}_0\bb1 + \bPhi_{\rm G} \hat{\balpha} - \bbg_{\rm G}) = \mathbf{0}$ and $(\hat{h}_0\bb1 + \bPhi_{\rm C} \hat{\bbeta} - \bbg_{\rm C}) = \mathbf{0}$, which satisfies the optimality condition \eqref{eq.cond}. 

Second, consider the general case that  $\lVert \bPhi_{\rm G} \bPhi_{\rm G}^\dagger - \bbI\lVert_F \neq 0$ and  $\lVert\bPhi_{\rm C} \bPhi_{\rm C}^\dagger -\bbI\lVert_F \neq 0$. By substituting the solution \eqref{eq.ls-2-solution} into the optimality condition \eqref{eq.cond}, we have 
\begin{equation} \label{eq.cond-2}
  \begin{cases}
    \begin{aligned} 
      \nabla_{h_0} J = & \big((\bPhi_{\rm G}\bPhi_{\rm G}^\dagger - \bbI) (\bbg_{\rm G} - g_0\bb1) \big)^\top \bb1 \\
      & + \big( (\bPhi_{\rm C}\bPhi_{\rm C}^\dagger - \bbI) (\bbg_{\rm C} - g_0\bb1  ) \big)^\top \bb1   ,
    \end{aligned} \\
    \nabla_{\balpha} J = \bPhi_{\rm G}^\top (\bPhi_{\rm G}\bPhi_{\rm G}^\dagger - \bbI   ) (\bbg_{\rm G} - g_0\bb1   )  ,\\    
    \nabla_{\bbeta} J = \bPhi_{\rm C}^\top (\bPhi_{\rm C}\bPhi_{\rm C}^\dagger - \bbI  ) (\bbg_{\rm C} - g_0\bb1  ).
  \end{cases}
\end{equation}

For any matrix $\bbA\in\setR^{m\times n}$ with singular values $\sigma_i$,  $i=1,\dots,\min\{m,n\}$, we have that $\lVert \bbA\lVert_F = \big(\sum_{i=1}^{\min\{m,n\}}\sigma_i^2\big)^{\frac{1}{2}}$. If it holds $\lVert \bPhi_{\rm G} \bPhi_{\rm G}^\dagger - \bbI    \lVert_F \rightarrow 0$ and  $\lVert \bPhi_{\rm C} \bPhi_{\rm C}^\dagger -\bbI   \lVert_F \rightarrow 0$, i.e., the Frobenius norms approach to zero, the number of trivial (zero) singular values of $\bPhi_{\rm G} \bPhi_{\rm G}^\dagger - \bbI   $ and $\bPhi_{\rm C} \bPhi_{\rm C}^\dagger -\bbI  $ increases. Accordingly, the number of trivial entries in $(\bPhi_{\rm G}\bPhi_{\rm G}^\dagger - \bbI   ) (\bbg_{\rm G} - g_0\bb1  )$ and $(\bPhi_{\rm C}\bPhi_{\rm C}^\dagger - \bbI  ) (\bbg_{\rm C} - g_0\bb1  )$ increases, which corresponds to a suboptimal condition of \eqref{eq.cond}, i.e., the gradients $\nabla_{h_0} J(\hat{h}_0,\hat{\balpha},\hat{\bbeta})\rightarrow 0$, $\nabla_{\balpha} J(\hat{h}_0,\hat{\balpha}) \rightarrow \mathbf{0}$ and $\nabla_{\bbeta} J(\hat{h}_0,\hat{\bbeta}) \rightarrow \mathbf{0}$. The proof completes.

\section{Proof of Proposition \ref{prop.chebyshev-bound}} \label{proof.chebyshev-bound}
Since the operator $\bbG$ corresponds to the desired continuous harmonic, gradient and curl frequency responses, it can be diagonalized by 
$\bbU_1$. Therefore, we have that 
\begin{equation}
  \begin{aligned}
    & \lVert \bbG - \bbH_{\rm{c}}\lVert_2 
    = \lVert \bbU_1 (g(\bLambda) - \tilde{H}_{1}(\bLambda)) \bbU_1^\top  \lVert_2 \\
    = & \lVert g(\bLambda) - \tilde{H}_{1}(\bLambda) \lVert_2 
    = \underset{i=1,\dots,N_1}{\max} |g(\lambda_i) - \tilde{H}_1(\lambda_i)|
  \end{aligned}
\end{equation}
where the diagonal matrix $g(\bLambda)$ has entries $g(\lambda_i) = g_0$ for $\lambda_i \in \ccalQ_{\rm{H}}$, $g(\lambda_i) = g_{\rm{G}}(\lambda_i)$ for $\lambda_i \in \ccalQ_{\rm{G}}$ and $g(\lambda_i) = g_{\rm{C}}(\lambda_i)$ for $\lambda_i \in \ccalQ_{\rm{C}}$. The frequency response $\tilde{H}_{1}(\lambda_i)$ for $\lambda_i\in\ccalQ$ is given in \eqref{eq.chebyshev-freq-response}. Moreover, based on the definition of $B_1(L_1)$ and $ B_2(L_2)$ we have that 
\begin{equation}
  \underset{i=1,\dots,N_1}{\max} |g(\lambda_i) - \tilde{H}_1(\lambda_i)| \leq \max \big\{ B_1(L_1), B_2(L_2) \big\} = B.
\end{equation}
The proof completes. 
 
\section{Proof of Lemma \ref{lemma.subcomponent-extraction-lemma}} \label{proof.subcomponent-extraction-lemma}
We first show the equivalence between the two projection operator forms. From the Hodge decomposition, the gradient, the curl and the harmonic components are in the subspaces $\im(\bbB_1^\top)$, $\im(\bbB_2)$ and $\ker(\bbL_1)$, respectively. Furthermore, from Proposition \ref{prop.correspondence}, we have that $\im(\bbB_1^\top) = \im(\bbU_{\rm{G}})$, $\im(\bbB_2) = \im(\bbU_{\rm{C}})$ and $\ker(\bbL_1) = \im(\bbU_{\rm{H}})$. Thus, each subcomponent can be obtained as the orthogonal projection of $\bbf$ onto the subspace spanned by the eigenbasis, i.e., the LS estimate. The gradient projection is $\bbP_{\rm{G}}:=\bbU_{\rm{G}}\bbU_{\rm{G}}^\top$, the curl one $\bbP_{\rm{C}}:=\bbU_{\rm{C}}\bbU_{\rm{C}}^\top$ and the harmonic one $\bbP_{\rm{H}}:=\bbU_{\rm{H}}\bbU_{\rm{H}}^\top$. This can be shown via the SFT as well, i.e., $\bbf_{\rm{G}} = \bbU_{\rm{G}}\tilde{\bbf}_{\rm{G}} = \bbU_{\rm{G}}\bbU_{\rm{G}}^\top\bbf$, likewise for the other two. 

Second, the simplicial filter $\bbH_1$ can implement the gradient projector $\bbP_{\rm{G}} = \bbU_{\rm{G}}\bbU_{\rm{G}}^\top$, if and only if we have that
\begin{equation} \label{eq.proof-projection-operator-1}
  \bbH_1 = \bbU_{\rm{G}}\bbU_{\rm{G}}^\top = [\bbU_{\rm{G}}^{\perp} \,\, \bbU_{\rm{G}}] 
    \begin{bmatrix}
      \mathbf{0} & \\ & \bbI_{N_{\rm{G}}} 
  \end{bmatrix}
     \begin{bmatrix}
    (\bbU_{\rm{G}}^\perp)^\top \\ \bbU_{\rm{G}}^\top 
  \end{bmatrix},
\end{equation}
with $\bbU_{\rm{G}}^\perp = [\bbU_{\rm{H}}\,\,\bbU_{\rm{C}}]$. We have that $\bbH_1 = \bbU_1 \tilde{\bbH}_1 \bbU_1^\top $ from \eqref{eq.freq-response-matrix}. Thus,  \eqref{eq.proof-projection-operator-1} is equivalent to problem \eqref{eq.ls-1} with $g_0=0$, $\bbg_{\rm{C}}=\mathbf{0}$ and $\bbg_{\rm{G}}=\bb1$. With $L_1=D_{\rm{G}}$ and $L_2=D_{\rm{C}}$, there admits a unique solution $\{h_0,\balpha,\bbeta\}$ to system \eqref{eq.ls-1}. Similar procedure can be followed for the implementation of the curl and harmonic projectors. The proof completes. 

\section{Proof of Corollary \ref{cor.subcomponent-extraction-cor}} \label{proof.cor.subcomponent-extraction-cor}
To implement the projector $\bbP_{\rm{G}}$ via $\bbH_1$ with $L_2=0$ ($\bbeta=\mathbf{0}$), from \eqref{eq.proof-projection-operator-1}, it is equivalent to set $g_0=0$ for $\lambda_i\in\ccalQ_{\rm{H}}\cup\ccalQ_{\rm{C}}$ and $g(\lambda_i)=1$ for $\lambda_{i}\in\ccalQ_{\rm{G}}$ in \eqref{eq.freq-response}, i.e., $\bbg = [0 \,\, \bb1_{D_{\rm{G}}}^\top]^\top$. This returns problem \eqref{eq.ls-1} without $\bPhi_{\rm{C}}$ and $\bbeta$, i.e.,
\begin{equation}\label{eq.ls-3}
  \underset{h_0,\balpha}{\min} \left\|  \begin{bmatrix} {\bf 1} \ \vline &  \begin{matrix} {\bf 0} \\  
  \bPhi_{\rm{G}} 
  \end{matrix}  \end{bmatrix} 
  \begin{bmatrix}
  h_0 \\ \balpha
  \end{bmatrix}
  - \bbg
  \right\|^2_2.
\end{equation}
If $L_1=D_{\rm{G}}$, the system matrix is square and any two rows are linearly independent, it admits a unique solution of $h_0=0$ and $\balpha=\bPhi_{\rm_G}^{-1}\bb1 $. Similar procedure can be followed for the curl projector $\bbP_{\rm{C}}$. The proof completes.

\end{document}